\documentclass[11pt,a4paper]{article}

\usepackage[utf8]{inputenc}
\usepackage[english]{babel}

\usepackage[table]{xcolor}
\usepackage{authblk}
\usepackage{geometry}
\usepackage{graphicx}
\usepackage{amssymb,amsmath,amsthm}
\usepackage{delarray}
\usepackage{enumitem}
\usepackage{tikz}
\usepackage{hyperref}

\graphicspath{{figures/}{pics/}{graphics/}}

\newtheorem{remark*}{\textbf{Remark}}

\newcommand{\N}{\mathbb{N}}
\newcommand{\Z}{\mathbb{Z}}
\newcommand{\R}{\mathbb{R}}

\newcommand{\Poly}{{\mathsf{P}}}
\newcommand{\NC}{{\mathsf{NC}}}

\newcommand{\degree}[1]{deg(#1)}

\newcommand{\stable}[1]{{#1}^{\circ}}
\newcommand{\transition}[1]{\overset{#1}{\rightarrow}}
\newcommand{\transitionsync}{\Rightarrow}
\newcommand{\odometer}[1]{{#1}}
\newcommand{\add}{{\,\oplus\,}}
\newcommand{\configs}{\mathcal{C}}
\newcommand{\stableconfigs}{\mathcal{C}_{\texttt{stab}}}
\newcommand{\recconfigs}{\mathcal{C}_{\texttt{rec}}}
\newcommand{\maxstable}{m}
\newcommand{\Tau}{\mathcal{T}}
\newcommand{\squaregrid}[2]{{#1}^{\tikz[scale=.2]{\draw (0,0) rectangle (1,1);}}_{{#2}}}
\newcommand{\sqrtthree}{1.7320508075688772}
\newcommand{\triangulargrid}[2]{{#1}^{\tikz[scale=.2]{\draw (0,0) -- (1,0) -- (.5,\sqrtthree/2) -- cycle;}}_{{#2}}}
\newcommand{\hexagonalgrid}[2]{{#1}^{\tikz[scale=.1]{\draw (-1,0) -- (-1/2,\sqrtthree/2) -- (1/2,\sqrtthree/2) -- (1,0) -- (1/2,-\sqrtthree/2) -- (-1/2,-\sqrtthree/2) -- cycle;}}_{{#2}}}
\newcommand{\coordinates}[1]{\texttt{coord}(#1)}
\newcommand{\surface}[1]{\texttt{body}(#1)}
\newcommand{\circleRadius}[1]{\texttt{B}(#1)}
\newcommand{\MSC}[1]{\texttt{MSC}(#1)}
\newcommand{\outerTiles}[1]{\texttt{outer}(#1)}
\newcommand{\innerTiles}[1]{\texttt{inner}(#1)}
\newcommand{\outerRadius}[1]{\underline{\texttt{r}}(#1)}
\newcommand{\innerRadius}[1]{\overline{\texttt{r}}(#1)}
\newcommand{\roundness}[1]{\texttt{r}(#1)}

\newcommand{\mytwolines}[2]{\begin{tabular}{@{}c@{}}{#1}\\[-.2em]{#2}\end{tabular}}

\title{Sandpile toppling on Penrose tilings:\\identity and isotropic dynamics}
\author[1]{J\'er\'emy Fersula}
\author[2]{Camille No\^us}
\author[1]{K\'evin Perrot}
\affil[1]{Universit\'e publique}
\affil[2]{Cogitamus laboratory}
\date{}

\begin{document}
\renewcommand{\labelitemi}{$\circ$}
\renewcommand{\labelitemii}{$\circ$}
\setlist[itemize,enumerate]{nosep}
\maketitle

\begin{abstract}
  We present experiments of sandpiles on grids (square, triangular, hexagonal)
  and Penrose tilings.
  The challenging part is to program such simulator;
  and our javacript code is available online, ready to play!
  We first present some identity elements of the sandpile group
  on these aperiodic structures, and then study the stabilization of the maximum
  stable configuration plus the identity, which lets a surprising circular shape
  appear. Roundness measurements reveal that the shapes are not approaching perfect
  circles, though they are close to be. We compare numerically this almost isotropic
  dynamical phenomenon on various tilings.
\end{abstract}

\section*{Preamble}

The experiments presented in this paper were conducted using {\em JS-Sandpile},
a javascript sandpile simulator we developed.
The code is available at\\
\centerline{\url{https://github.com/huacayacauh/JS-Sandpile},}
and it is ready for the reader to play at\\
\centerline{\url{https://huacayacauh.github.io/JS-Sandpile/}.}
The color codes of all our pictures are a gradation,
from {\em almost white} = 0 grain,
to {\em almost black} = $\degree{v}-1$ grains (for any vertex $v$),
{\em i.e.} the darker it is, the closer to the stability threshold.
Stable vertices have greyscale colors, and unstable vertices have flashy colors.
Table~\ref{table:colors} presents the color codes.

\begin{table}
  \definecolor{c3-0}{HTML}{eeeeee}
  \definecolor{c3-1}{HTML}{aaaaaa}
  \definecolor{c3-2}{HTML}{555555}
  \definecolor{c3-3}{HTML}{ff1a1a}
  \definecolor{c3-4}{HTML}{ff751a}
  \definecolor{c3-5}{HTML}{ffbb33}
  \definecolor{c3-6}{HTML}{ffff4d}
  \definecolor{c3-7}{HTML}{99ff66}
  \definecolor{c4-0}{HTML}{eeeeee}
  \definecolor{c4-1}{HTML}{bfbfbf}
  \definecolor{c4-2}{HTML}{7f7f7f}
  \definecolor{c4-3}{HTML}{3f3f3f}
  \definecolor{c4-4}{HTML}{ff1a1a}
  \definecolor{c4-5}{HTML}{ff751a}
  \definecolor{c4-6}{HTML}{ffbb33}
  \definecolor{c4-7}{HTML}{ffff4d}
  \definecolor{c6-0}{HTML}{eeeeee}
  \definecolor{c6-1}{HTML}{d4d4d4}
  \definecolor{c6-2}{HTML}{aaaaaa}
  \definecolor{c6-3}{HTML}{7f7f7f}
  \definecolor{c6-4}{HTML}{555555}
  \definecolor{c6-5}{HTML}{2a2a2a}
  \definecolor{c6-6}{HTML}{ff1a1a}
  \definecolor{c6-7}{HTML}{ff751a}
  \centerline{
    \begin{tabular}{|c|c|c|c|c|c|c|c|c|}
      \hline
      $\degree{v}$ & 0 {\tiny grain} & 1 {\tiny grain} & 2 {\tiny grains} & 3 {\tiny grains} & 4 {\tiny grains} & 5 {\tiny grains} & 6 {\tiny grains} & 7 {\tiny grains}\\
      \hline
      3 & \cellcolor{c3-0} & \cellcolor{c3-1} & \cellcolor{c3-2} & \cellcolor{c3-3} & \cellcolor{c3-4} & \cellcolor{c3-5} & \cellcolor{c3-6} & \cellcolor{c3-7}\\
      \hline
      4 & \cellcolor{c4-0} & \cellcolor{c4-1} & \cellcolor{c4-2} & \cellcolor{c4-3} & \cellcolor{c4-4} & \cellcolor{c4-5} & \cellcolor{c4-6} & \cellcolor{c4-7}\\
      \hline
      6 & \cellcolor{c6-0} & \cellcolor{c6-1} & \cellcolor{c6-2} & \cellcolor{c6-3} & \cellcolor{c6-4} & \cellcolor{c6-5} & \cellcolor{c6-6} & \cellcolor{c6-7}\\
      \hline
    \end{tabular}
  }
  \caption{
    Color codes of the pictures in this article,
    according to the degree of the vertices
    (all our tilings have the same degree for all its tiles/vertices).
  }
  \label{table:colors}
\end{table}

\section{Introduction}

It all started in 1987 with the work of Bak, Tang and Wiesenfled~\cite{btw87}.
Sandpiles have initially been introduced as number conserving cellular automata
on the two dimensional square grid, defined by the local toppling of sand grains
to neighboring cells, with statistics on chain reactions presenting
scale invariance typical of phase transition phenomena in physics,
the so called {\em self-organized criticality}~\cite{btw88,knwz89}.
Soon after, Dhar realized that sandpiles have a beautiful algebraic structure,
which generalizes to graphs and digraphs~\cite{d90}.
Since then, sandpiles have raised great interests for their
simple local definitions exhibiting complex global behaviors.
Following the work of Goles~\cite{g92},
numerous researches have been conducted on one-dimensional
models under sequential update mode~\cite{gk93,glmmp04,gmp02,p98},
parallel update mode~\cite{dl98},
and some variants such as symmetric~\cite{fmp06,fppt14,ppp11,p08}
or Kadanoff rule~\cite{p13}.

On the two-dimensional side,
the identity of the sandpile group on square grids retains its mysteries,
though the relaxation of hourglasses (toppling from a single site)
begin to reveal its structure through involved partial differential equation
developments~\cite{lps16,lps17,lp17,ps13,ps20}.

The sandpile model on graphs is general enough to embed arbitrary computation,
as attested by its Turing-universality~\cite{gm97}.
In order to analyse the apparent complexity of the model in a formal framework,
particular interest has been raised on the computational complexity of predicting the
dynamics of sandpiles. Despite strong efforts started with Moore and Nilsson
in~\cite{mn99} and important contributions from Goles, the computational
complexity of the problem on two dimensional grids remains
open between $\NC$ and $\Poly$.
The one-dimensional model is in $\NC$~\cite{m05,mn99},
as well as the Kadanoff rule~\cite{fgm10,fpr14}
and more general variants~\cite{fpr18}.
It is $\Poly$-complete from dimension three~\cite{mn99},
and even allows for some undecidable problems~\cite{c18}.
The (im)possibility of building crossing gates in two dimensions
is studied in~\cite{gg06,np18},
and similar issues on closely related threshold automata such as
the majority rule have been
fruitful~\cite{gmmo17,gm14,gm16,gmpt17,gmt13,m97}.
See~\cite{fp19} for a survey of the results on lattices.

Achieving isotropy in a cellular automaton is not a trivial matter,
as cells evolve on intrinsically anisotropic supports (grids).
The authors of~\cite{dmt99} manage to build parabolas and circles in a two-dimensional
cellular automaton with $5^{13}$ states.
Though optimizing the number of state was not an objective of their work,
it reflects the difficulty of the task.
Approximative (and simpler) approaches to build circles
(isotropic diffusion) include:
probabilistic methods~\cite{mh90,sm92,s97,w97},
using a continuous state space~\cite{m13},
or a large neighborhood to alleviate the
anisotropy~\cite{fe91,skt05,wtw92}.

We give the definition of the sandpile model on general graphs in
Section~\ref{s:model}, and present its algebraic structure at the heart of
the experiments conducted in this article. We also define the tilings
considered as supports for the sandpile dynamics: square, triangular and
hexagonal grids, Penrose tilings (kite-dart and rhombus) obtained by
substitution, and Penrose tilings (rhombus) obtained by the cut and project
(multigrid) method. Identity elements of the sandpile group on Penrose
tilings are exposed in Section~\ref{s:penroseid}. Our main contribution is
in Section~\ref{s:iso}, where we study numerically the isotropy observed during the
stabilization process from the maximum stable configuration plus the
identity.

\section{Model}
\label{s:model}

Bak, Tang and Wiesenfeld first defined in~\cite{btw87} the sandpile model on
two-dimensional grids with von Neumann neighborhood.
We present in Subsection~\ref{ss:sandpile} a general definition on any
undirected multi graph, in Subsection~\ref{ss:group} the algebraic structure first
revealed by Dhar in~\cite{d90}, and in Subsection~\ref{ss:tilings}
we introduce various grids and more generally tilings on which the sandpile
model can be studied.

\subsection{Sandpiles on graphs}
\label{ss:sandpile}

Given a (connected) finite undirected multi graph $G=(V,E)$ with a distinguished {\em sink} $s \in V$,
let $\tilde{V} = V \setminus \{s\}$. A {\em configuration} $c : \tilde{V} \to
\N$ assigns a number of sand grains to each non-sink vertex of $G$. The
basic local evolution rule is the {\em toppling} at some vertex $v \in
\tilde{V}$, which may occur when $c(v) \geq \degree{v}$, and consists in vertex
$v$ giving as many grains as it has edges to each of its neighbors.
When $c(v) \geq \degree{v}$ we say that vertex $v$ is {\em unstable},
and a configuration with no unstable vertex is called {\em stable}.
Formally, let $\Delta=D-A$ be the {\em graph Laplacian} of $G$, {\em i.e.} with
$D$ the degree matrix having $\degree{v}$ on the diagonal and $A$ the adjacency
matrix of $G$, and let $\tilde{\Delta}$ be the {\em reduced graph Laplacian}
obtained from $\Delta$ by deleting the row and column corresponding to the sink
$s$. With $\tilde{\Delta}_v$ the row of $\tilde{\Delta}$ corresponding to
vertex $v \in \tilde{V}$, performing a toppling at $v$ corresponds to going
from $c$ to $c'=c-\tilde{\Delta}_v$, which we denote $c \transition{v} c'$,
or simply $c \transition{} c'$, and $\transition{}^*$ its reflexive-transitive
closure.

As such the system is non-deterministic on configuration space $\N^{\tilde{V}}$,
but it is straightforward to notice that any configuration $c$ converges
(because of the absorbing sink vertex $s$) to a unique (because topplings
commute) stable configuration, denoted $\stable{c}$.
This is the so called {\em abelian property} of sandpiles,
or {\em fundamental theorem} of sandpiles.
The vector $\odometer{x}$ such that $c-\tilde{\Delta} \cdot \odometer{x} = \stable{c}$
is called the {\em shot vector} or {\em odometer function} of configuration $c$,
it records how many times each vertex has toppled during the {\em stabilization}
of $c$. The fundamental theorem of sandpiles furthermore states that
the odometer function of any configuration is unique.

\begin{figure}
  \centerline{
    \includegraphics[width=.12\textwidth]{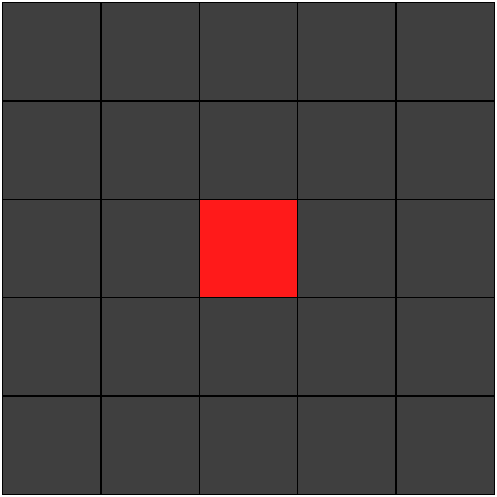}
    \raisebox{.8cm}{$\transitionsync$}
    \includegraphics[width=.12\textwidth]{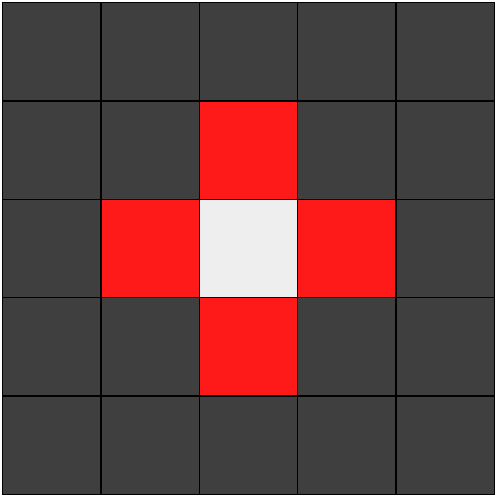}
    \raisebox{.8cm}{$\transitionsync$}
    \includegraphics[width=.12\textwidth]{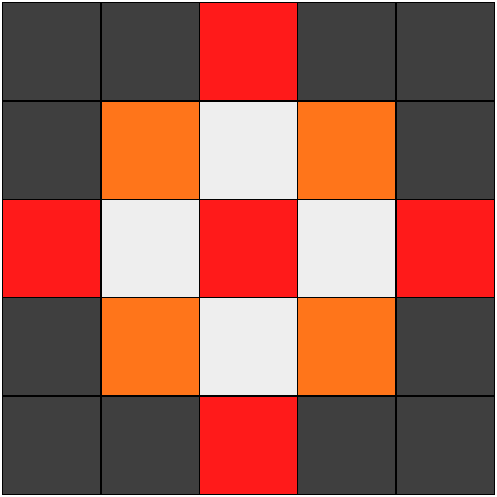}
    \raisebox{.8cm}{$\transitionsync$}
    \includegraphics[width=.12\textwidth]{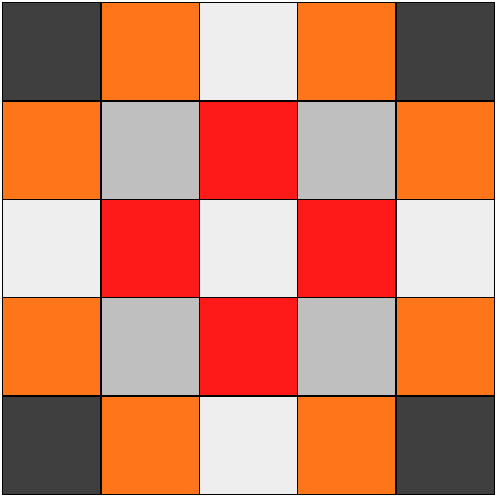}
    \raisebox{.8cm}{$\transitionsync$}
    \includegraphics[width=.12\textwidth]{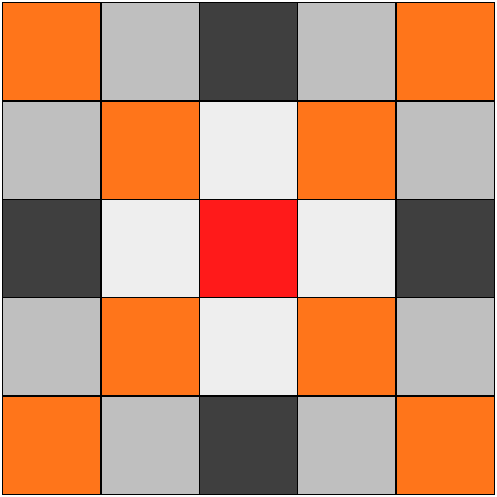}
    \raisebox{.8cm}{$\transitionsync$}
    \includegraphics[width=.12\textwidth]{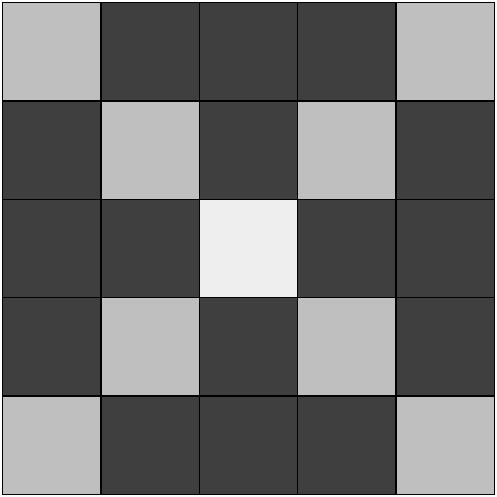}
  }
  \caption{
    Five steps of the sandpile model
    on a square grid of size $5 \times 5$, corresponding to
    the graph with one vertex per square and north-east-south-west
    adjacencies (all vertices have degree $4$). The sink $s$ is not
    pictured, squares/vertices on the border have one edge connected
    to $s$, and squares/vertices on the corners have two edges
    connected to $s$.
  }
  \label{fig:sandpile}
\end{figure}

For the purpose of simulations that take some importance in the present work,
we define the deterministic {\em synchronous} dynamics as the evolution
toppling synchronously all unstable vertices, at each step.
Formally, we define $c \transitionsync c'$ with
$$
  c'=c-\sum_{\substack{v \in \tilde{V}\\c(v) \geq \degree{v}}} \tilde{\Delta}_v.
$$
An example is given on Figure~\ref{fig:sandpile}.
For the sake of simplicity, in the following {\em graph} will stand for
{\em finite undirected multi graph}.

\subsection{Abelian group structure}
\label{ss:group}

Given a graph $G=(V,E)$ with sink $s$, let
$\configs=\N^{\tilde{V}}$ denote its set of configurations. The set of {\em
stable configurations} $\stableconfigs$ is naturally equipped with the {\em
operation} $\add$ defined as
$$c \add c' = \stable{(c + c')}$$
where $c + c'$ is the componentwise addition of two configurations,
{\em i.e.} $(c+c')(v)=c(v)+c'(v)$ for all $v \in \tilde{V}$.
It is straightforward to notice that $(\stableconfigs,\add)$ is a
{\em commutative monoid} (closure, associativity, identity, commutativity),
the identity being the configuration $z$ such that $z(v)=0$
for all $v \in \tilde{V}$.

\begin{figure}
  \centerline{$
    \vcenter{\hbox{\includegraphics[scale=.09]{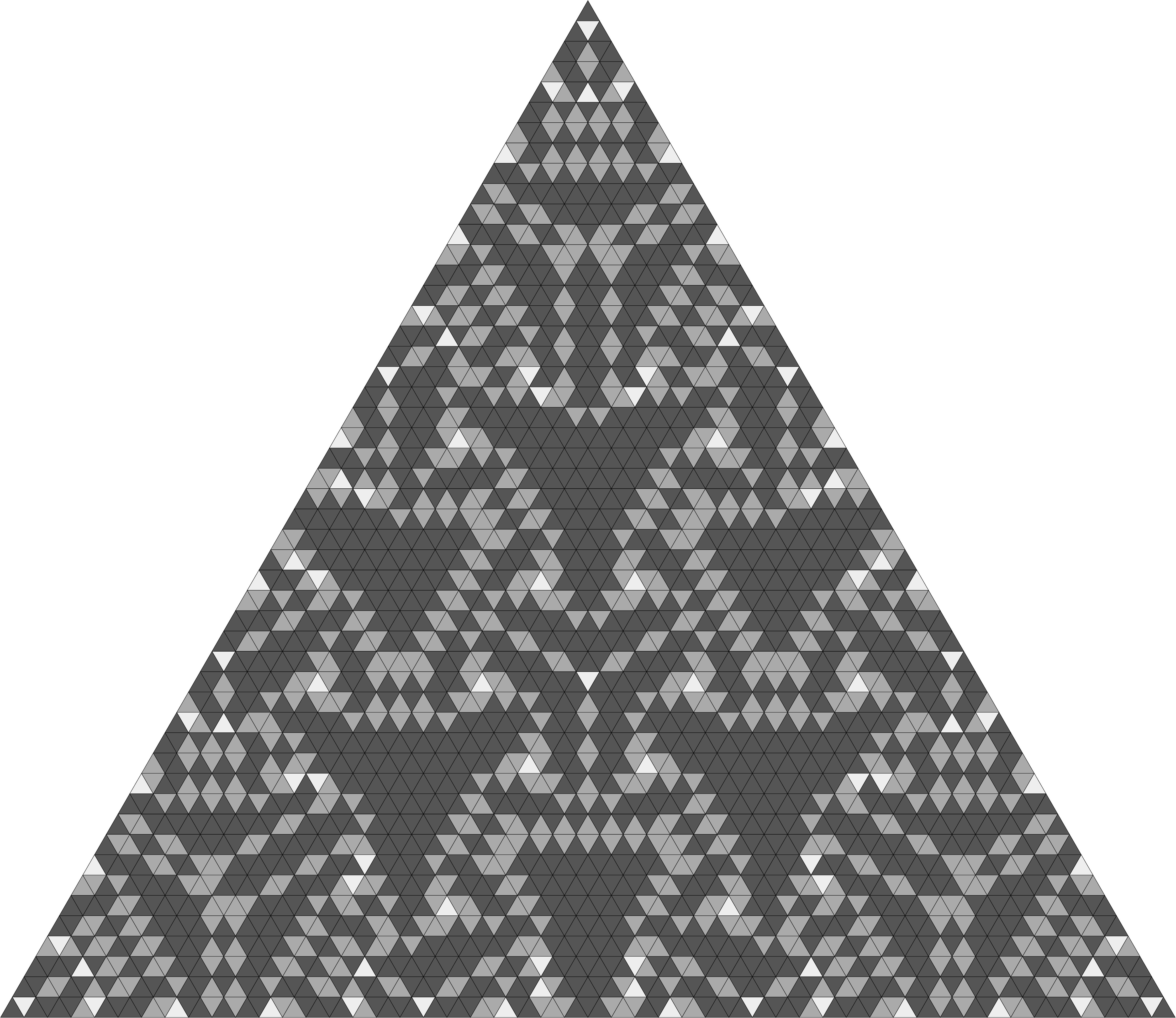}}}\quad
    \vcenter{\hbox{\includegraphics[scale=.09]{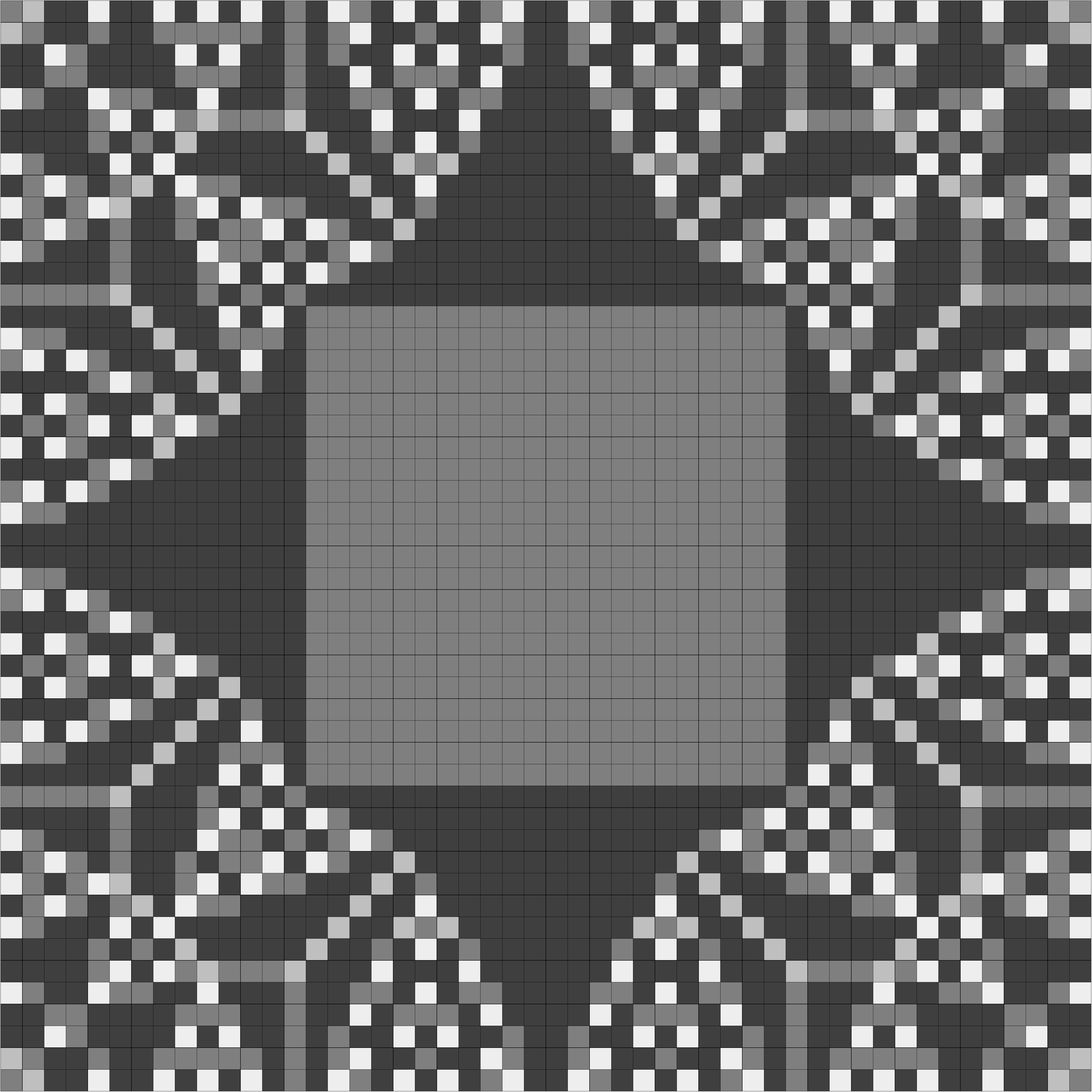}}}\quad
    \vcenter{\hbox{\includegraphics[scale=.09]{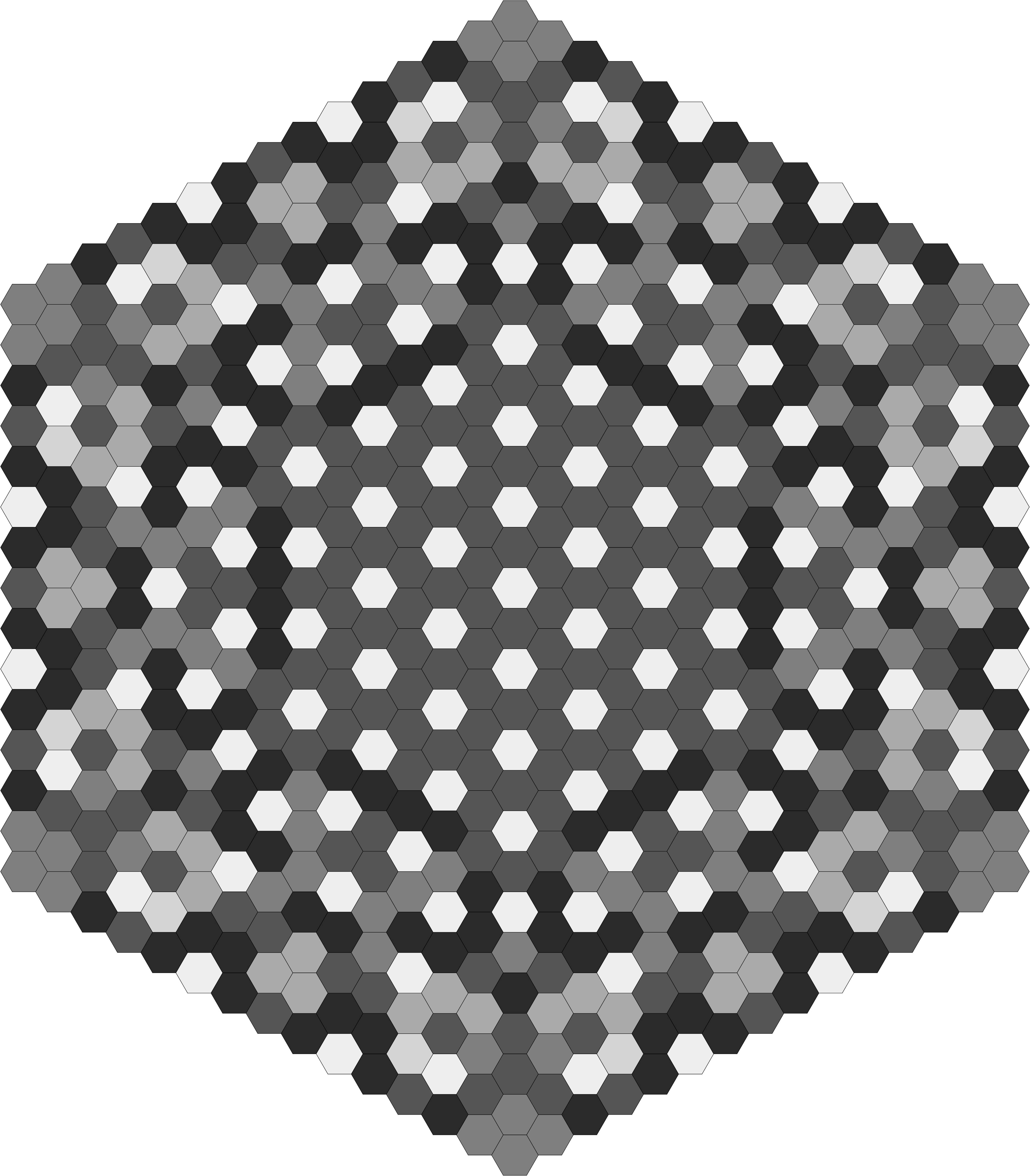}}}
  $}
  \caption{
    Identity elements of the sandpile group on graphs corresponding to
    the triangular grid of side length $50$ (left, $\degree{v}=3$ for any vertex $v$),
    the square grid of side length $50$ (center, $\degree{v}=4$ for any vertex $v$),
    the hexagonal grid of side length $14$ (right, $\degree{v}=6$ for any vertex $v$). All at the same scale.
  }
  \label{fig:identities}
\end{figure}

Here comes the magics of sandpiles. The set of {\em recurrent configurations}
\begin{align*}
  \recconfigs
  =& \{ c \in \stableconfigs \mid \forall c' \in \configs : \exists c'' \in \configs : c' \add c'' = c \}\\
  =& \{ c \in \stableconfigs \mid \forall c' \in \stableconfigs : \exists c'' \in \stableconfigs : c' \add c'' = c \}
\end{align*}
corresponds to the intersection of ideals of $\stableconfigs$, {\em i.e.}
$$
  \recconfigs=\bigcap_{\substack{I \subseteq \stableconfigs\\ I \text{ ideal of } \stableconfigs}} I
$$
with $I$ (right) {\em ideal} of $\stableconfigs$ if and only if
$I \add \stableconfigs \subseteq \stableconfigs$ where
$I \add \stableconfigs = \{ i \add c \mid i \in I \text{ and } c \in \stableconfigs \}$.
Moreover, it is a classical result of algebra that the intersection of all ideals of
a {\em commutative semigroup} (closure, associativity, commutativity)
gives an {\em abelian group} (closure, associativity, identity, inverse, commutativity).
So $(\recconfigs,\add)$ is an abelian group, called the {\em sandpile group} on graph $G$
with sink $s$. Now remark that the configuration $z$ containing no grain is
(except on very restricted cases) not an element of $\recconfigs$, hence
the\footnote{The identity element of the sandpile group is unique.}
identity element $e \in \recconfigs$ of the sandpile group is {\em a priori}
not obvious to construct, and it turns out that few is know about its structure
on numerous interesting cases, as presented on Figure~\ref{fig:identities}.
It can be proven that
\begin{equation}
  \label{eq:id}
  e=\stable{(2\maxstable - \stable{(2\maxstable)})}
\end{equation}
with $m$ the {\em maximum stable} configuration defined as $m(v)=\degree{v}-1$
for all $v \in \tilde{V}$, and $(2m)(v)=2\,c(v)$ for all $v \in \tilde{V}$.
Indeed, $2m-\stable{(2\maxstable)}$ contains at least $\degree{v}-1$ grains at
each vertex $v \in \tilde{V}$ hence its stabilization is recurrent, and
furthermore it corresponds to subtracting two configurations from the same
equivalence class according to relation $\leftrightarrow^*$
(the symmetric closure of $\transition{}^*$), therefore to a
configuration in the class of the identity (see~\cite{drsv95} for details).
The equality follows since the identity element of $(\recconfigs,\add)$ is unique.

\subsection{Tilings}
\label{ss:tilings}

We will concentrate on finite tilings, but still introduce general definitions.
An {\em infinite tiling} $\Tau$ by $\tau$ is a covering of $\R^2$ by finitely
many polygonal {\em tiles} and their images by isometry (translation, rotation,
flip), {\em i.e.} copies of the tiles from $\tau$ cover the plane without gaps
nor overlaps. Let $\tau$ be a finite set of polygonal tiles called a {\em tile
set}, then $\Tau$ is a partition of $\R^2$ into countably many isometries of
the elements from $\tau$.

An infinite tiling $\Tau$ is {\em periodic} when it has a non-null periodicity
vector $\vec{p} \in \R^2$, such that $\Tau+\vec{p}=\Tau$. A tile set $\tau$ is
{\em aperiodic} when it admits at least one infinite tiling (we say that $\tau$
{\em tiles the plane}), and none are periodic. Aperiodicity is fundamentally
related to the uncomputability of the {\em domino problem}: given a (finite)
tile set, does it tile the plane? Let us quickly mention the seminal
contributions of Wang~\cite{w61}, Berger~\cite{b66} and Robinson~\cite{r71},
along with the book {\em Tilings and Patterns} by Gr\"{u}nbaum and
Shephard~\cite{gs86}.

In order to consider sandpiles on tilings, we explain now how to construct
finite undirected multi graphs with a distinguished sink.
A {\em finite tiling} is simply a subset of some infinite
tiling $\Tau$. Remark that this requires a finite tiling to be {\em extensible}
into an infinite tiling, which will be the case for all our finite tilings.
Indeed, given some tile set $\tau$ we will generate arbitrarily large finite
tilings, which implies the existence of an infinite tiling by compactness of
the set of infinite tilings by $\tau$.
Two tiles are {\em adjacent} in $\Tau$ when they share an edge.
All the tilings we consider will see their adjacent tiles share full sides,
{\em i.e.} no tile will share a partial side or more than one side with another
tile. These are called {\em edge-to-edge} tilings.
Given a finite tiling, each tile corresponds to a vertex, plus an additional
sink vertex corresponding to the {\em outside face}. The tile to tile
adjacencies are given by the edge-to-edge connections, and the number of edges
connecting a tile to the sink is equal to its number of sides connected to the
outside face, so that the degree of every tile is equal to its number of sides.
Tiles connected to the sink are said to be on the {\em border} of the Tiling.
Note that the sandpile dynamics is given by this underlying graph, but
tiles furthermore have coordinates in $\R^2$.

We now present the finite tilings considered in this article.
The size of tiles and the position of coordinate $(0,0)$
will be important for the observations presented in Section~\ref{s:iso}.
Grids are illustrated on Figure~\ref{fig:identities}.

\paragraph{Square grids.} They correspond to the original two-dimensional
sandpile model by Bak, Tang and Wiesenfled~\cite{btw87}. Given size $n$,
it basically consists in a $n \times n$ square grid with
adjacencies given by von Neumann neighborhood (north, east, south, west). Tiles
on the borders have one edge connected to the sink, and tiles on the corners have
two edges connected to the sink.
Each tile is a square of side length $1$, and the finite tiling is centered
with coordinate $(0,0)$ in the middle of the grid: if $n$ is even then
coordinate $(0,0)$ corresponds to four tiles corners; otherwise
coordinate $(0,0)$ corresponds to the center of a tile.

\paragraph{Triangular grids.} Given size $n$, it is made of equilateral
triangles of side length $1$, arranged up and down to form an equilateral
triangle of side length $n$, where the three outer sides of the tiling are made
of $n$ triangular tiles. Tiles on the border have one edge connected to the
sink, and tiles on the corners have two edges connected to the sink. The finite
tiling is centered with coordinate $(0,0)$ at the barycenter of its three
corners.

\paragraph{Hexagonal grids.} Given size $n$, it is made of regular hexagons of
side length $1$, arranged to form an hexagonal grid (orientation is {\em flat})
with six sides each made of $n$ tiles.
Tiles on the border have two edges connected to the sink, and tiles on corners
have three edges connected to the sink.
The finite tiling is centered with coordinate $(0,0)$ at the center of the central hexagon.

\begin{figure}
  \centerline{
    \includegraphics{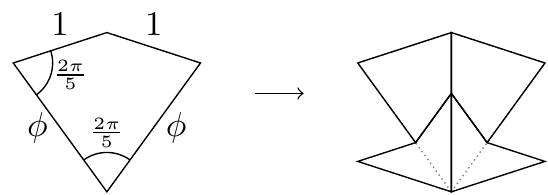}
    \hspace*{1.5cm}
    \includegraphics{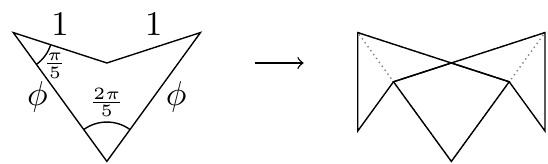}
  }
  \caption{
    Substitutions of P2 kite (left) and dart (right) tiles.
  }
  \label{fig:P2-subst}
  \vspace*{1em}
  \centerline{
    \includegraphics{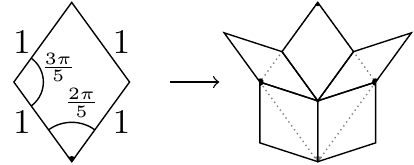}
    \hspace*{1.5cm}
    \includegraphics{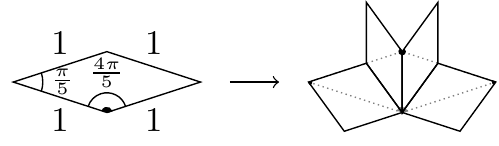}
  }
  \caption{
    Substitutions of P3 fat (left) and thin (right) rhombi tiles.
    Due to the symmetry of P3 tiles,
    we highlight the origin point of each tile.
  }
  \label{fig:P3-subst}
\end{figure}

\begin{figure}
  \centerline{$
    \vcenter{\hbox{\includegraphics[scale=.3]{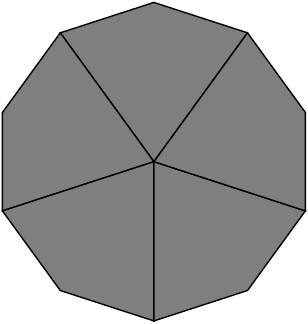}}}\quad
    \vcenter{\hbox{\includegraphics[scale=.3]{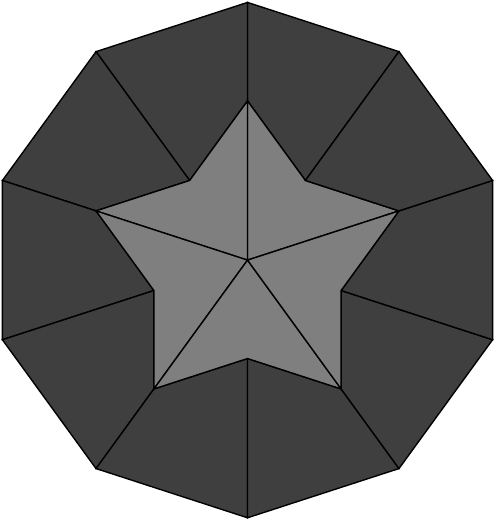}}}\quad
    \vcenter{\hbox{\includegraphics[scale=.3]{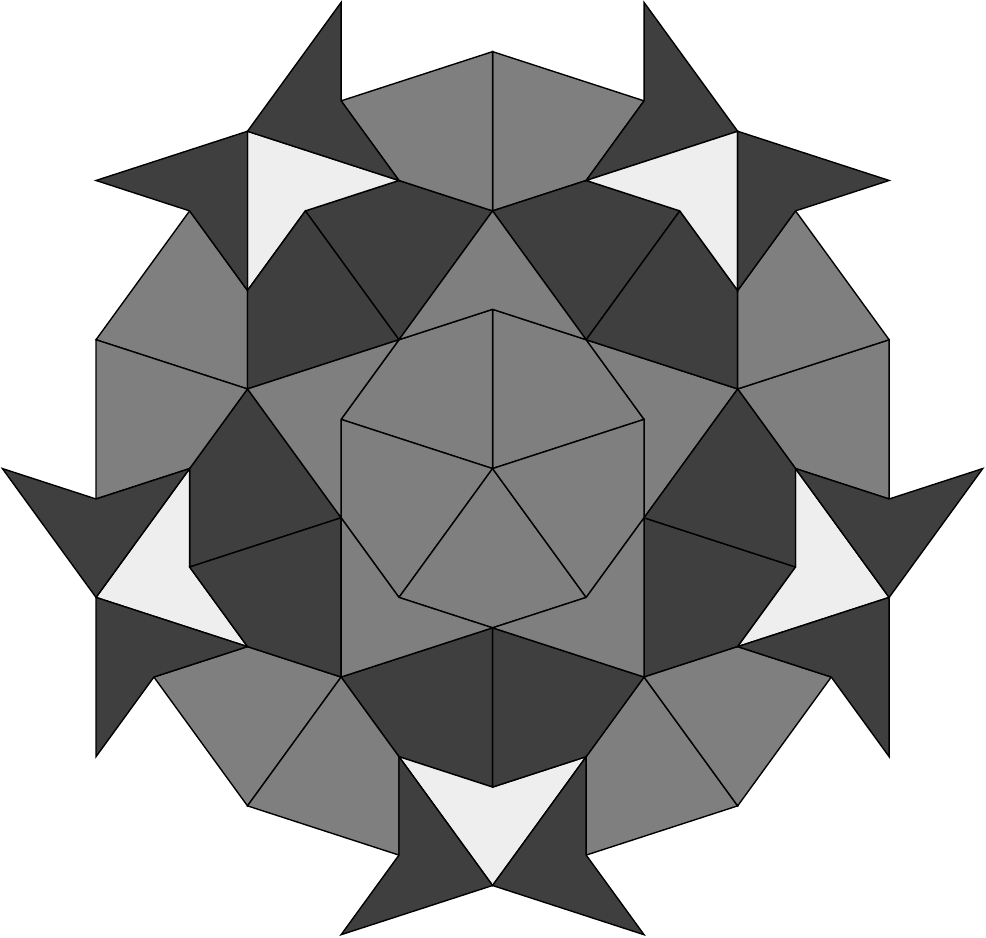}}}\quad
    \vcenter{\hbox{\includegraphics[scale=.3]{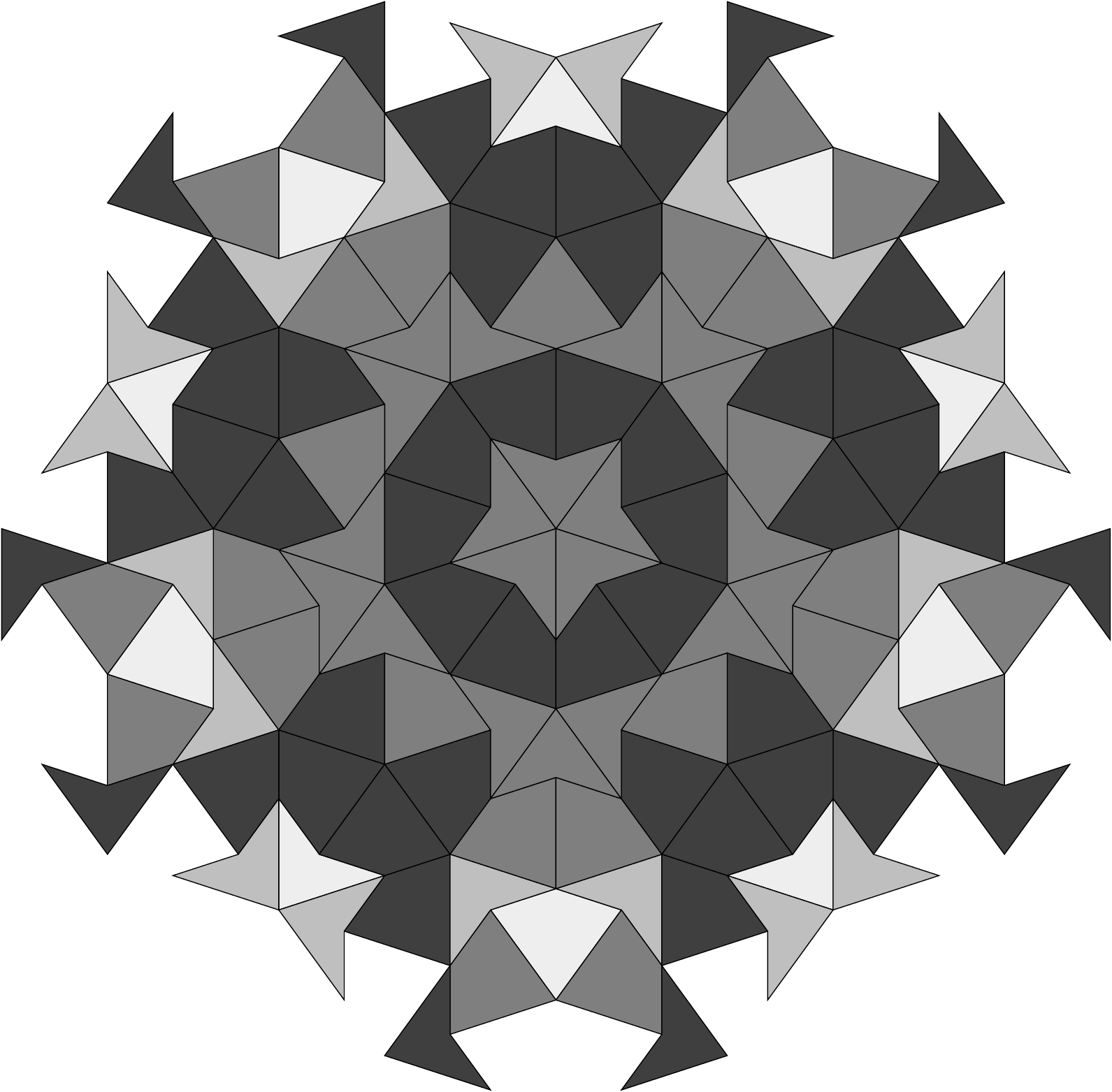}}}
  $}
  \caption{
    Three iterations of the susbtitution from a {\em P2 Sun},
    with identity elements of the sandpile group pictured.
  }
  \label{fig:P2-sun}
  \vspace*{1em}
  \centerline{$
    \vcenter{\hbox{\includegraphics[scale=.3]{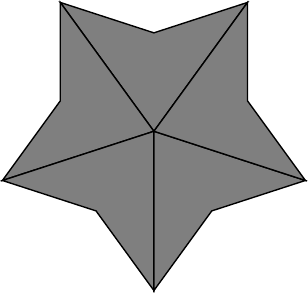}}}\quad
    \vcenter{\hbox{\includegraphics[scale=.3]{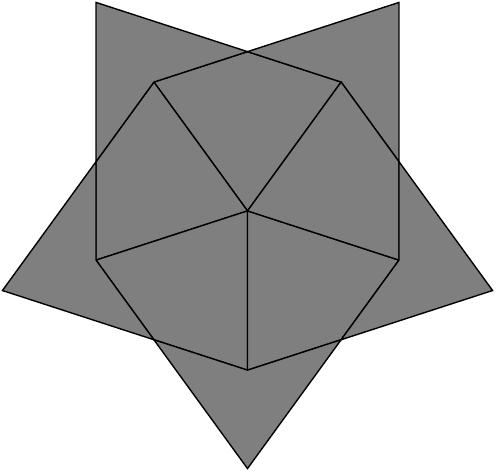}}}\quad
    \vcenter{\hbox{\includegraphics[scale=.3]{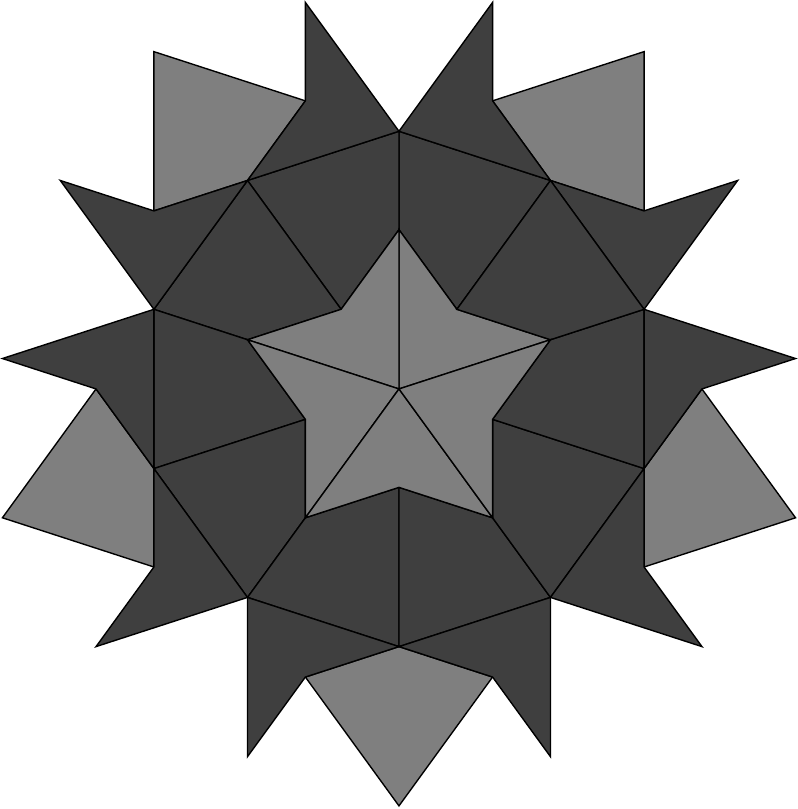}}}\quad
    \vcenter{\hbox{\includegraphics[scale=.3]{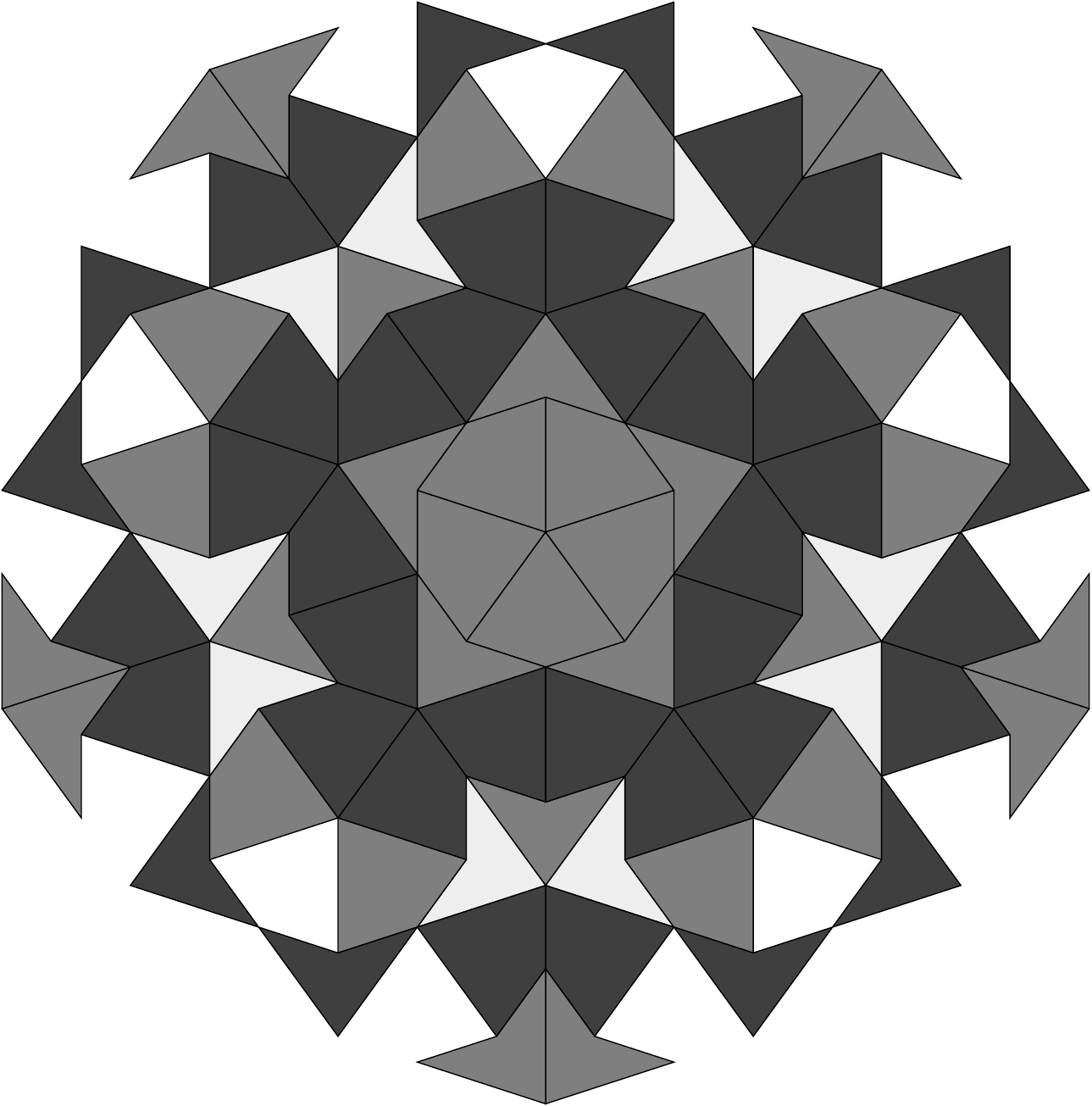}}}
  $}
  \caption{
    Three iterations of the susbtitution from a {\em P2 Star},
    with identity elements of the sandpile group pictured.
  }
  \label{fig:P2-star}
  \vspace*{1em}
  \centerline{$
    \vcenter{\hbox{\includegraphics[scale=.3]{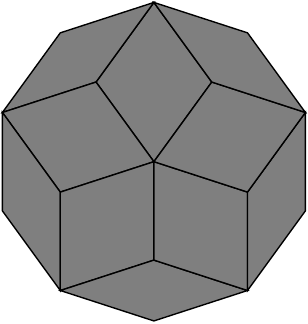}}}\quad
    \vcenter{\hbox{\includegraphics[scale=.3]{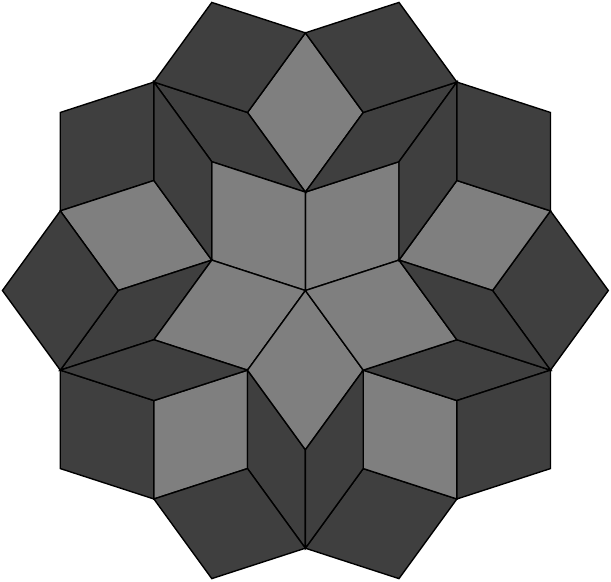}}}\quad
    \vcenter{\hbox{\includegraphics[scale=.3]{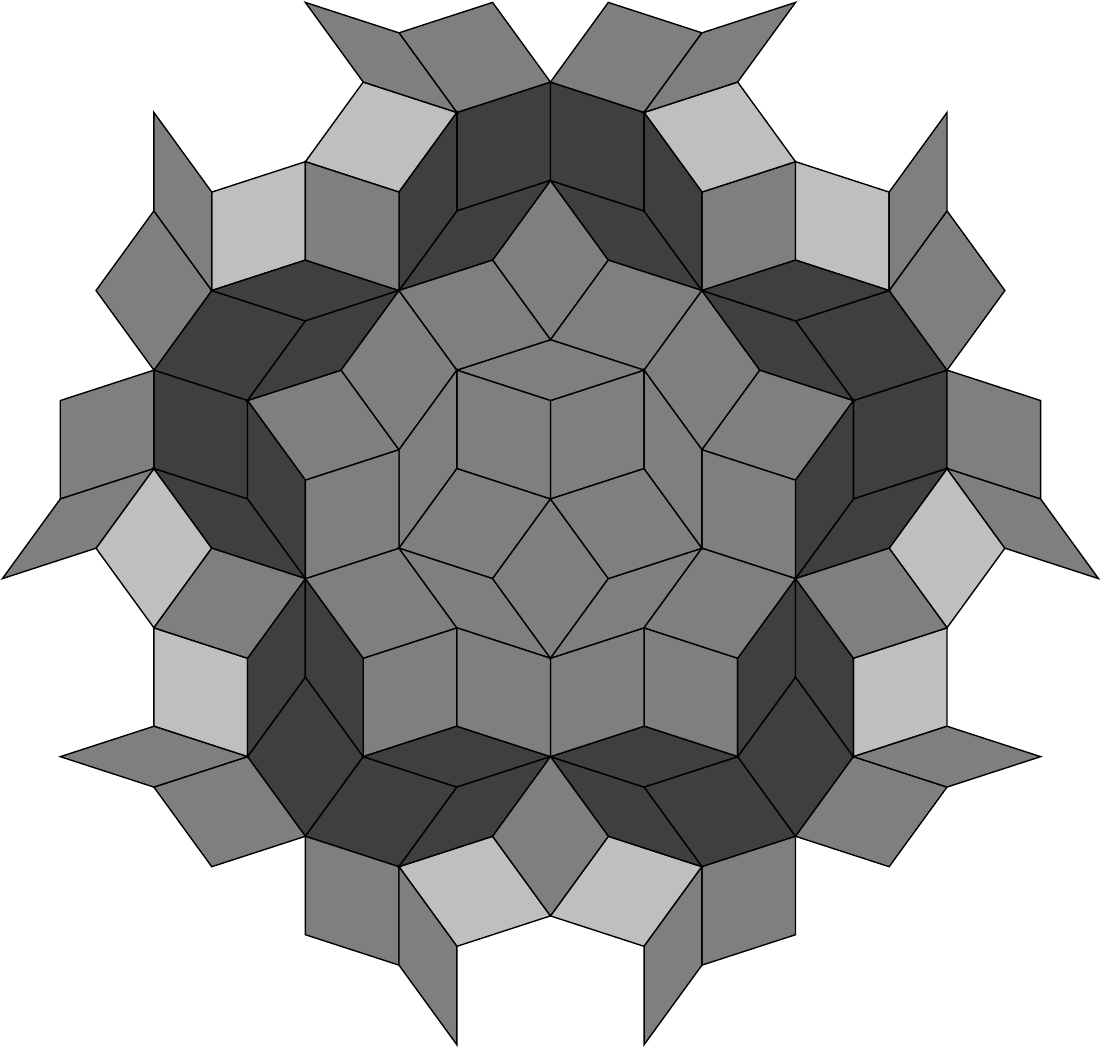}}}\quad
    \vcenter{\hbox{\includegraphics[scale=.3]{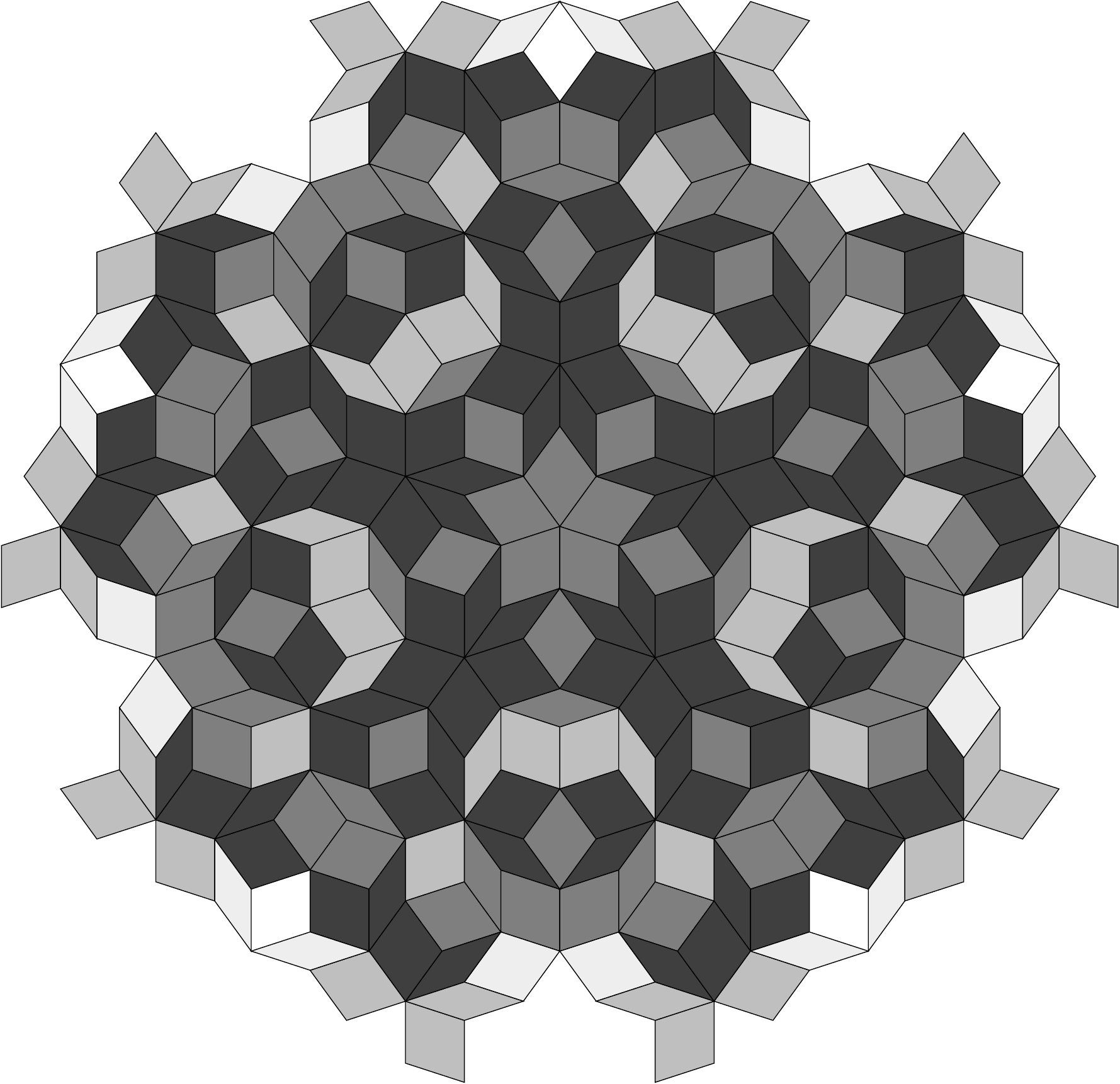}}}
  $}
  \caption{
    Three iterations of the susbtitution from a {\em P3 Sun},
    with identity elements of the sandpile group pictured.
  }
  \label{fig:P3-sun}
\end{figure}

\paragraph{Penrose tilings.} Penrose developed in~\cite{p74,p79}
a series of elegant aperiodic tile sets.
We consider {\em P2} ({\em kite-dart}), and {\em P3} ({\em rhombus})
(tilings by P2 and P3 are {\em mutually locally derivable},
see~\cite{bsj91}).
Penrose tilings may be obtained by
{\em substitution}\footnote{Note that in order to enforce aperiodicity in P2 and P3,
matching constraints should be added on tile edges,
for example via notches, but the finite tiling generation methods we employ
do not require such considerations.}
as described on
Figure~\ref{fig:P2-subst} for P2, and Figure~\ref{fig:P3-subst} for P3.
After substituting, we rescale all tiles up by the substitution factor
$\phi=\frac{1+\sqrt{5}}{2}$, so that the tiles of the final tiling have the exact
same size as the base tiles (kite, dart, fat, thin).
Figure~\ref{fig:P2-sun} presents three iterations of the substitution from a {\em P2 Sun},
Figure~\ref{fig:P2-star} presents three iterations of the substitution from a {\em P2 Star},
and Figure~\ref{fig:P3-sun} presents three iterations of the substitution from a {\em P3 Sun}.
Coordinate $(0,0)$ is at the center of the initial Suns and Stars, and remains at the symmetry center of tilings obtained by subsequent substitutions.

\begin{figure}
  \centerline{$
    \vcenter{\hbox{\includegraphics[scale=.25]{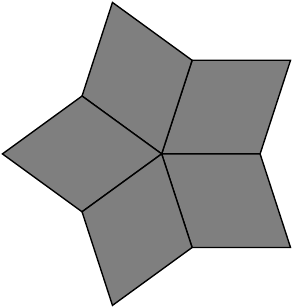}}}\quad
    \vcenter{\hbox{\includegraphics[scale=.25]{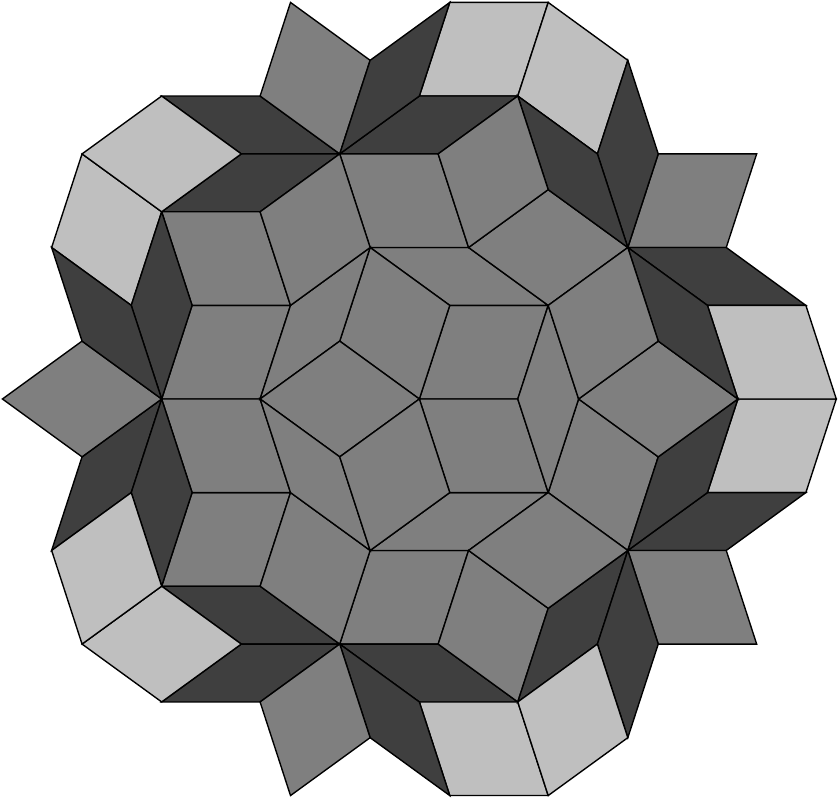}}}\quad
    \vcenter{\hbox{\includegraphics[scale=.25]{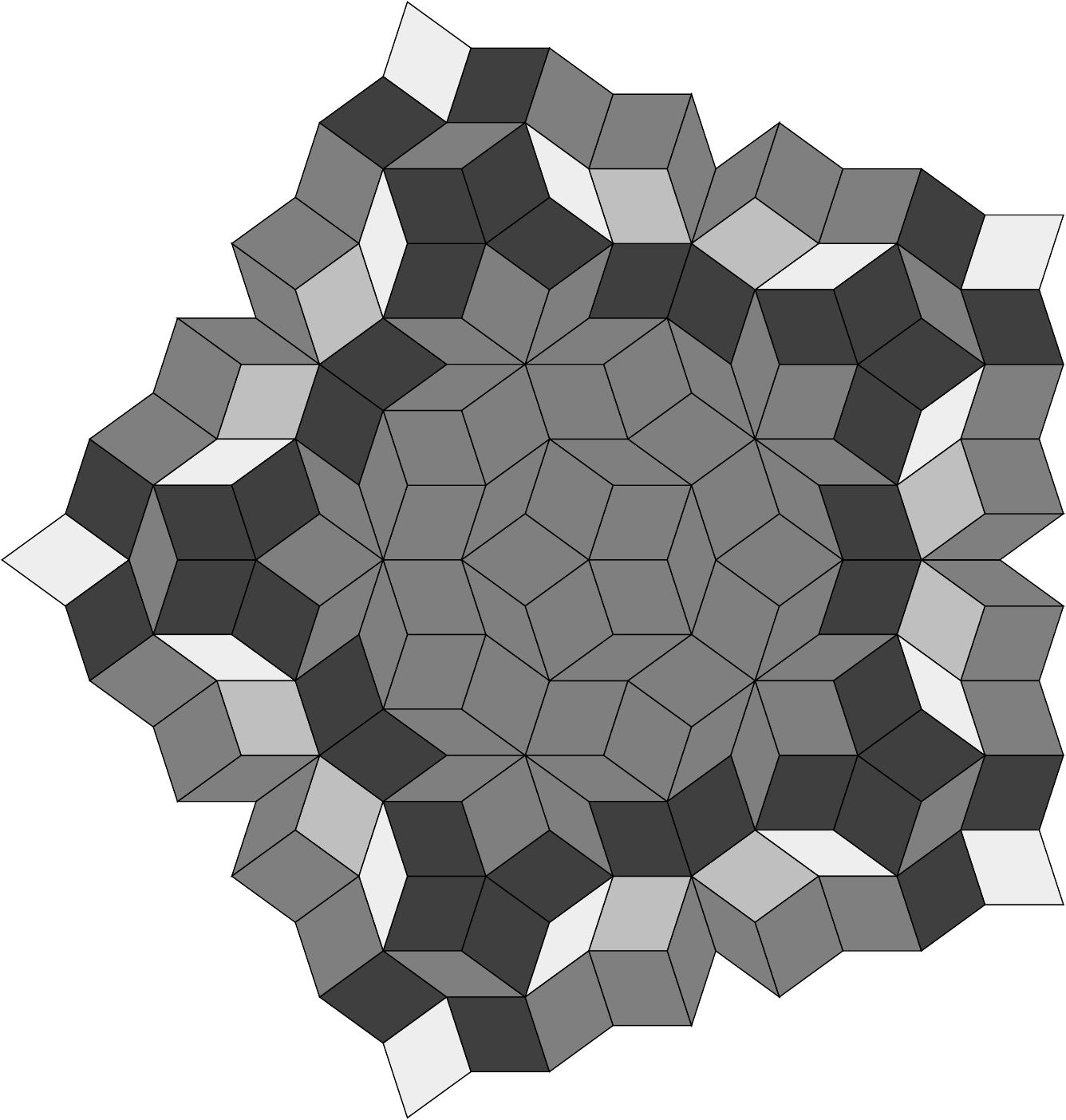}}}\quad
    \vcenter{\hbox{\includegraphics[scale=.25]{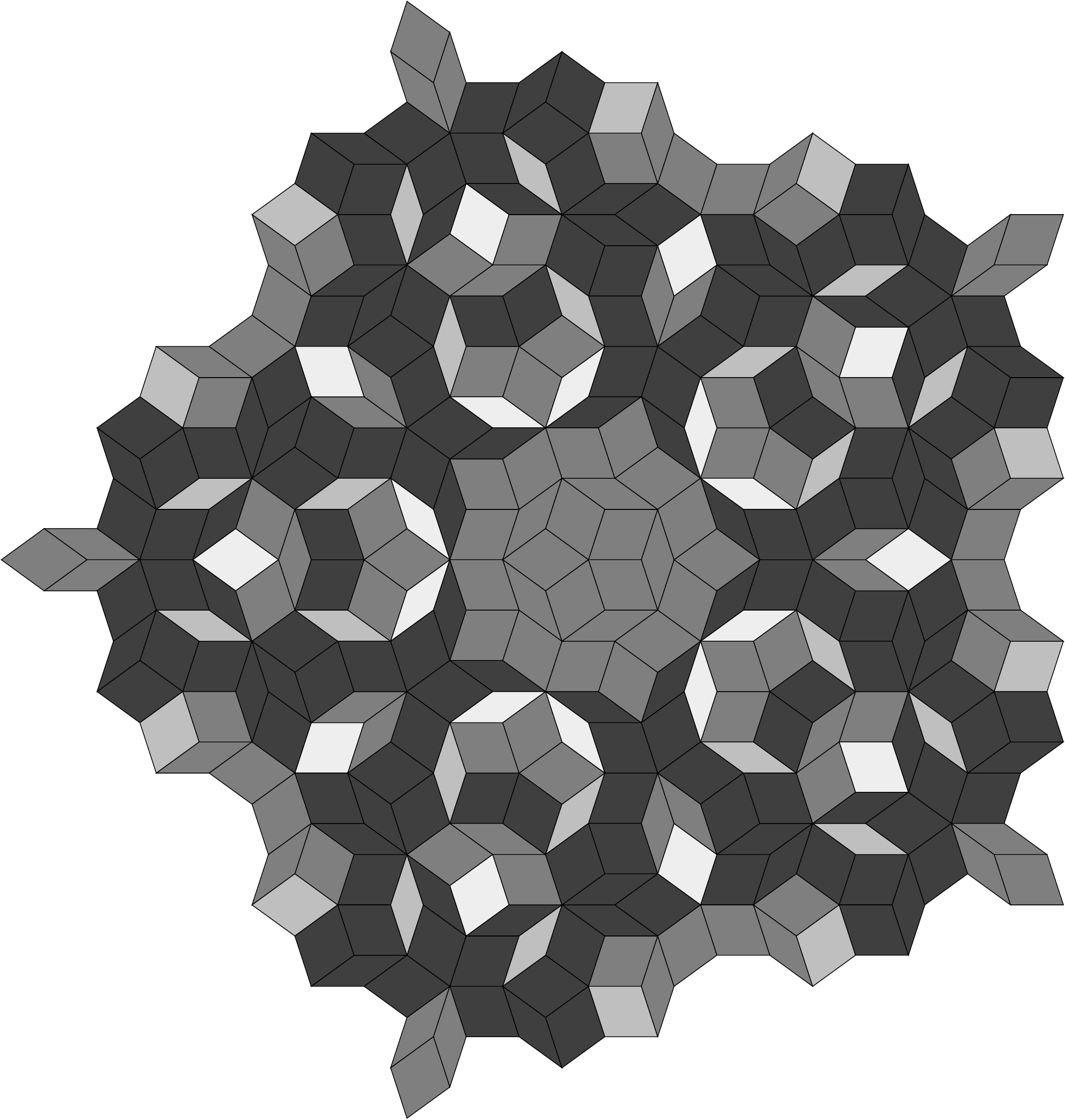}}}
  $}
  \caption{
    P3 cut and project tilings of sizes $1$, $2$, $3$ and $4$,
    with identity elements of the sandpile group pictured.
  }
  \label{fig:P3-cutproject}
\end{figure}

Penrose P3 tilings may alternatively be generated by {\em cut and project} method:
consider the 5-dimensional plane $E$ spaned by vectors
$( \cos \frac{2k\pi}{5} )_{0 \leq k < 5}$
and
$( \sin \frac{2k\pi}{5} )_{0 \leq k < 5}$,
passing through the point
$(\frac{1}{5},\frac{1}{5},\frac{1}{5},\frac{1}{5},\frac{1}{5})$.
The orthogonal projection of the 5-dimensional grid lines (1-simplices)
contained in $E+[0,1]^5$ onto $E$ gives a tiling by P3 tiles.
We implement it via the dual {\em multigrid} method of de Bruijn~\cite{db81},
where one considers a 2-dimensional space and five line families (our {\em pentagrid})
given by the intersections of $E$ with the five 5-dimensional hyperplanes
$G_i = \{ x \in \R^5 \mid x \cdot e_k \in \Z \}$ for $0 \leq k < 5$,
where $e_k$ is the $k$-th unit vector of $\R^5$.
Pentagrid lines of a family are {\em indexed} by the value of $x \cdot e_k$,
and we denote $\ell^k_i$ the line from family $0 \leq k < 5$ of index $i \in \Z$.
Note that with this setting, no more than two pentagrid lines 
intersect at a given position.
The 2-dimensional space is divided into polygonal {\em cells}
delimited by pentagrid lines. Each cell $p$ is labeled by a $5$-tuple of integers
$(p_k)_{0 \leq k <5}$ such that cell $p$ lies in between lines
$\ell^k_{p_k}$ and $\ell^k_{p_k+1}$.
To each cell $p$ corresponds a point (a tile's bound coordinate)
in the 2-dimensional space,
at $\sum_{0 \leq k < 5} p_k(\cos \frac{2k\pi}{5},\sin \frac{2k\pi}{5})$.
It follows that to each intersecting pair of pentagrid lines corresponds a
tile, whose two edge orientations are given by the two line families.
Details can be found in~\cite{f07}.
We bound this process to a finite {\em P3 cut and project tiling of size $n$}
by considering only tiles corresponding to points of intersection lying in between
$\ell^i_{-n}$ and $\ell^i_{n}$ for all families $0 \leq i < 5$.
The coordinate $(0,0)$ of this tiling is given by the cell labeled $(0,0,0,0,0)$,
and is therefore a tile bound (shared by multiple tiles).
Rhombus tiles have side length $1$ as on Figure~\ref{fig:P3-subst}.
Figure~\ref{fig:P3-cutproject} presents some P3 cut and project tilings.

\section{Sandpile identity on Penrose tilings}
\label{s:penroseid}

\begin{figure}
  \centerline{\includegraphics[width=\textwidth]{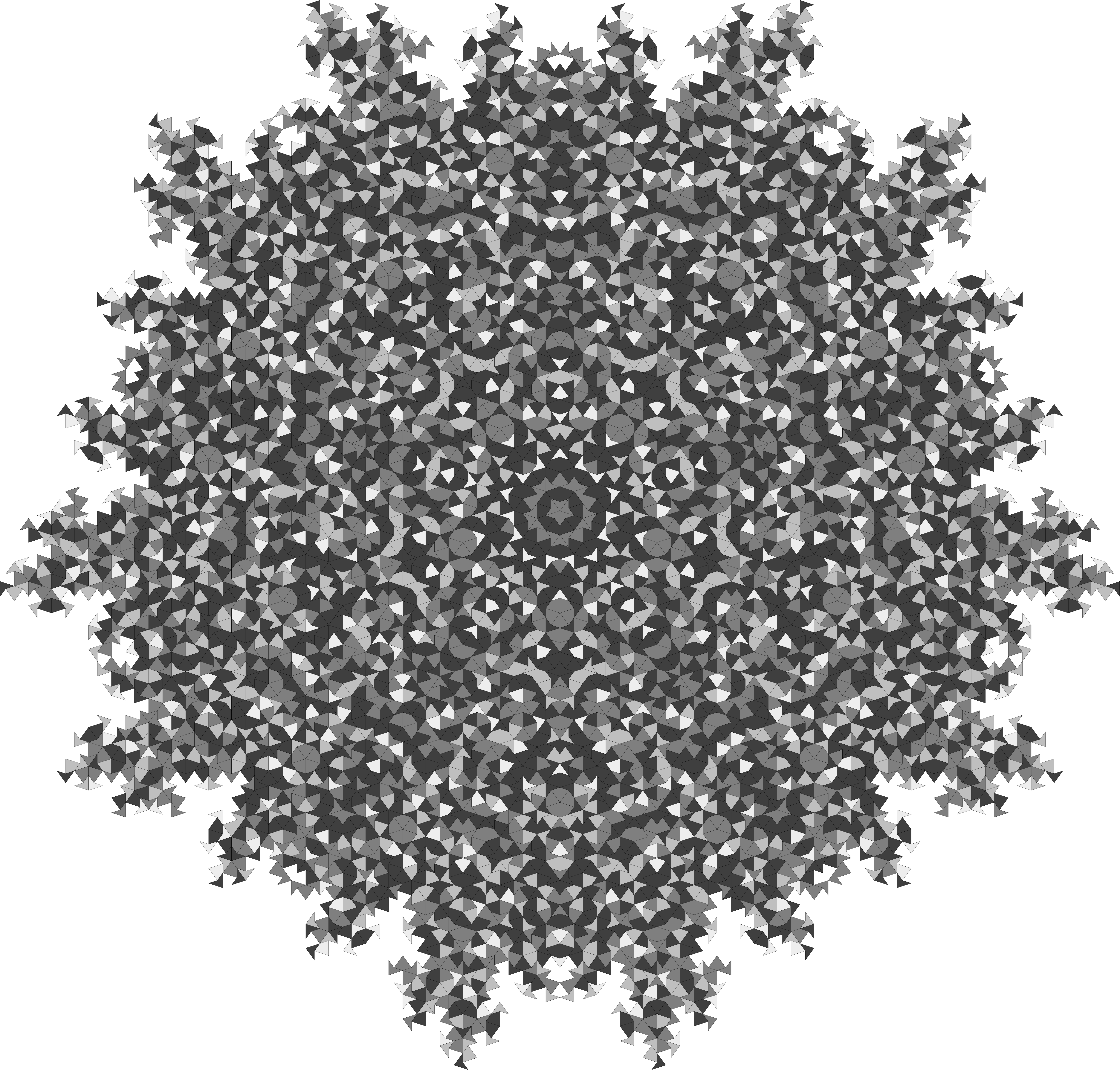}}
  \caption{
    Identity element of the sandpile group on
    the tiling obtained after 7 iterations of the substitution
    from a P2 Sun ($6710$ tiles).
  }
  \label{fig:id-P2-sun-7}
\end{figure}

\begin{figure}
  \centerline{\includegraphics[width=\textwidth]{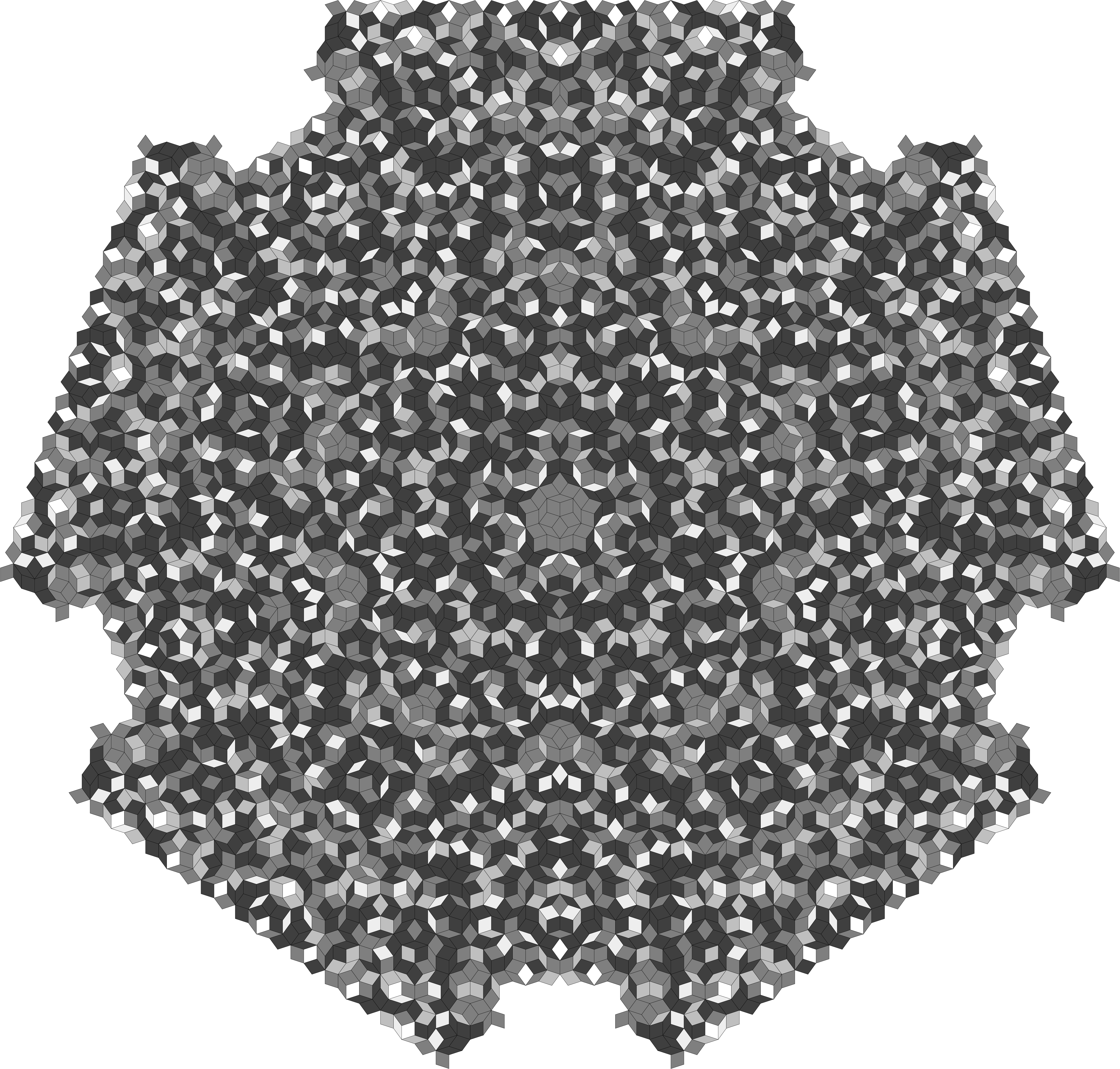}}
  \caption{
    Identity element of the sandpile group on
    the tiling obtained after 6 iterations of the substitution
    from a P3 Sun ($5415$ tiles).
  }
  \label{fig:id-P3-sun-6}
\end{figure}

\begin{figure}
  \centerline{\includegraphics[width=\textwidth]{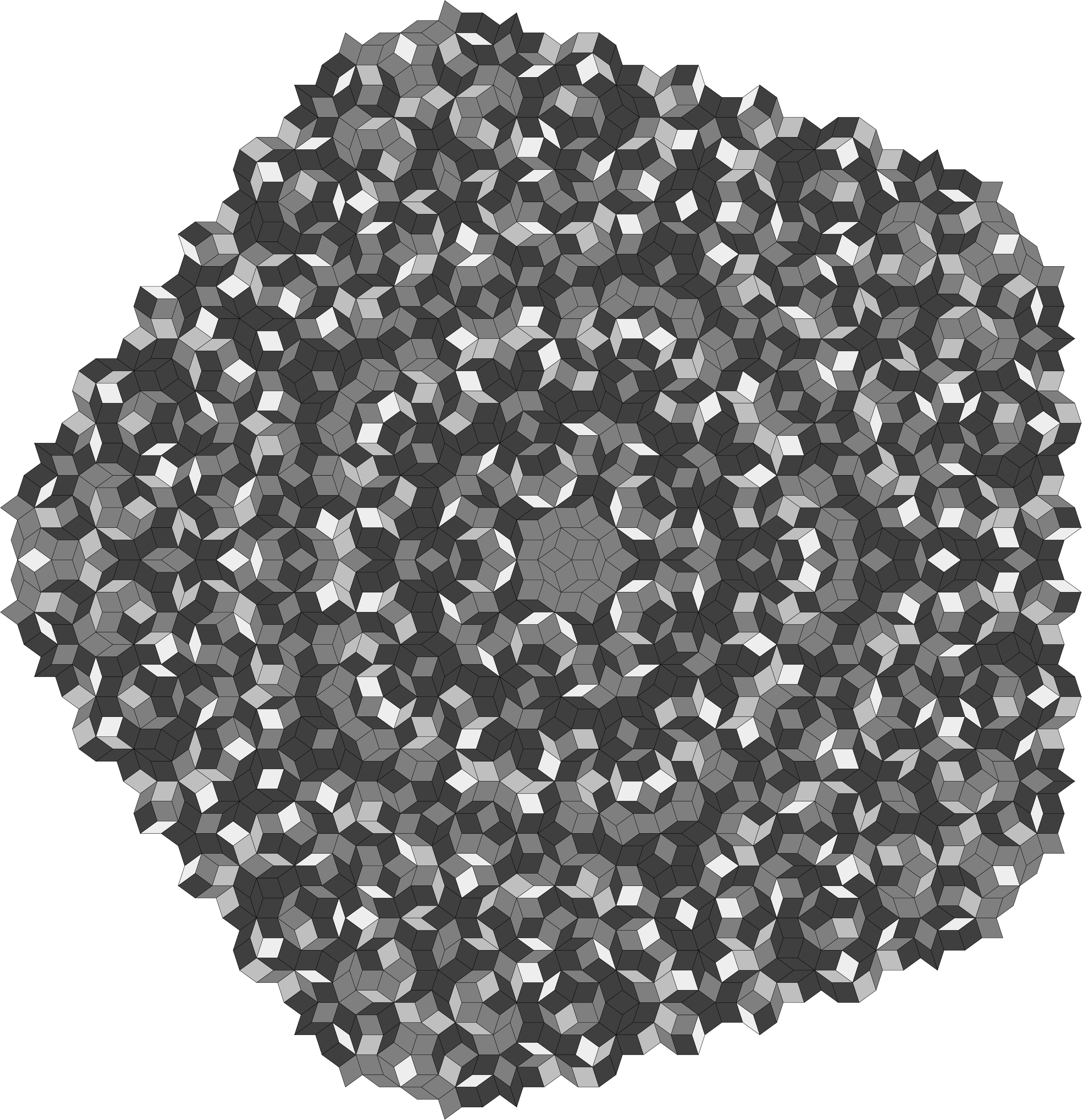}}
  \caption{
    Identity element of the sandpile group on
    a P3 cut and project tiling of size $10$ ($2440$ tiles).
  }
  \label{fig:id-P3-cutproject-10}
\end{figure}

We were curious to see what the identity element of the sandpile group would
look like on (finite) Penrose tilings, and it seems that no particular
structure appears\footnote{Well, this is a bit disappointing, but we think that
it is worth showing that it does not appear to be a fruitful research direction,
or maybe a more insightful reader would encounter something out there\dots}.
Examples are presented on Figures~\ref{fig:id-P2-sun-7}, \ref{fig:id-P3-sun-6}
and~\ref{fig:id-P3-cutproject-10},
{\em JS-Sandpile} computes them from Formula~\ref{eq:id}.

\begin{figure}
  \centerline{$
    \vcenter{\hbox{\includegraphics[width=.33\textwidth]{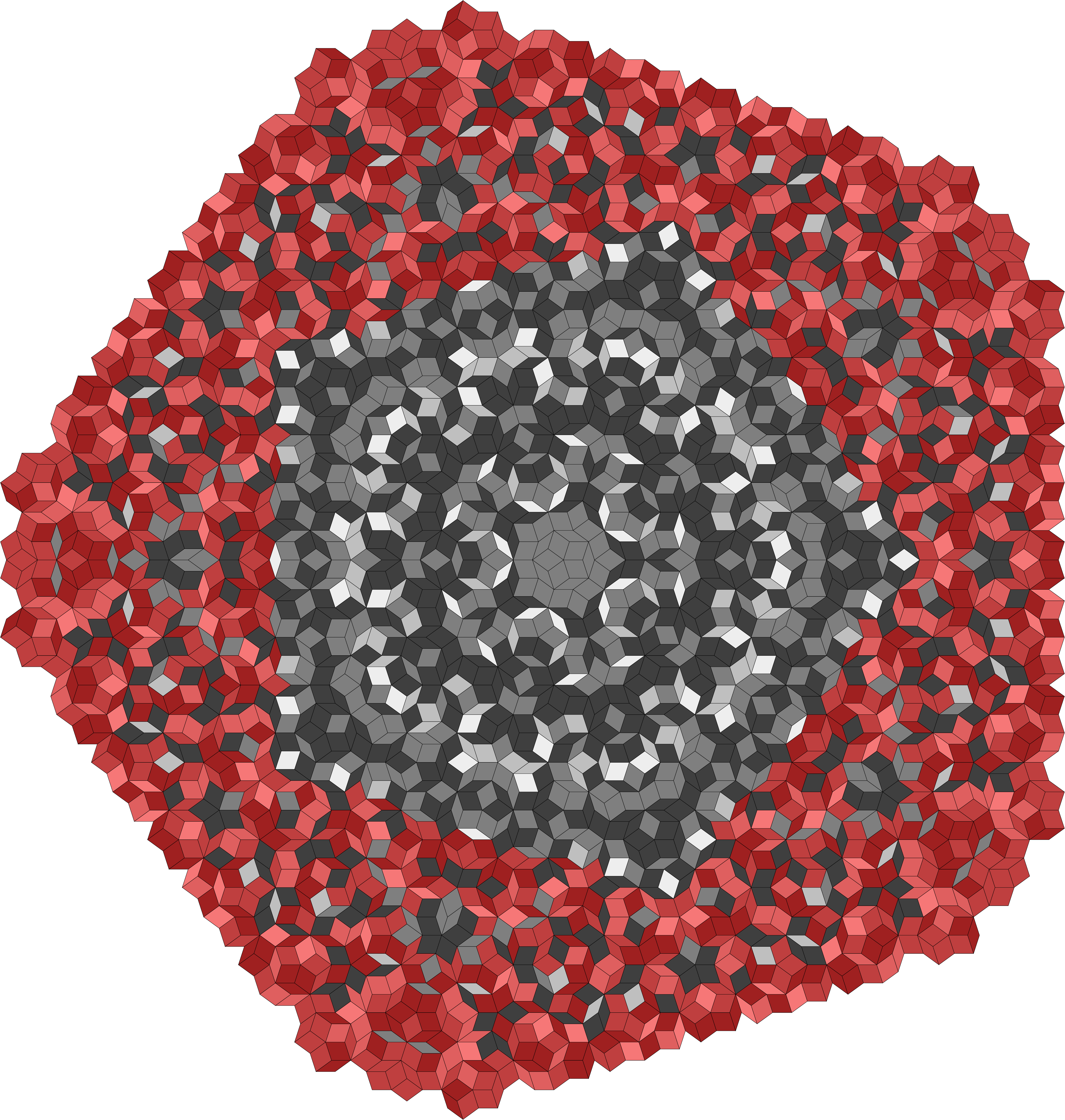}}}~
    \vcenter{\hbox{\includegraphics[width=.33\textwidth]{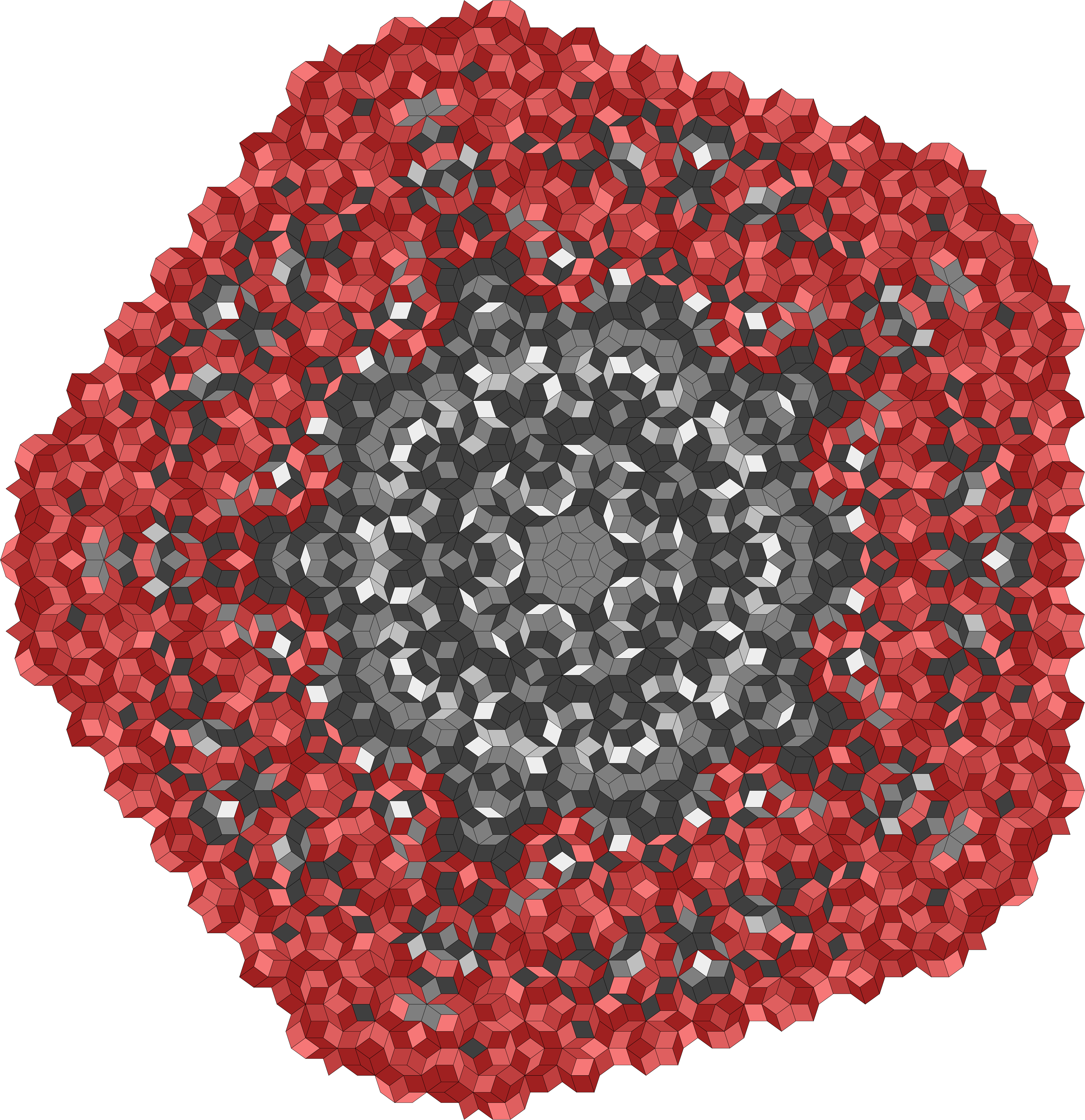}}}~
    \vcenter{\hbox{\includegraphics[width=.33\textwidth]{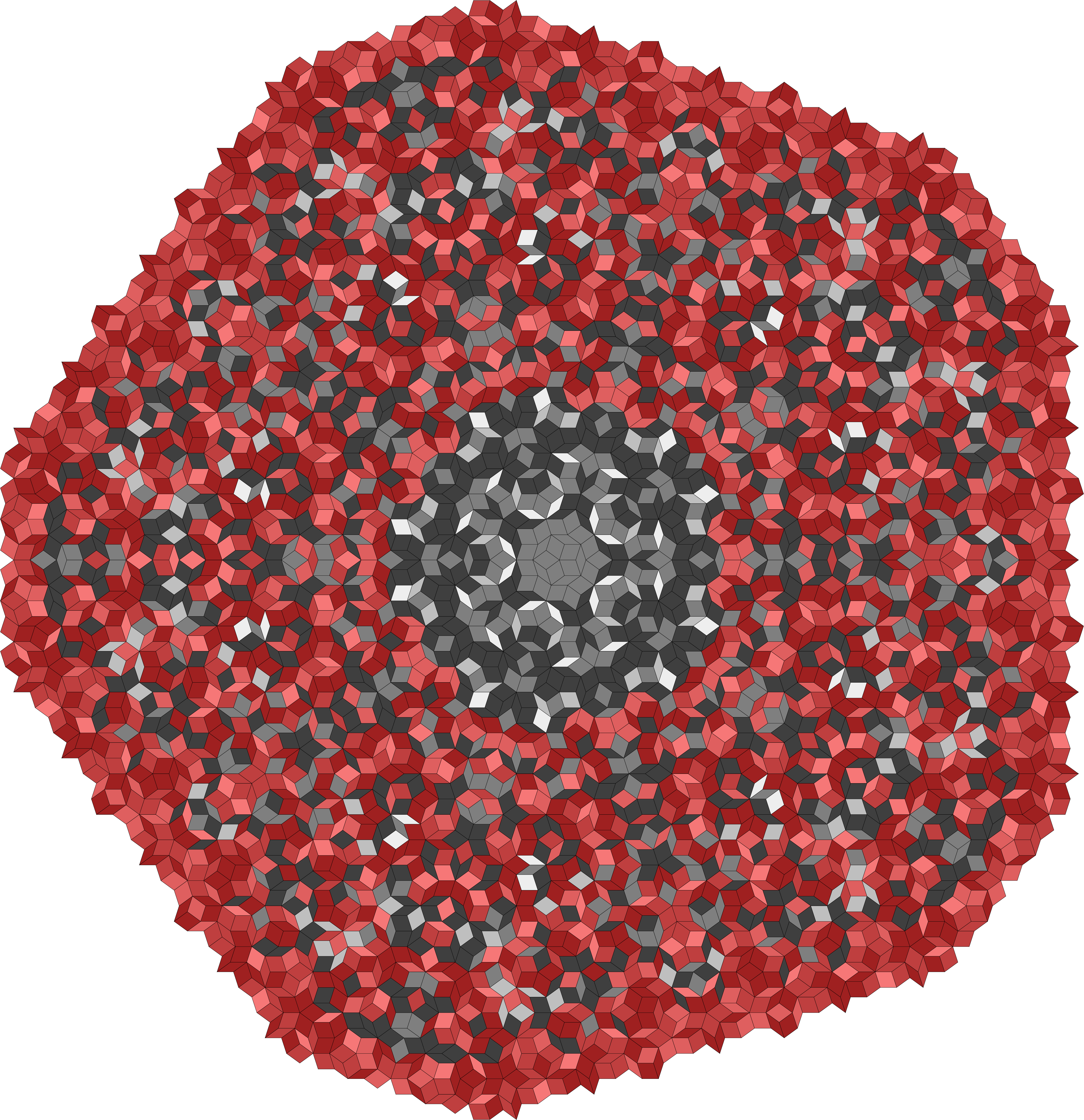}}}
  $}
  \centerline{$
    \vcenter{\hbox{\includegraphics[width=.33\textwidth]{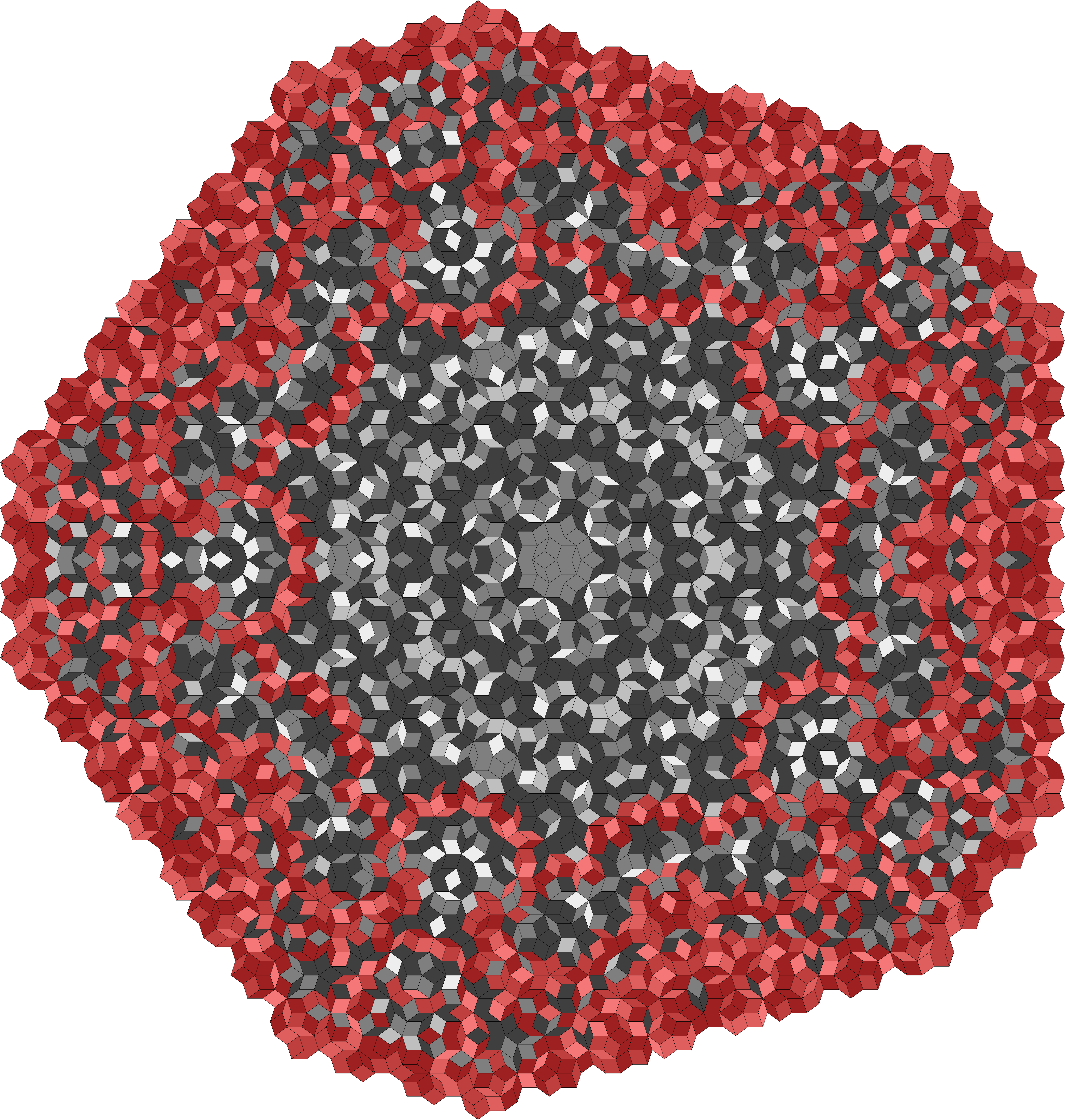}}}~
    \vcenter{\hbox{\includegraphics[width=.33\textwidth]{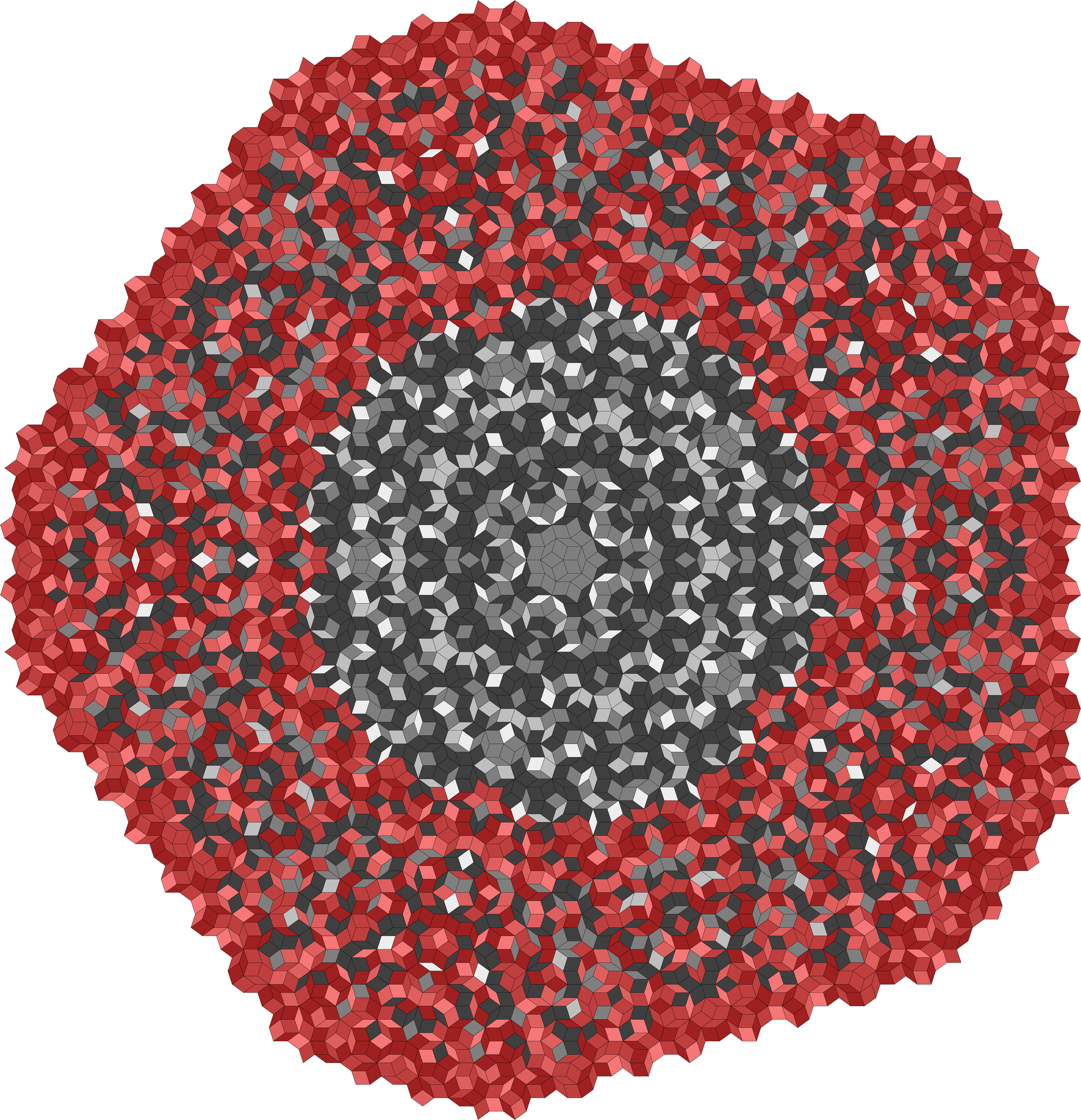}}}~
    \vcenter{\hbox{\includegraphics[width=.33\textwidth]{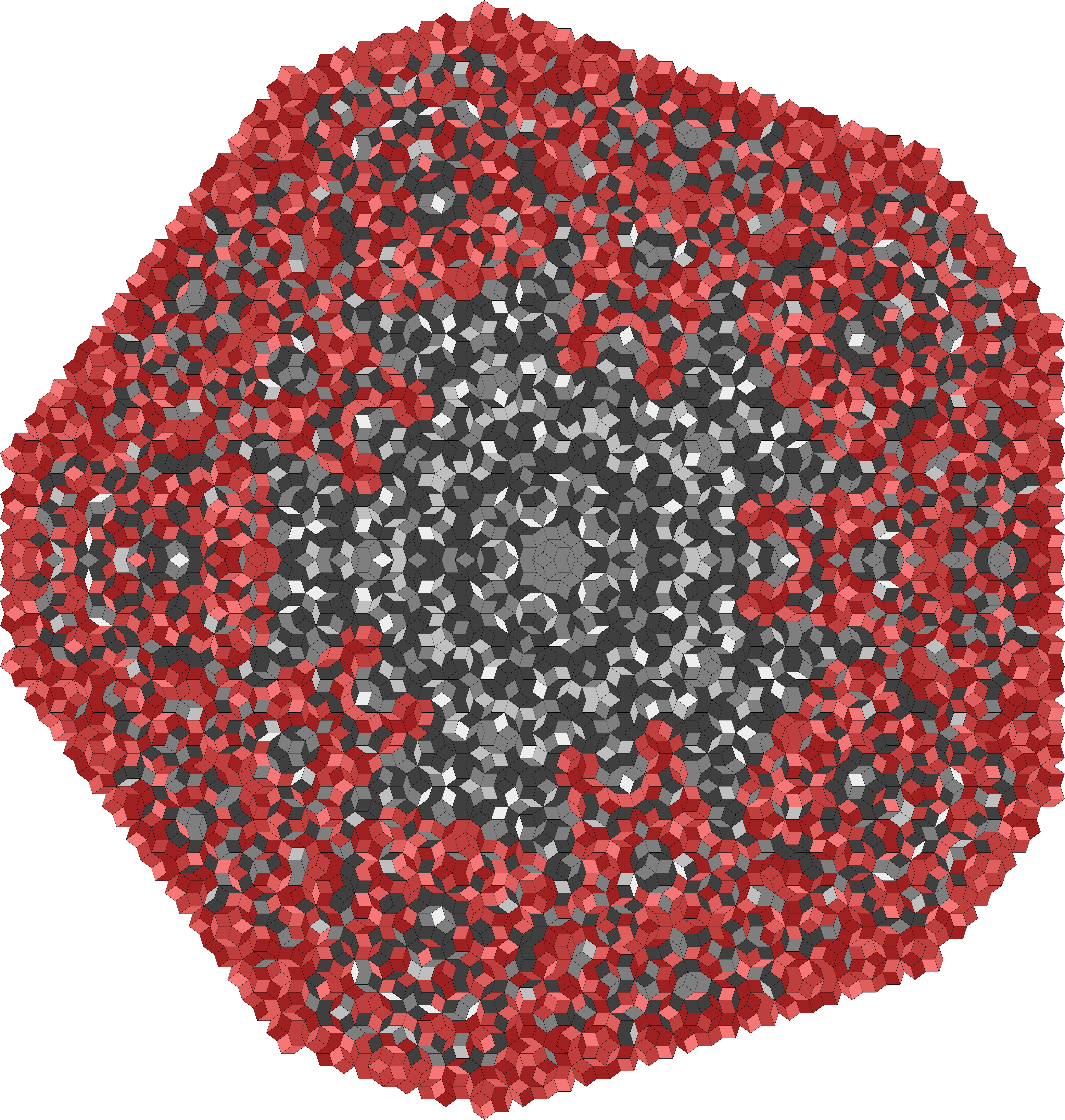}}}
  $}
  \centerline{$
    \vcenter{\hbox{\includegraphics[width=.33\textwidth]{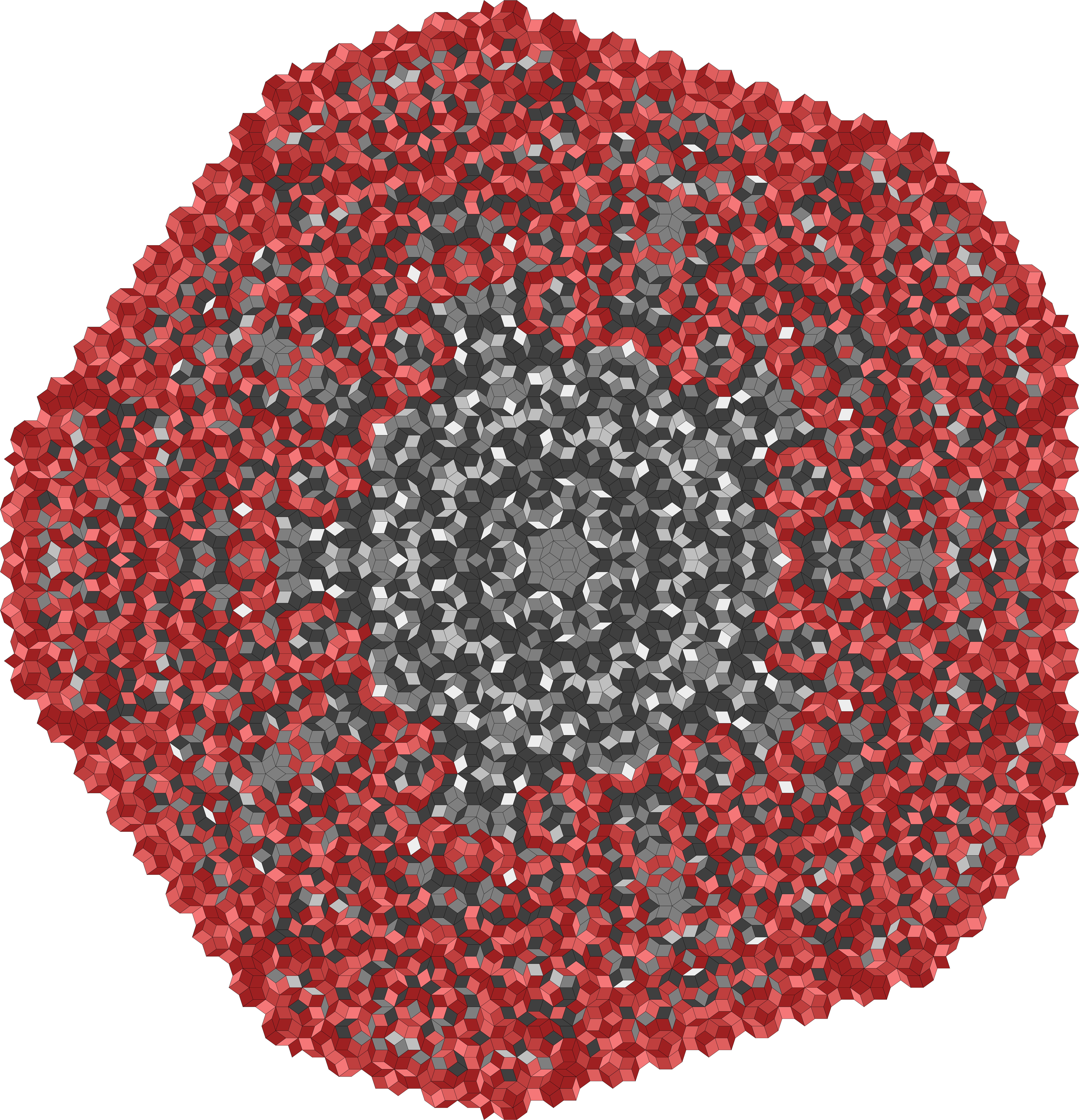}}}~
    \vcenter{\hbox{\includegraphics[width=.33\textwidth]{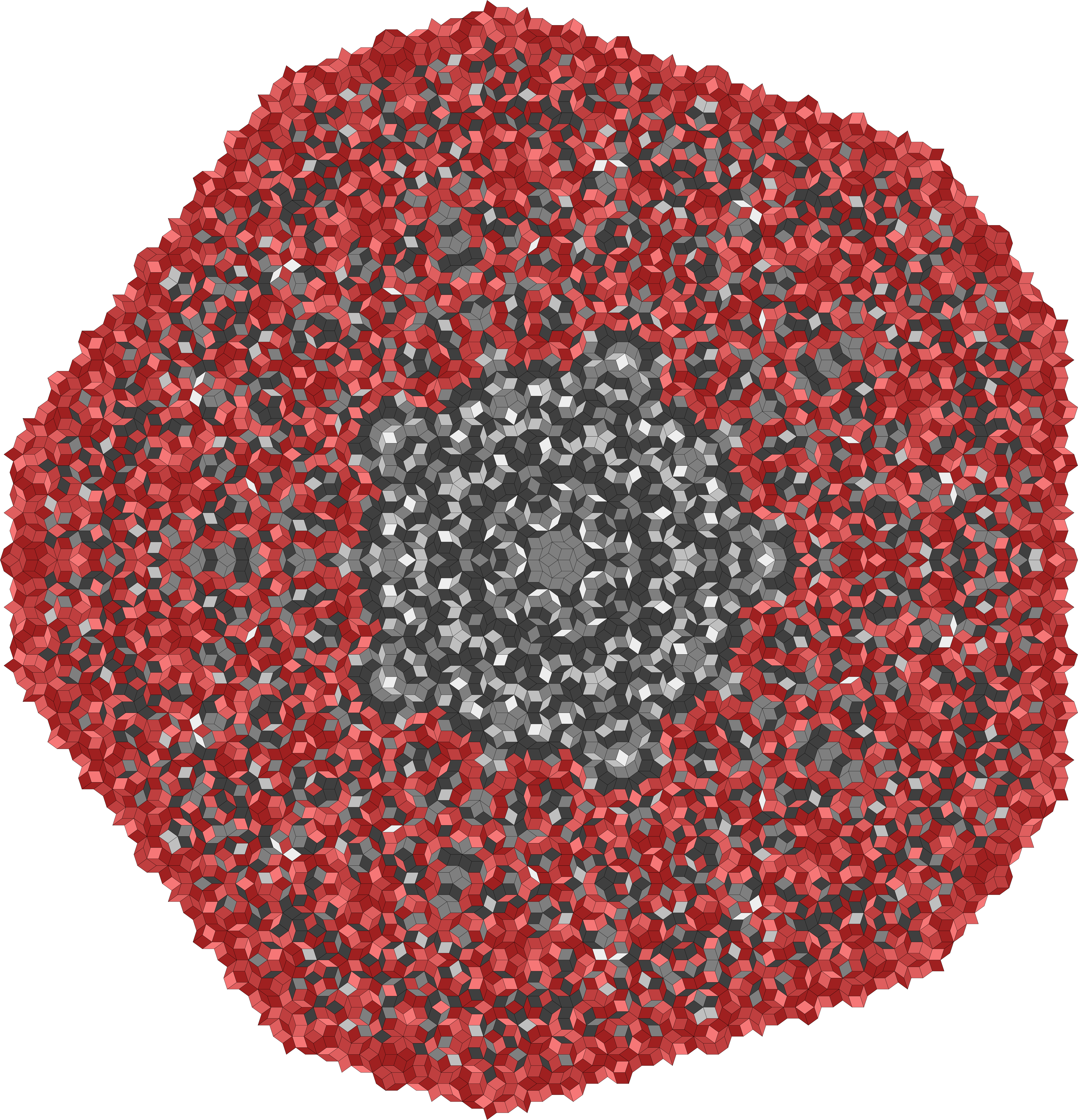}}}~
    \vcenter{\hbox{\includegraphics[width=.33\textwidth]{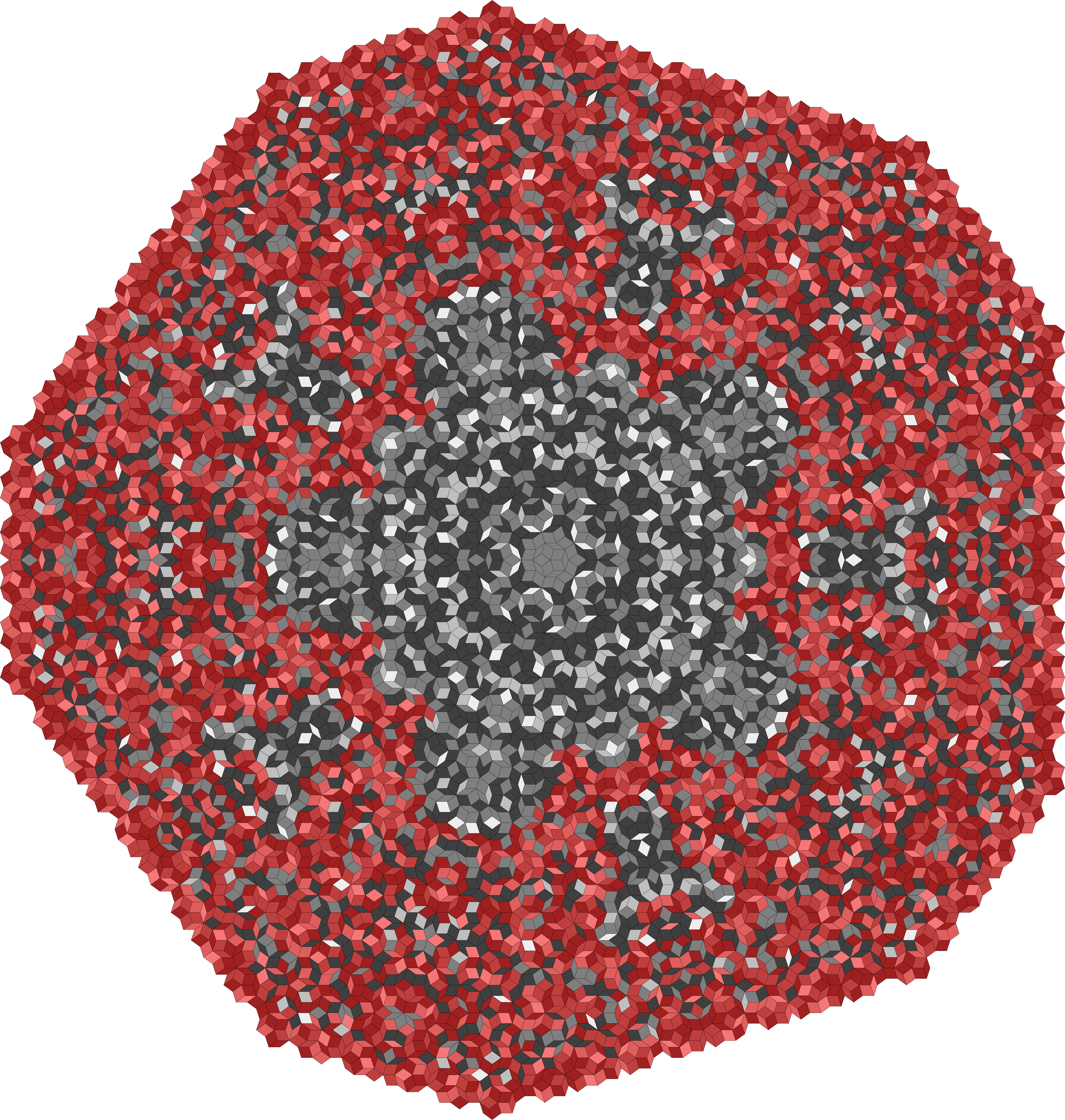}}}
  $}
  \caption{
    Differences between successive identity elements of the sandpile group on
    P3 cut and project tilings. From left to right, top to bottom, are displayed
    the identity elements on P3 cut and project tilings of sizes
    $11$ ($2900$ tiles),
    $12$ ($3520$ tiles),
    $13$ ($4155$ tiles),
    $14$ ($4790$ tiles),
    $15$ ($5570$ tiles),
    $16$ ($6250$ tiles),
    $17$ ($7140$ tiles),
    $18$ ($8080$ tiles),
    $19$ ($8890$ tiles) and
    $20$ ($9940$ tiles),
    where the part of the configuration which differs from the previous size
    (sizes $n$ is compared to size $n-1$) are highlighted with redshifted colors.
  }
  \label{fig:id-P3-cutproject-diff}
\end{figure}

We nevertheless discovered an interesting phenomenon on P3 cut and project tilings,
where the identity elements seem to display some stability.
Indeed, identity elements on successive sizes have a somewhat large central part of
the configuration in common. This obeservation is presented on
Figure~\ref{fig:id-P3-cutproject-diff}.
We may therefore conjecture a convergence of the sandpile identity on P3 cut and project tilings:
as the size $n$ increases,
a larger part of the identity is fixed
(remains the same for all sizes $n' \geq n$).
Let us remark that this phenomenon does not seem to
take place on P2 nor P3 tilings obtained by subsitution.

\section{Isotropic dynamics}
\label{s:iso}

\begin{figure}
  \centerline{
    \includegraphics[width=.24\textwidth]{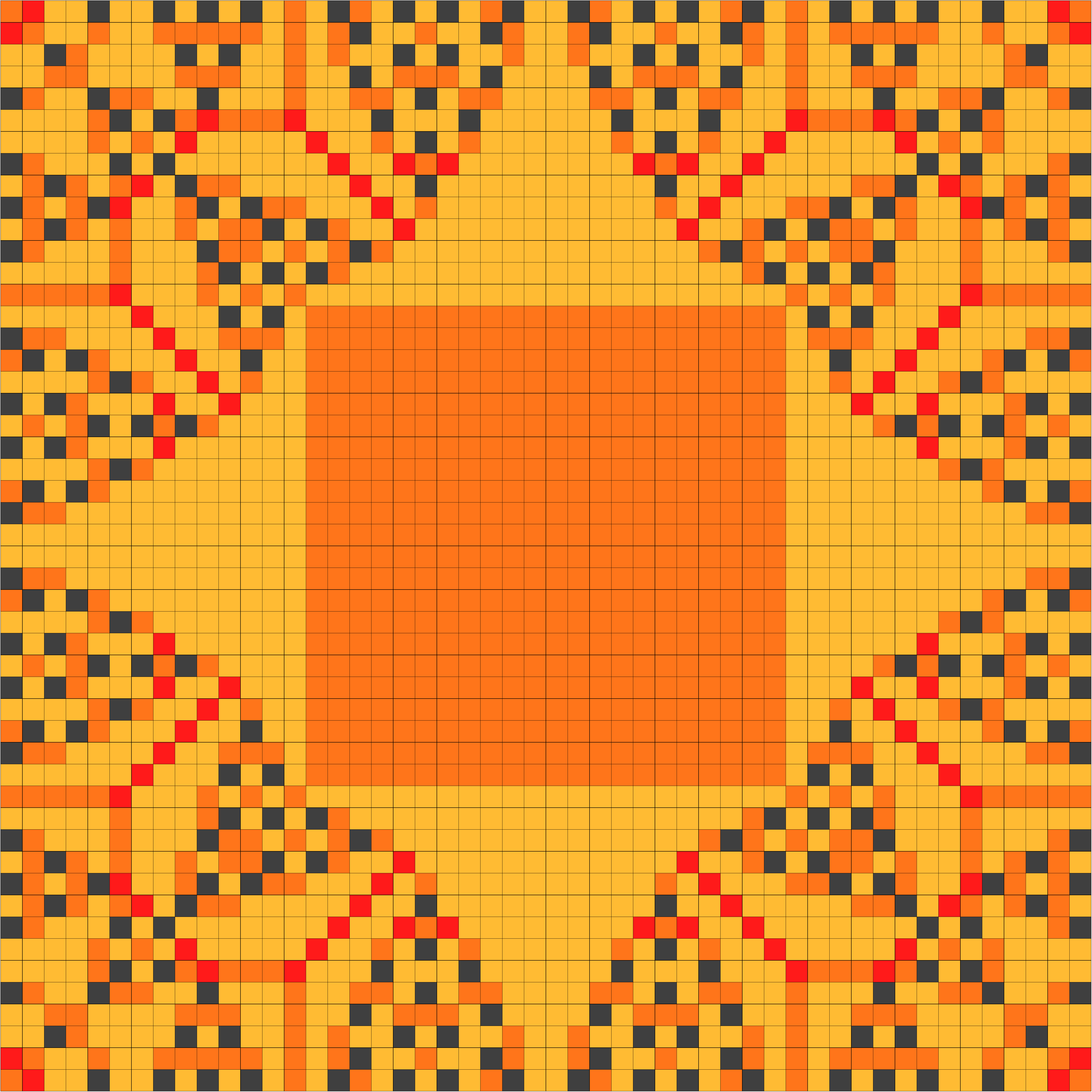}
    \includegraphics[width=.24\textwidth]{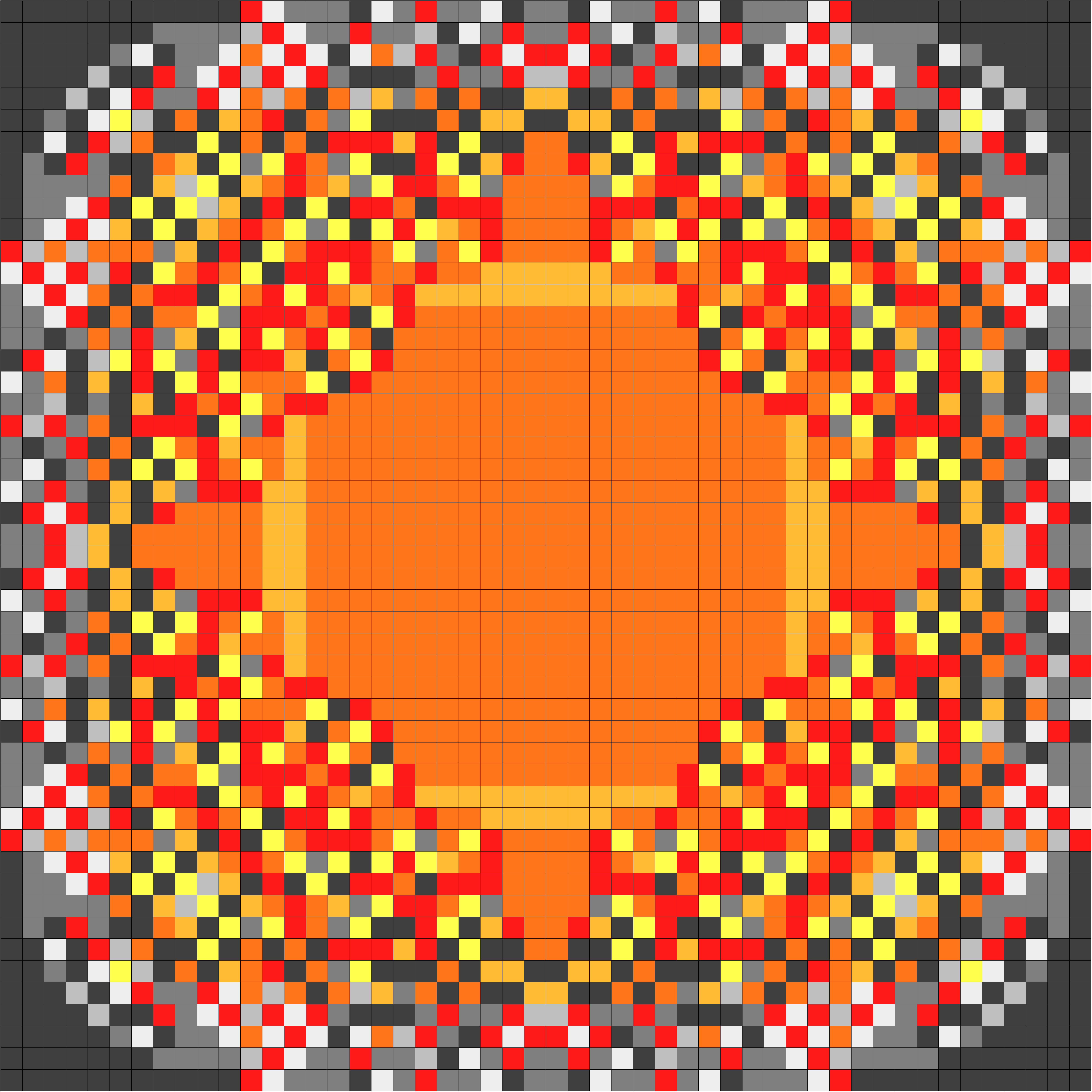}
    \includegraphics[width=.24\textwidth]{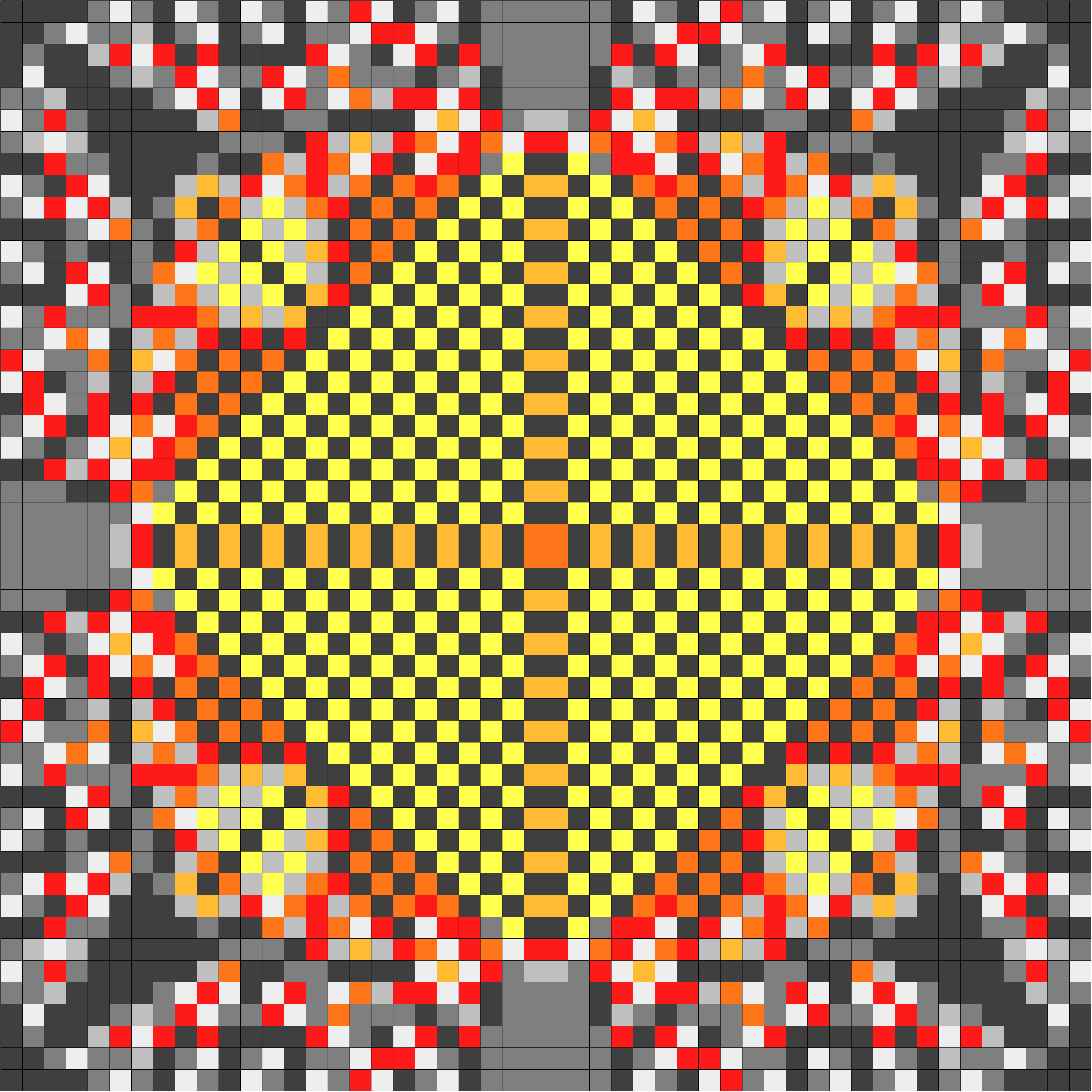}
    \includegraphics[width=.24\textwidth]{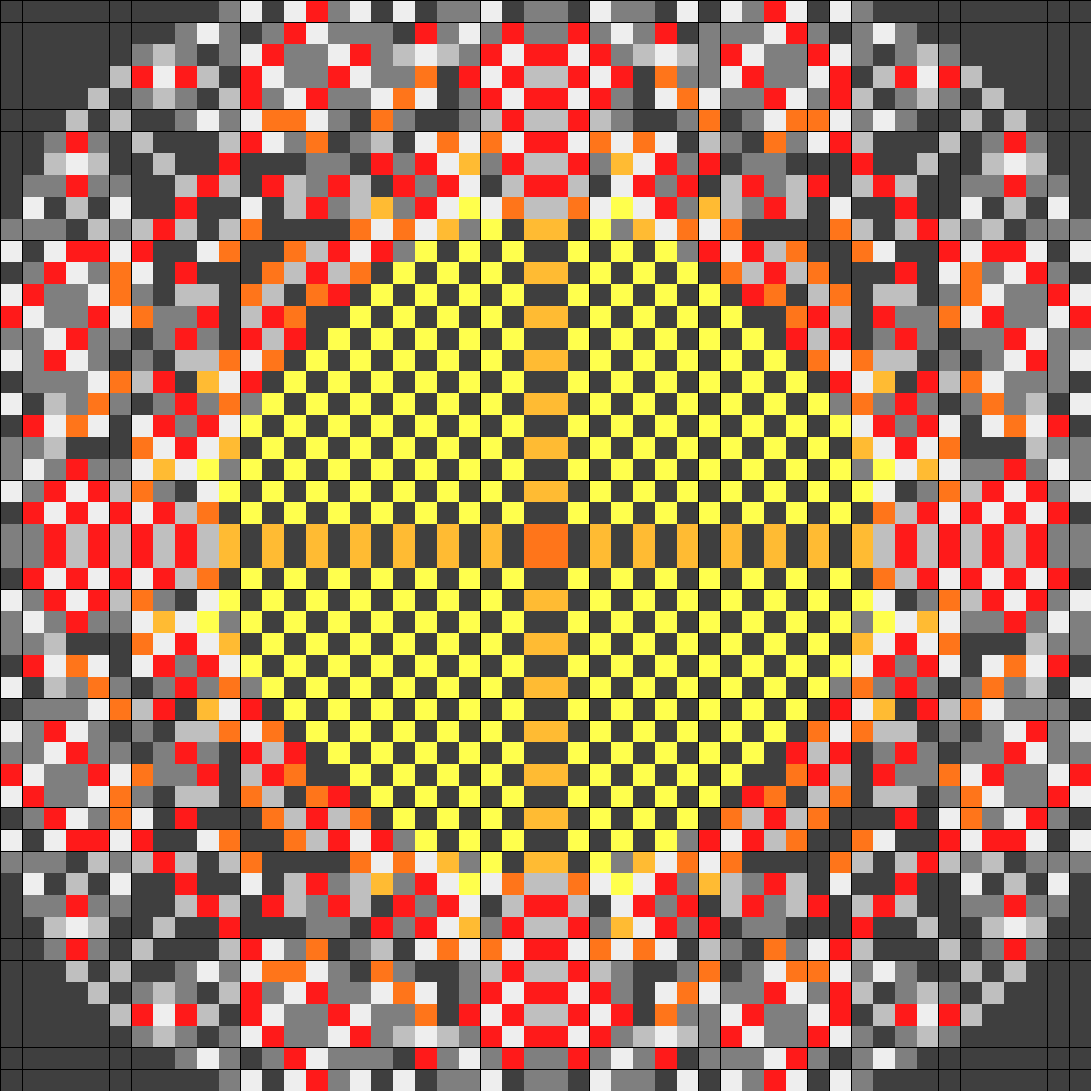}
  }
  \vspace*{3pt}
  \centerline{
    \includegraphics[width=.24\textwidth]{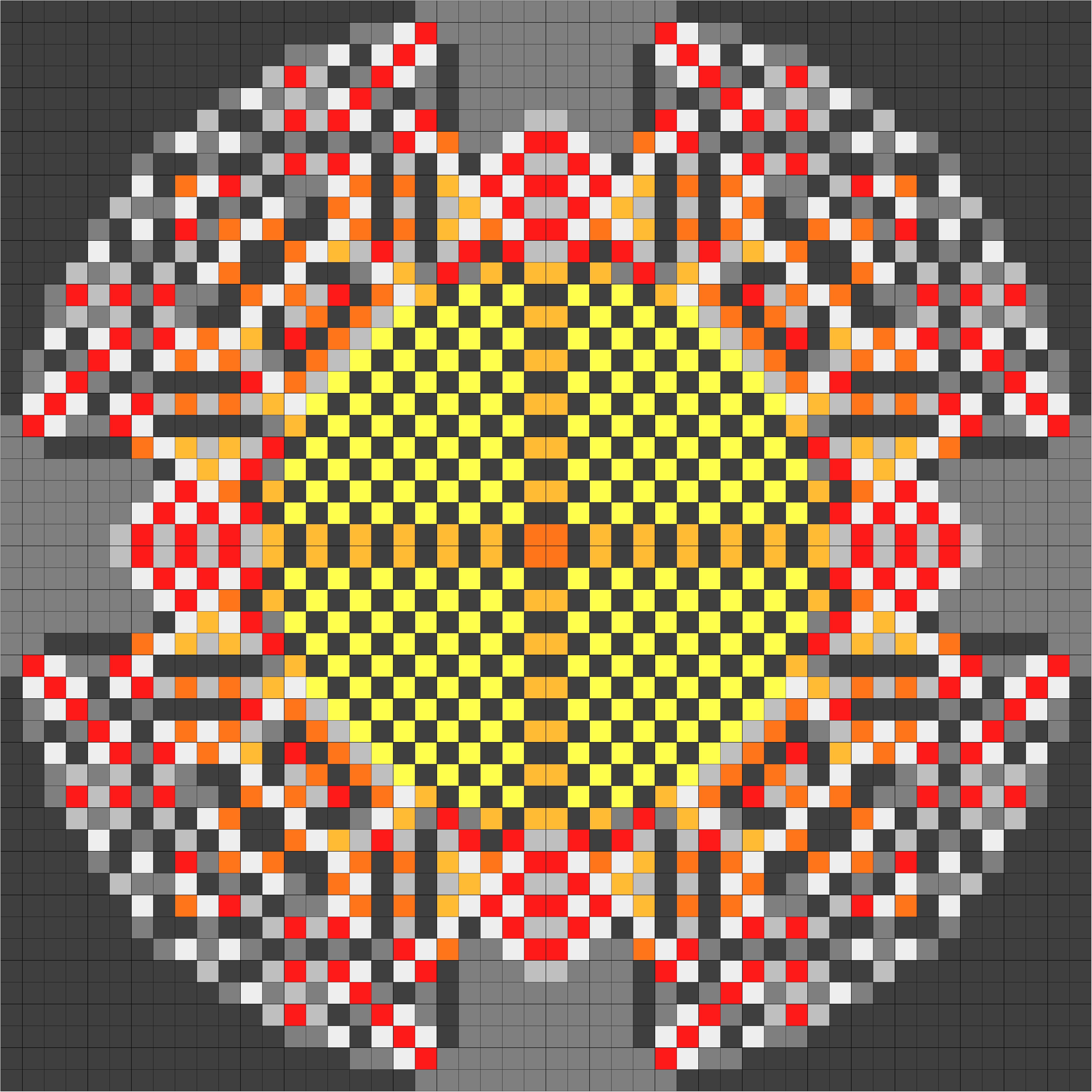}
    \includegraphics[width=.24\textwidth]{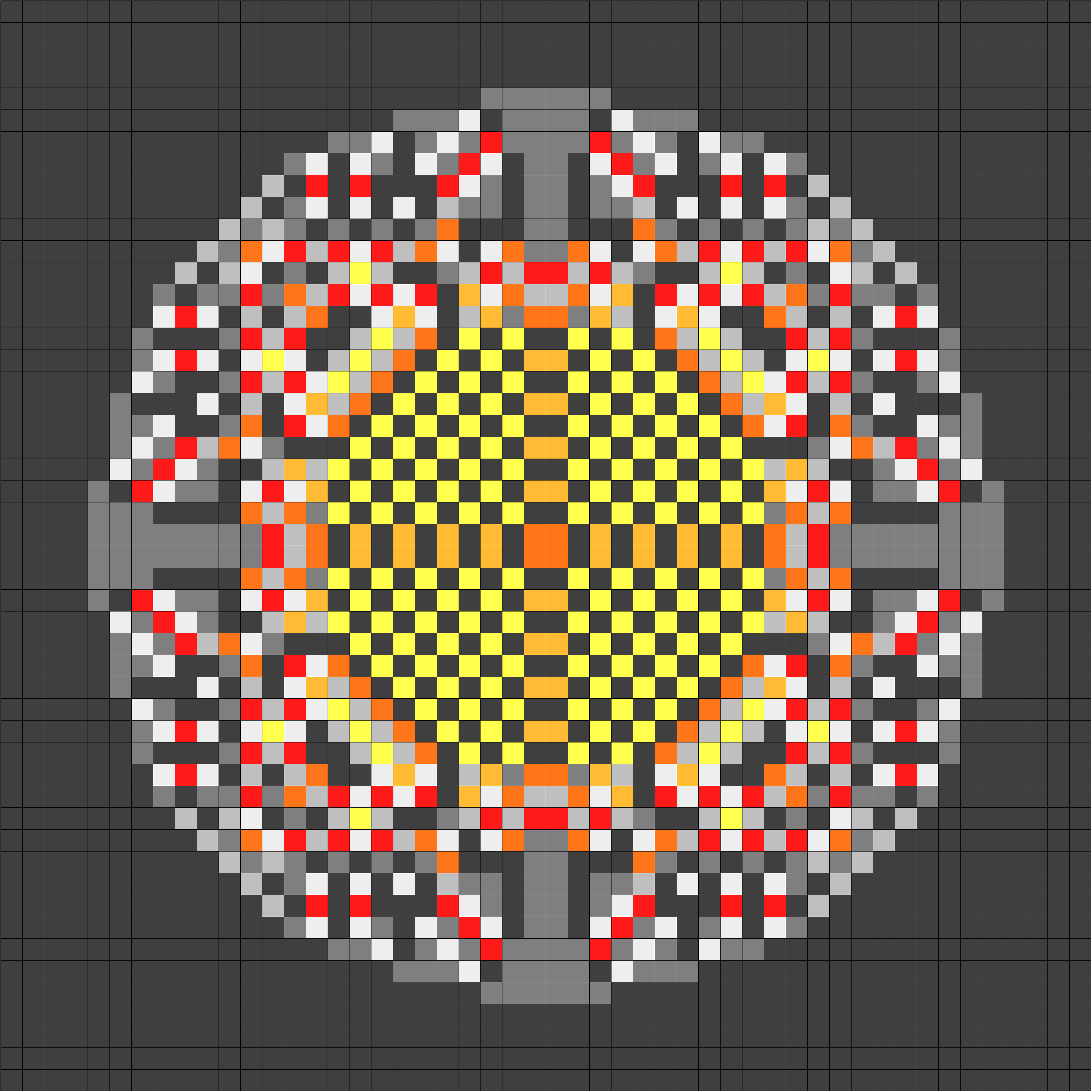}
    \includegraphics[width=.24\textwidth]{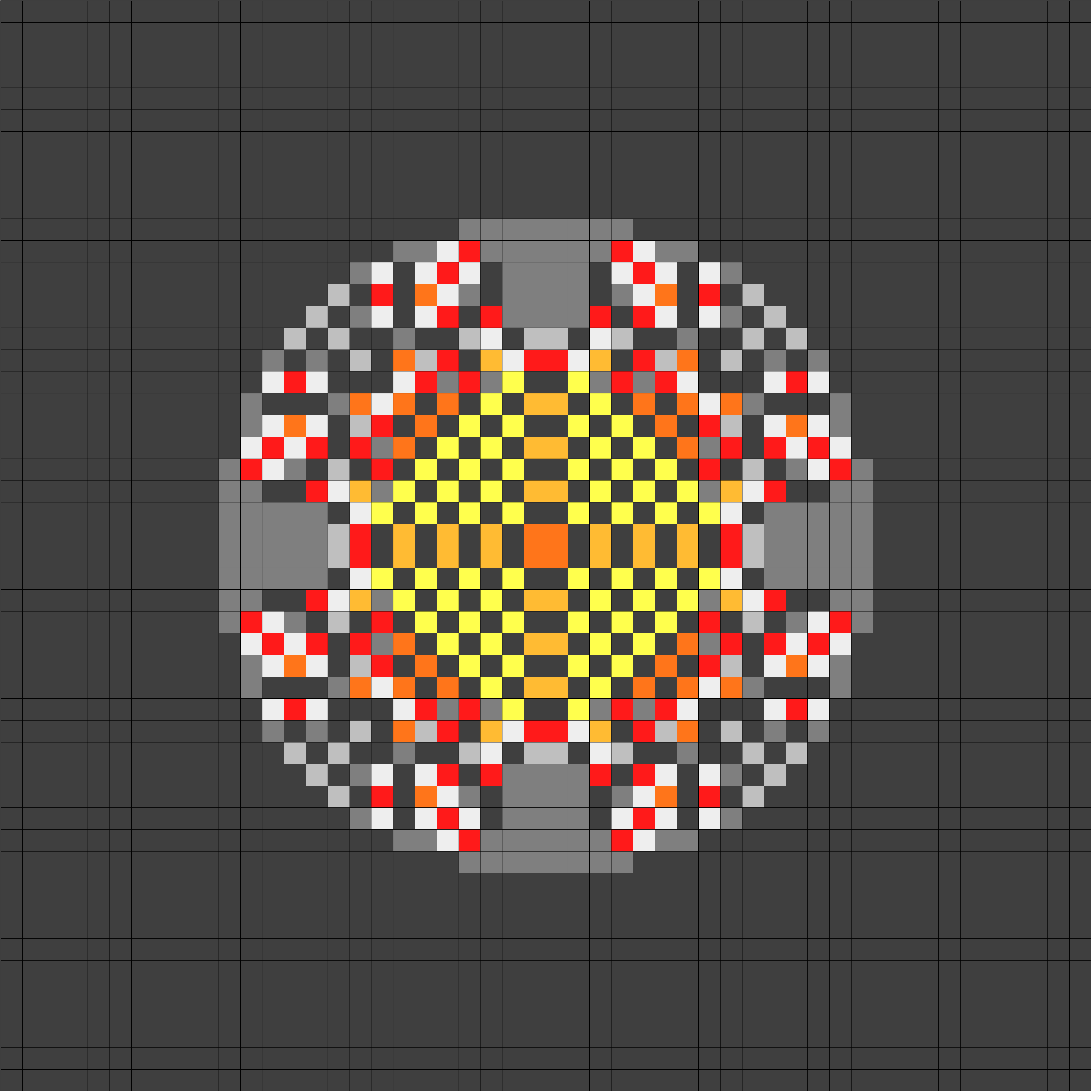}
    \includegraphics[width=.24\textwidth]{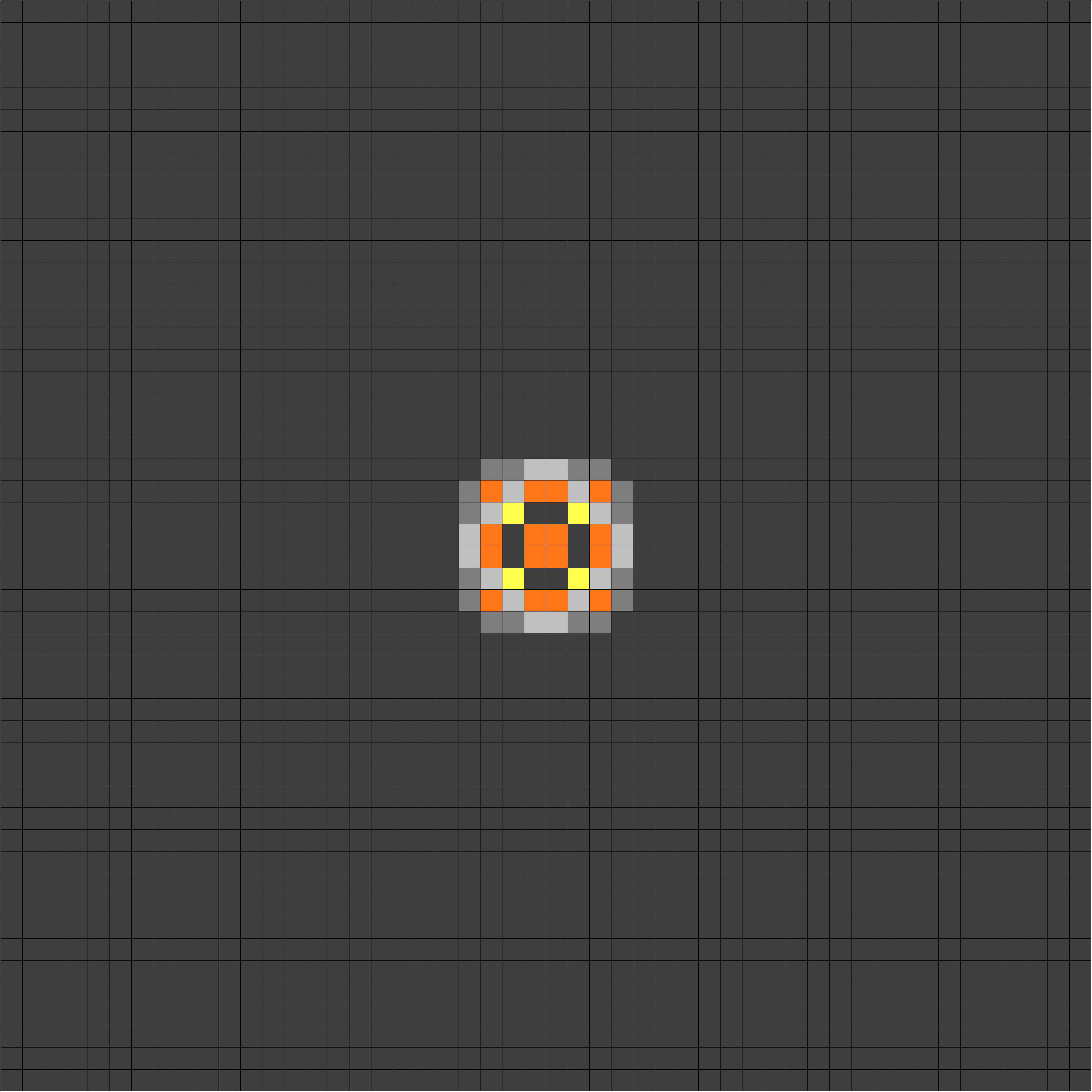}
  }
  \caption{
    Stabilization of $\stable{(m+e)}=m$ on a square grid of side length $50$. 
    From left to right, top to bottom, are displayed the configurations every $100$ time steps
    (starting with $m+e$ at step $0$, ending with time step $700$).
    The process converges to $m$ at step $707$.
  }
  \label{fig:circle-squaregrid}
\end{figure}

\begin{figure}
  \centerline{
    \includegraphics[width=.24\textwidth]{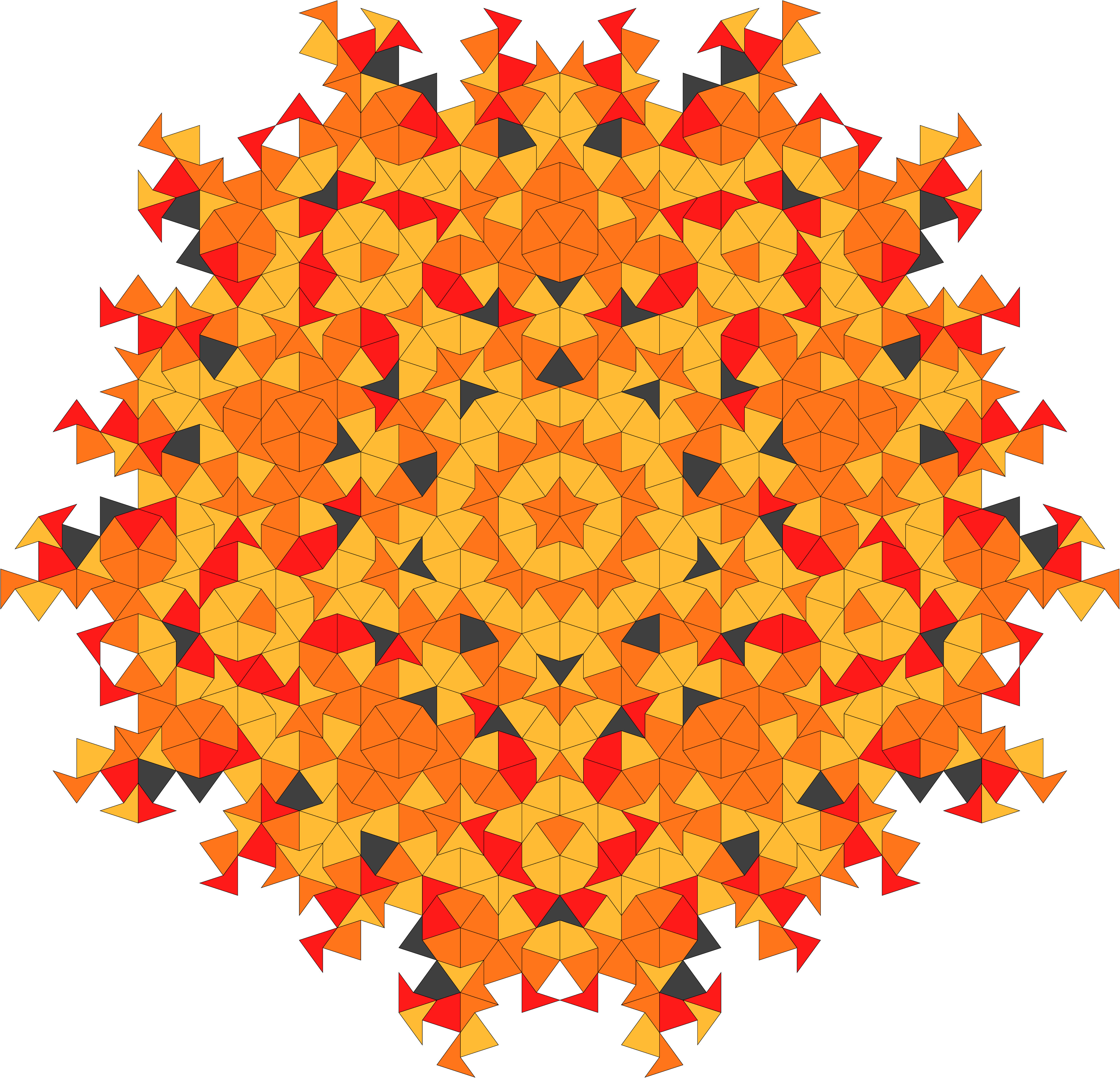}
    \includegraphics[width=.24\textwidth]{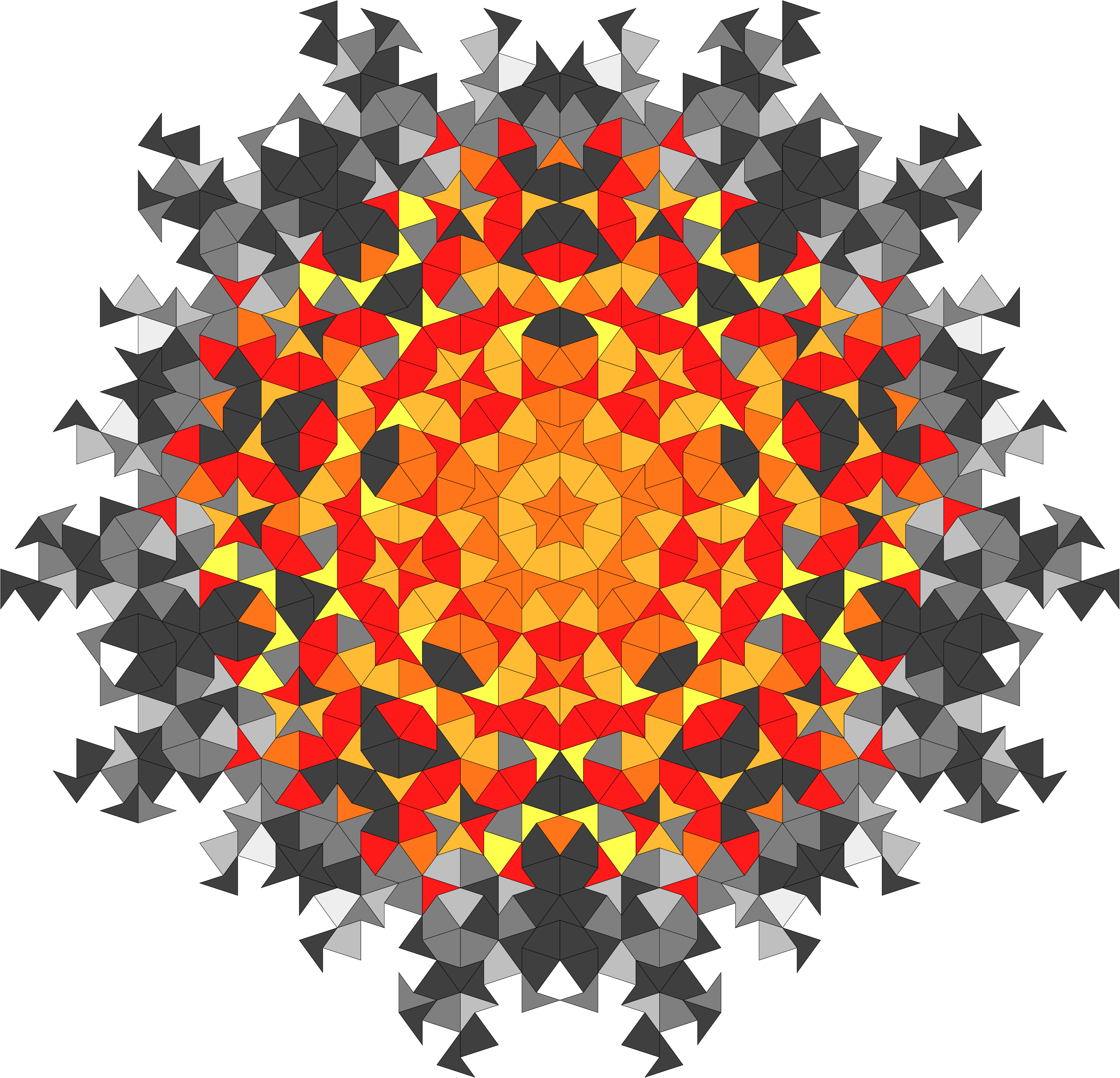}
    \includegraphics[width=.24\textwidth]{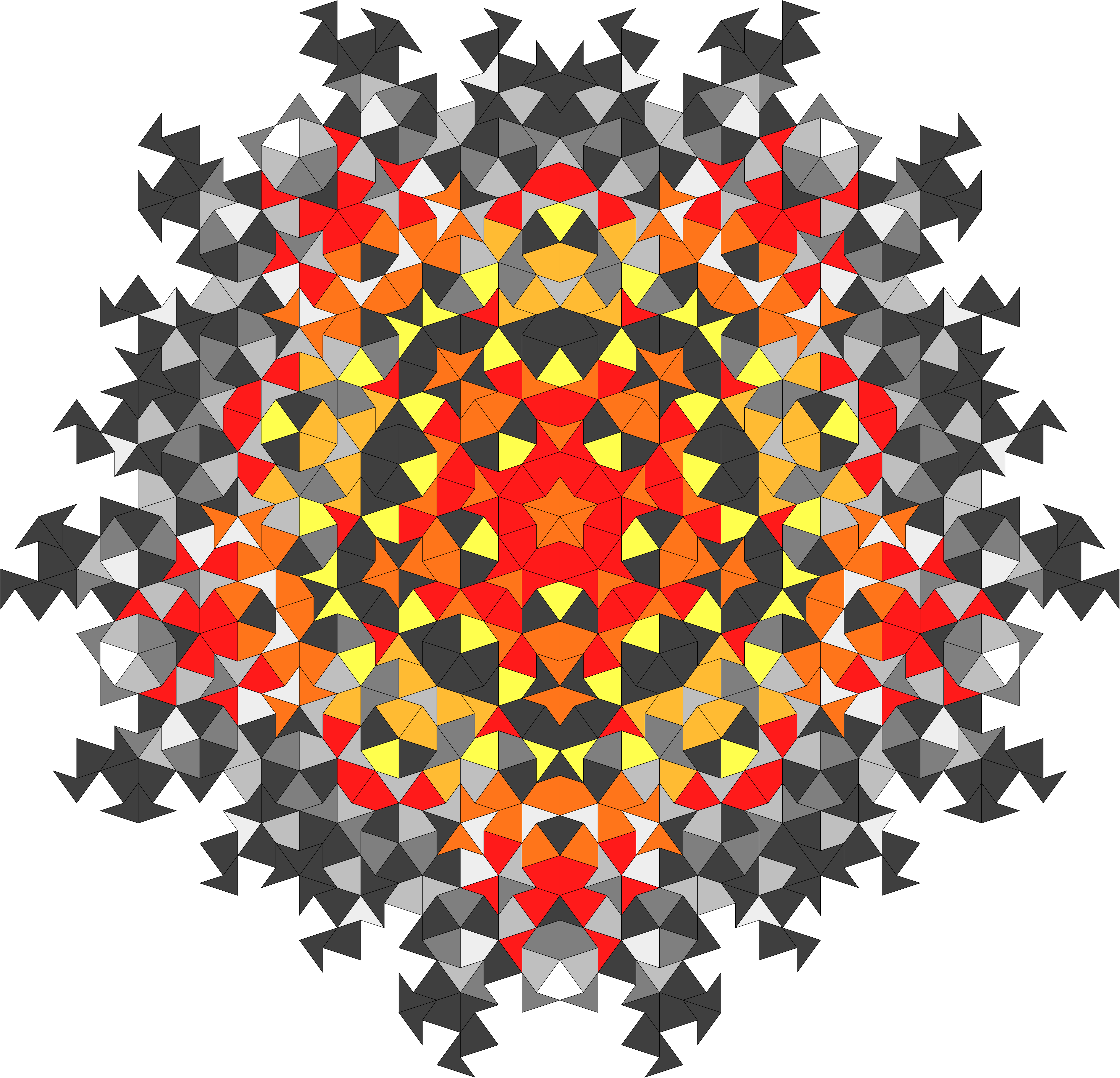}
    \includegraphics[width=.24\textwidth]{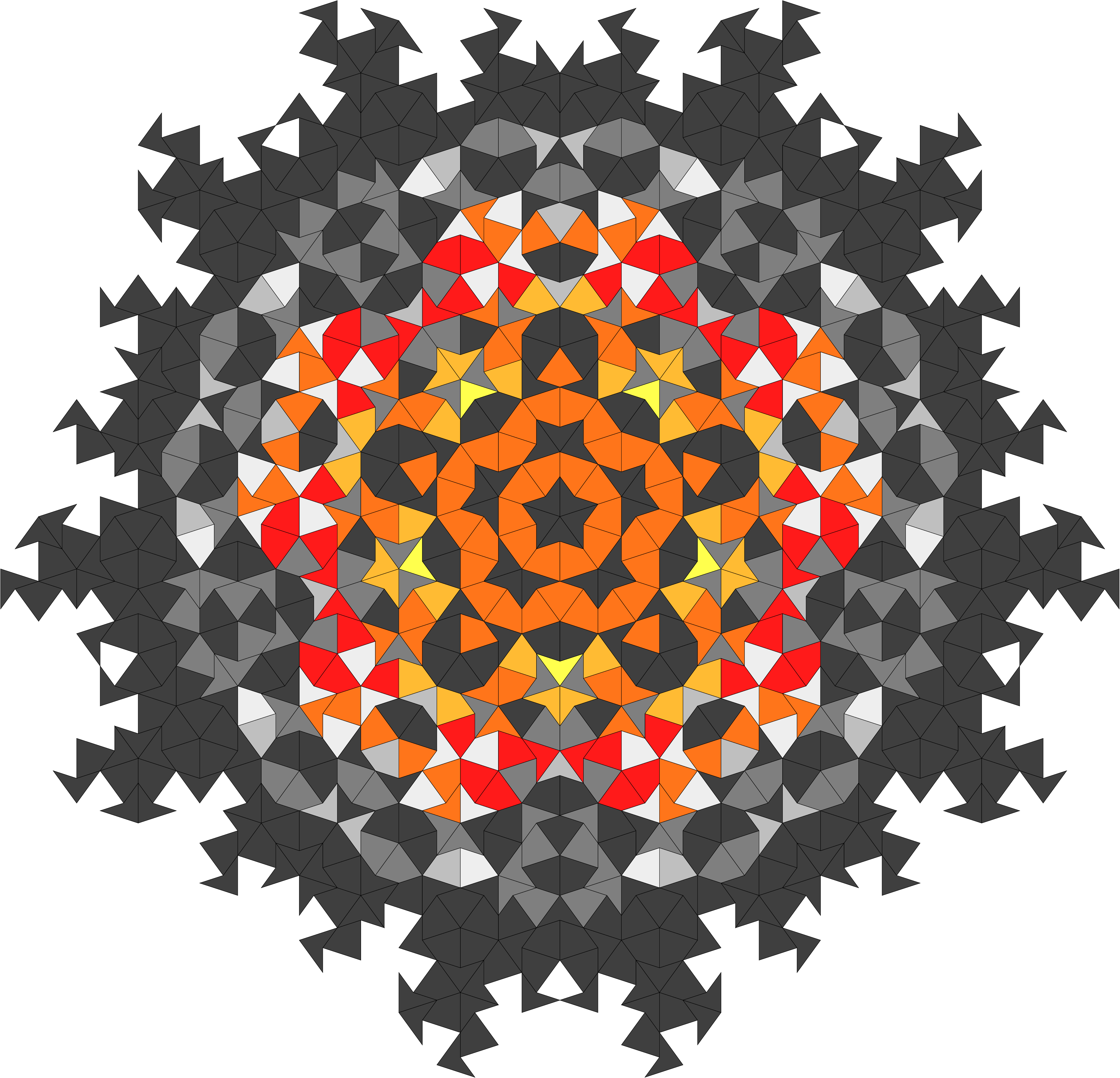}
  }
  \vspace*{3pt}
  \centerline{
    \includegraphics[width=.24\textwidth]{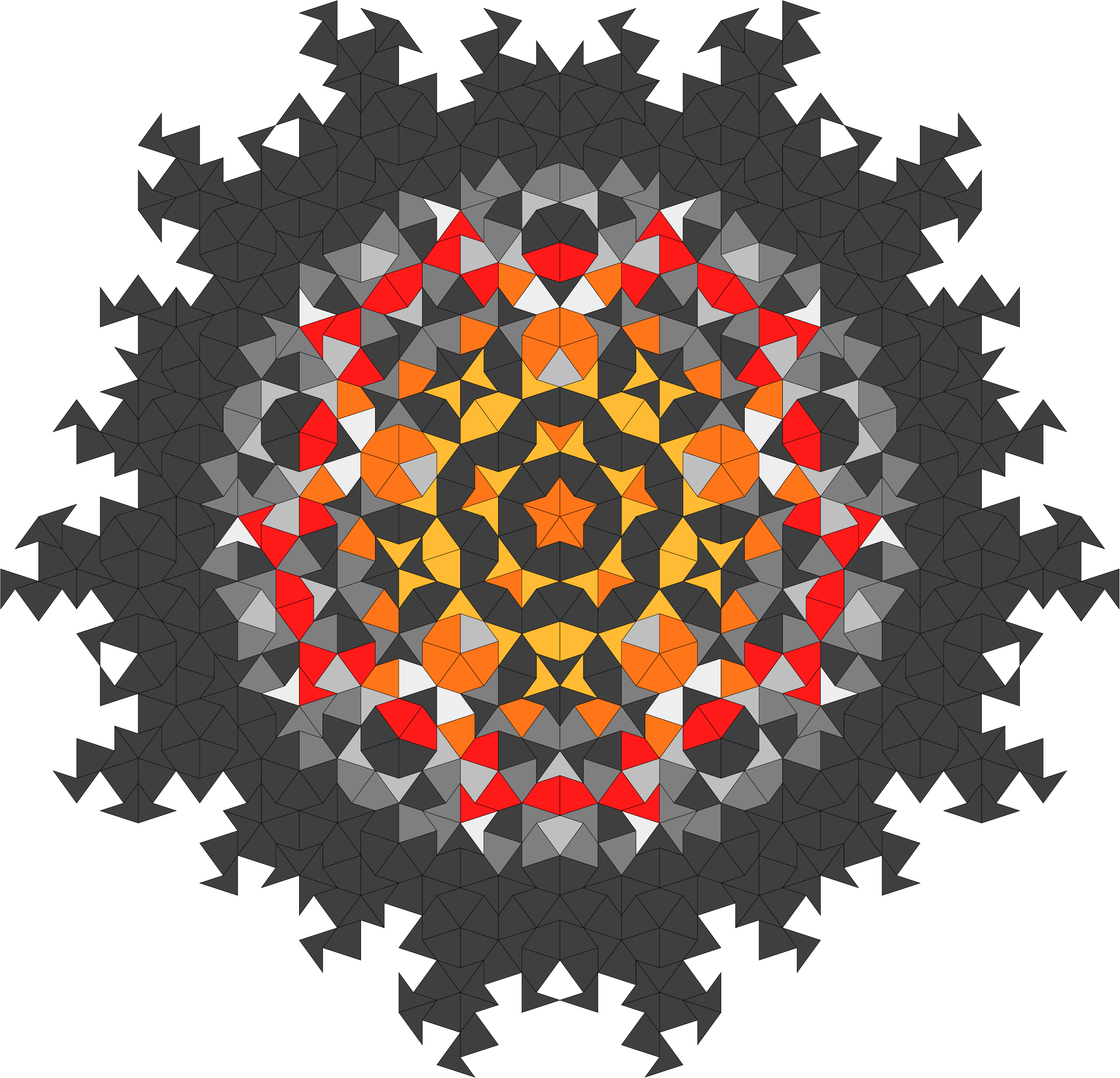}
    \includegraphics[width=.24\textwidth]{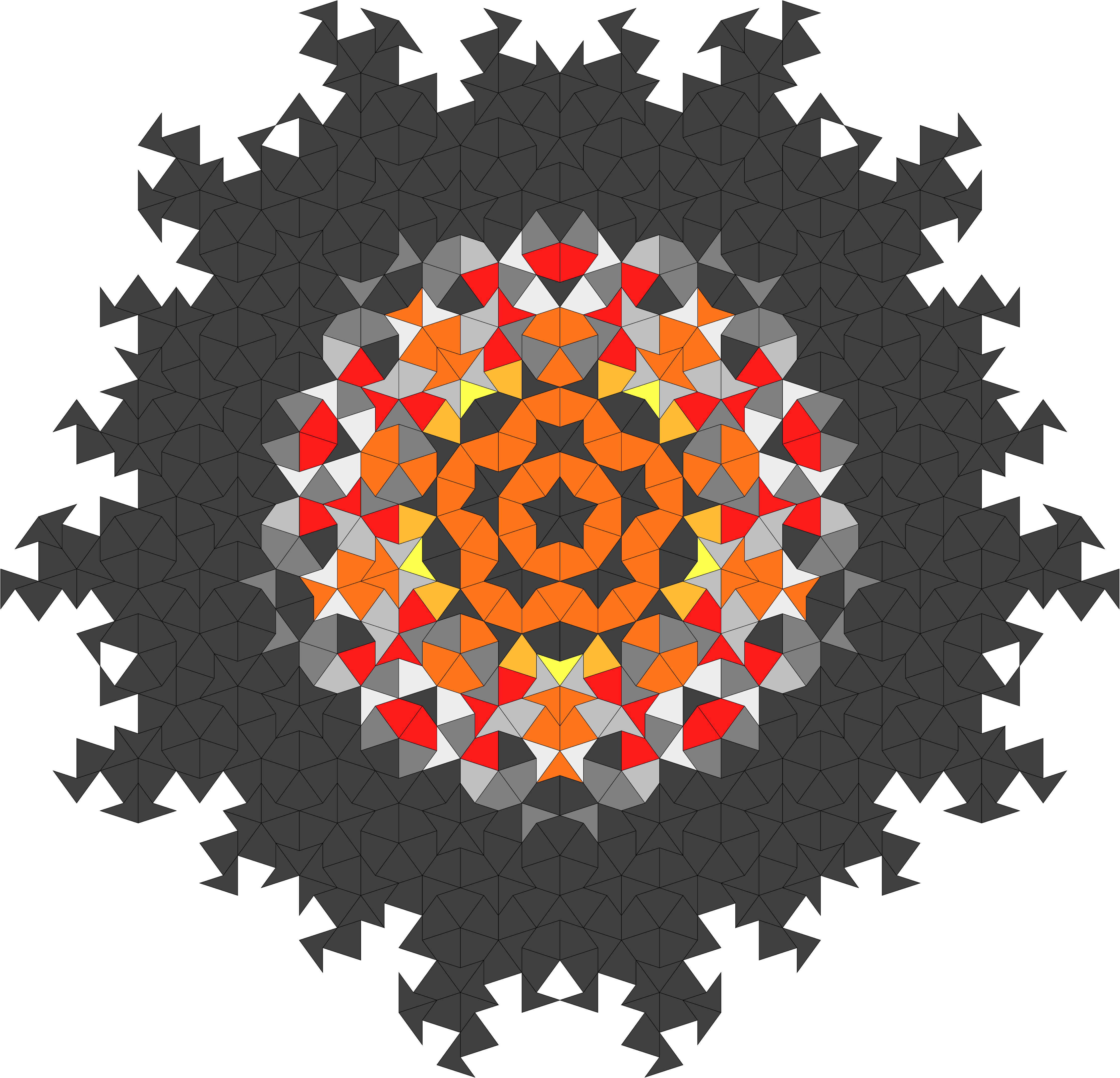}
    \includegraphics[width=.24\textwidth]{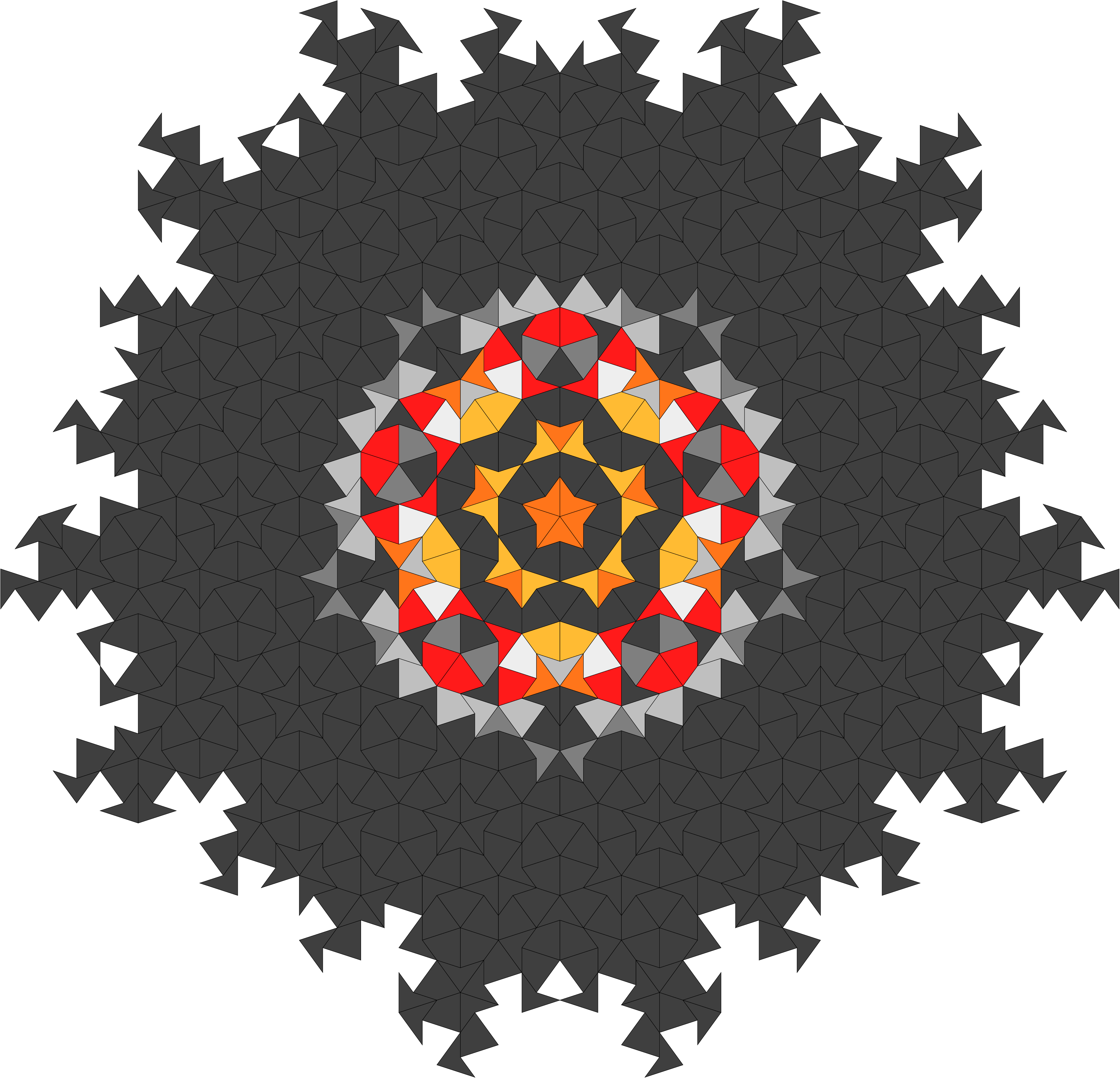}
    \includegraphics[width=.24\textwidth]{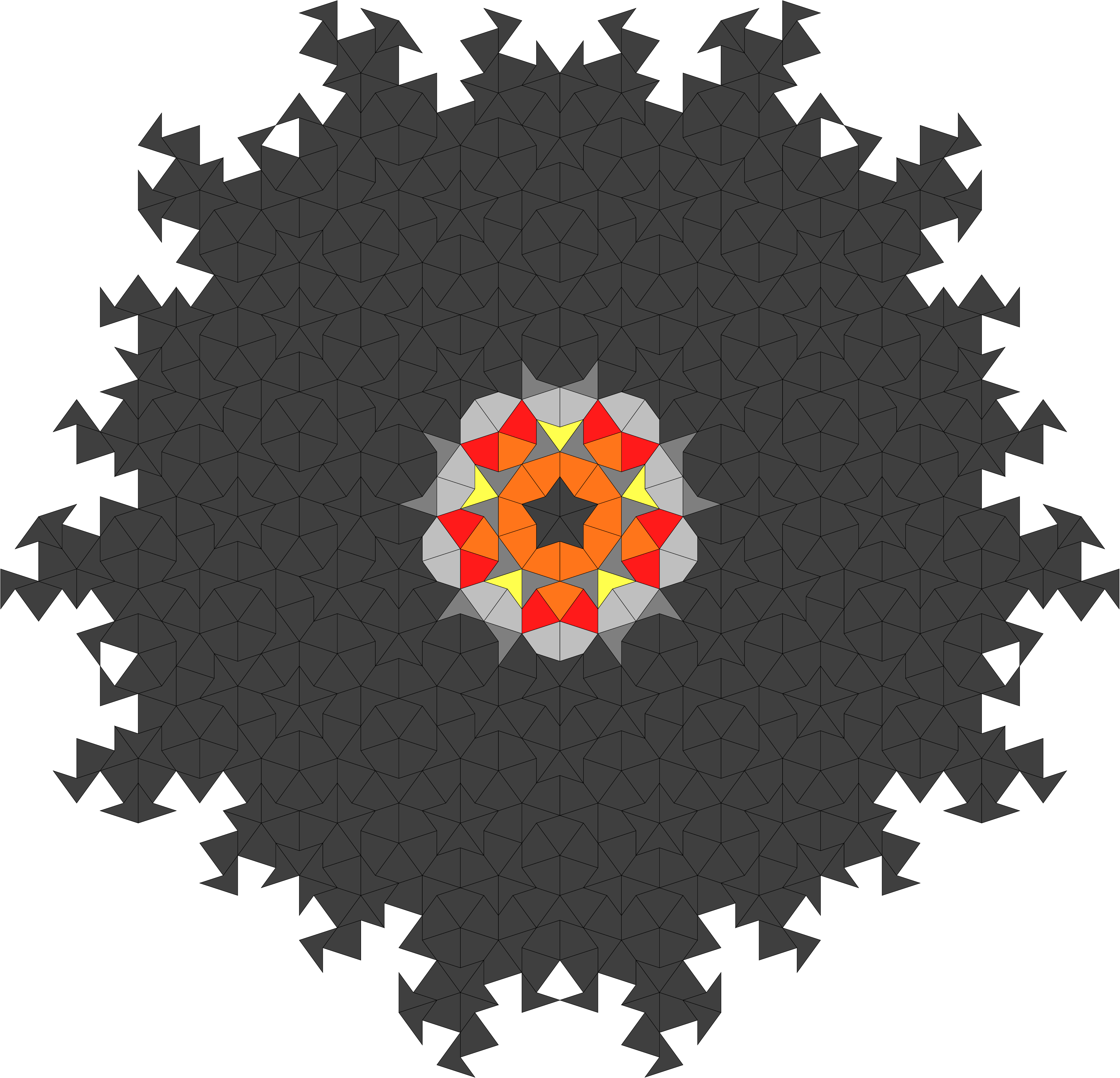}
  }
  \caption{
    Stabilization of $\stable{(m+e)}=m$ on $5$ iterations of the substitution from a P2 Sun.
    From left to right, on top are displayed time steps
    $0, 50, 100, 150$
    and at the bottom time steps
    $165, 180, 195, 210$.
    The process converges to $m$ at step $220$.
  }
  \label{fig:circle-penrose}
\end{figure}

On square, triangular, hexagonal grids and Penrose tilings,
a very interesting phenomenon appears\footnote{This
also takes place on other tilings,
outside the scope of the present work.}
during the stabilization process
\[
  \stable{(m+e)}=m
\]
where $m$ is the maximum stable configuration
($m(v)=\degree{v}-1$ for all tile $v$)
and $e$ is the identity element of the sandpile group.
Indeed, during the last phase of the stabilization process leading back to $m$
(which is a uniform configuration in these cases since the number of neighbors is identical for all tiles),
one can see the configuration $m$ reappear from the outside (near the border) towards the center,
outside a shrinking circular shape, in a process step by step covering the whole configuration with tiles
containing $\degree{v}-1$ grains.
The first part of the stabilization process is quite involved.
Two illustrations are given on Figures~\ref{fig:circle-squaregrid} and~\ref{fig:circle-penrose}.

\subsection{Roundness}

\begin{figure}
  \centerline{
    \includegraphics[width=.45\textwidth]{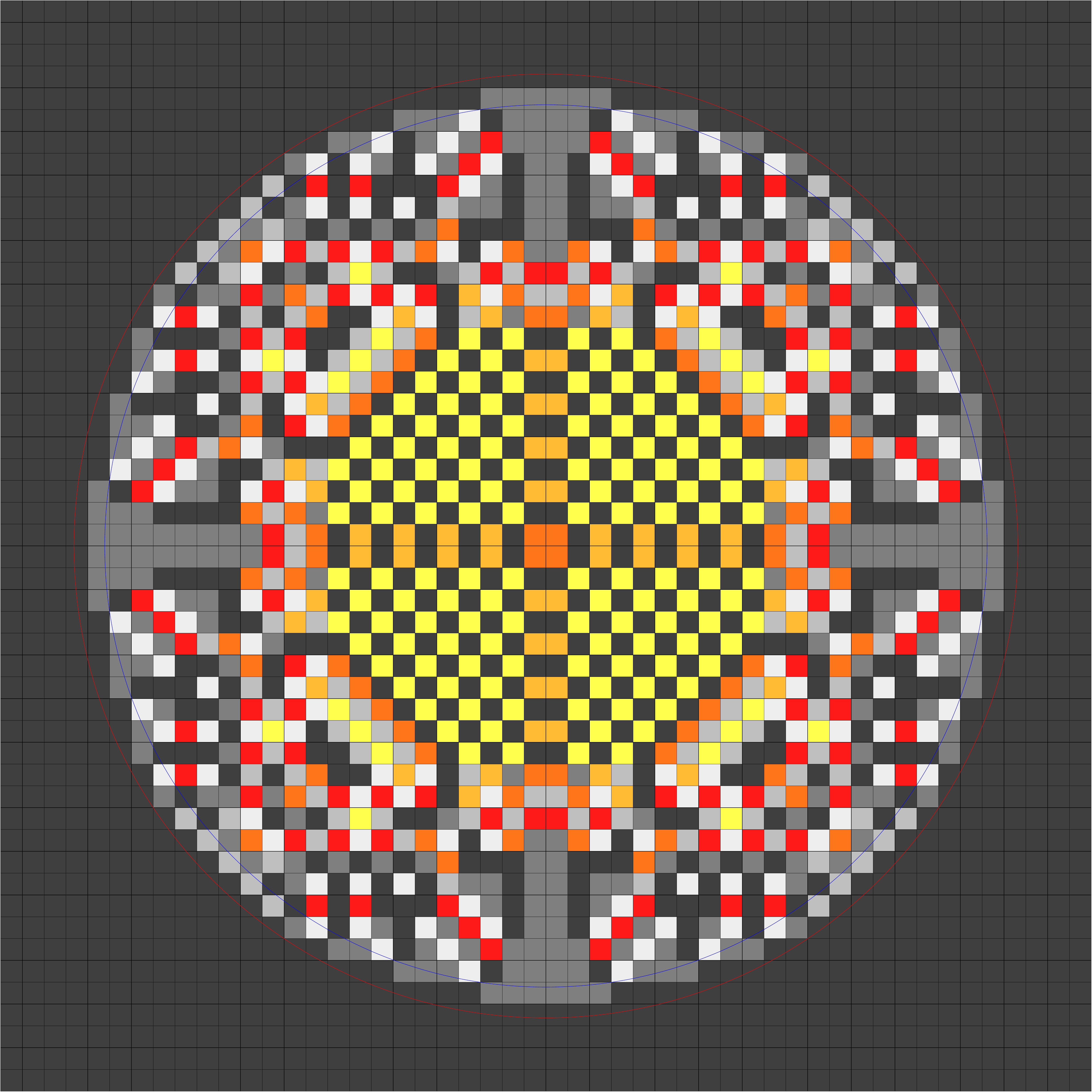}
    \hspace*{.5cm}
    \includegraphics[width=.45\textwidth]{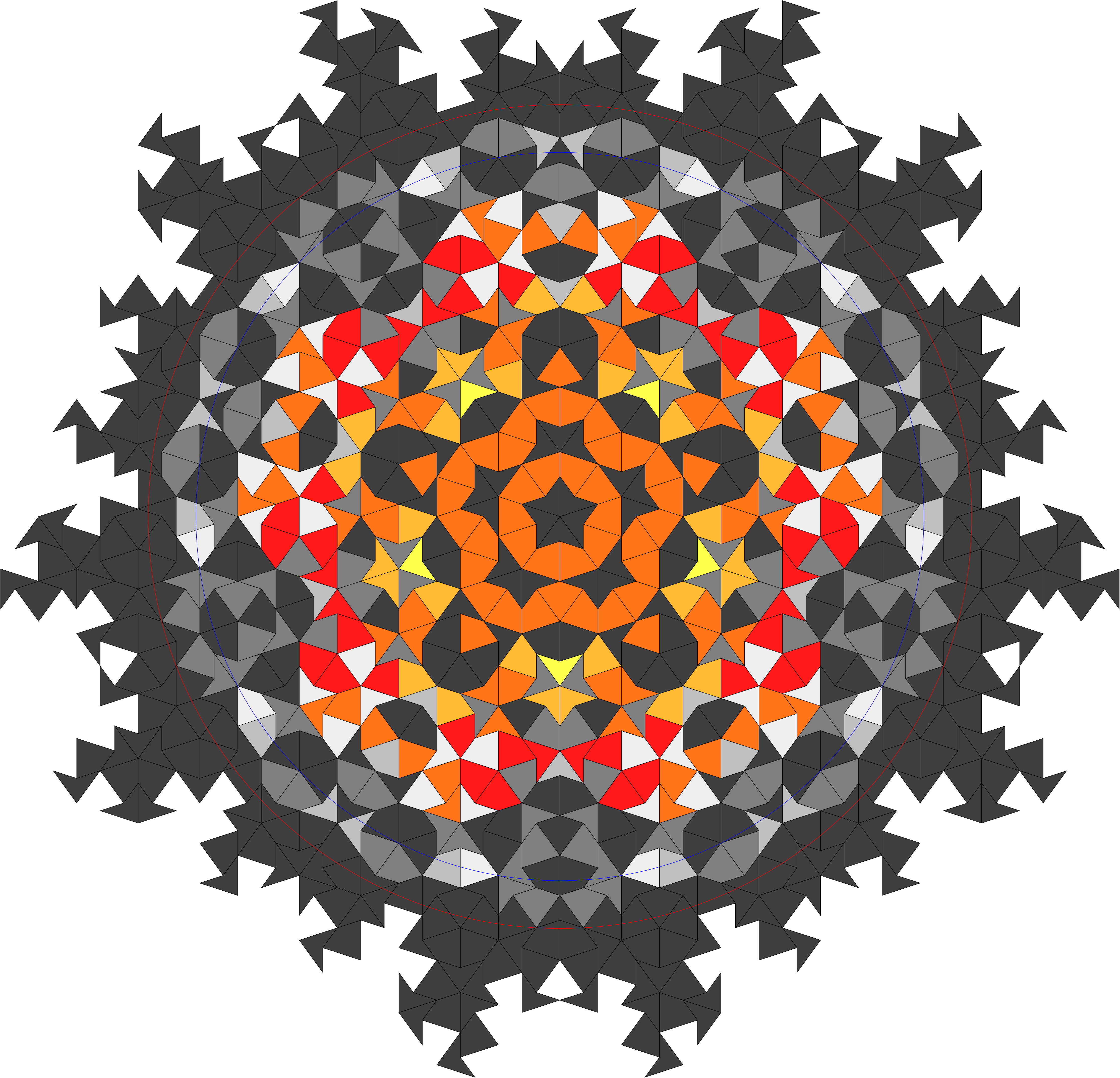}
  }
  \caption{
    Examples of roundness measures, circles of
    outer radius $\outerRadius{c}$ in blue,
    inner radius $\innerRadius{c}$ in red.
    Left: square grid of side length $50$, after $500$ steps from $(m+e)$ (converges to $m$ in $707$ steps), $\roundness{c} \approx 21.633 - 20.224 = 1.409$.
    Right: $5$ iterations of the substition from a P2 Sun, after $150$ steps from $(m+e)$ (converges to $m$ in $220$ steps), $\roundness{c} \approx 16.662 - 14.729 = 1.933$.
  }
  \label{fig:roundness}
\end{figure}

In order to measure this phenomenon, we introduce the {\em roundness} as follows.
First, we partition a configuration into two parts:
the outside part with all tiles having $\degree{v}-1$ grains connected to the border,
and the inner part.
Given a tiling $G=(V,E)$ with sink $s$
and a configuration $c:\tilde{V}\to\N$,
let the {\em maximum stable components},
$\MSC{c}= \{ V_1, V_2, \dots, V_k \}$,
be the connected components of tiles (from $\tilde{V}$) having $\degree{v}-1$ grains.
Then the {\em outer tiles} is the set
\[
  \outerTiles{c}=\bigcup\limits_{\substack{
    V_i \in \MSC{c}\\
    \exists v \in V_i: \{v,s\} \in E
  }} V_i
\]
and the {\em inner tiles} is the set $\innerTiles{c}=\tilde{V}\setminus\outerTiles{c}$.
From this partition of the set of tiles, we are interested in the
frontier between the outer and inner tiles, and how close it is
from a perfect circle.
This has to do with the coordinates of tiles in the Euclidean space $\R^2$,
so let us denote $\coordinates{v}$ the set of coordinates of the bounds
of some tile $v \in \tilde{V}$,
and $\surface{v}$ the subset of $\R^2$ covered by the tile.
Our convention regarding the sink $s$ is discussed below.
Regarding the circle, let us denote $\circleRadius{r}$ the inside of the circle
of radius $r \in \R$ centered at coordinate $(0,0)$, {\em i.e.}
\[
  \circleRadius{r} = \{ (x,y) \in \R^2 \mid \sqrt{x^2+y^2} \leq r \}.
\]
We may therefore denote 
$\surface{v} \cap \circleRadius{r} = \emptyset$
to state that tile $v$ is entirely outside the circle of radius $r$
centered at $(0,0)$, and
$\surface{v} \subseteq \circleRadius{r}$
(or equivalently $\coordinates{v} \subseteq \circleRadius{r}$
since we deal with polygons and the circle is convex)
to state that tile $v$ is entirely inside the same circle.
We define the {\em outer radius} as the maximum scalar $r$ such
that all outer tiles are outside the circle of radius $r$ around
the origin $(0,0)$ of the tiling,
\[
  \outerRadius{c}=\max\{ r \in \R_+ \mid
    \forall v \in \outerTiles{c} :
    \surface{v} \cap \circleRadius{r} = \emptyset
  \}
\]
and the {\em inner radius} as the minimum scalar $r$ such that all
inner tiles are inside the circle of radius $r$ around the origin
$(0,0)$ of the tiling,
\[
  \innerRadius{c}=\min\{ r \in \R_+ \mid 
    \forall v \in \innerTiles{c} :
    \surface{v} \subseteq \circleRadius{r}
  \}.
\]
In order to deal with the case $\outerTiles{c}=\emptyset$,
which may for example be the case on some configurations $(m+e)$,
we add the convention that the sink $s$ is an infinite tile covering
all the space outside the tiling, whose coordinates are the union of
all bounds from tile edges adjacent to the sink,
and that it always belongs to $\outerTiles{c}$.
As a consequence the outer radius is upper bounded by
the radius of the inscribed circle (inside the finite tiling)
with center at $(0,0)$.
The case $\innerTiles{c}=\emptyset$ is not problematic since the minimum
defining $\innerRadius{c}$ is taken on $\R_+$,
and would therefore equal $0$ as expected.
The inner radius is upper bounded by
the radius of the circumscribed circle (outside the finite tiling)
with center at $(0,0)$.
The {\em roundness of a configuration} $c$ is then measured
as the difference between the inner and outer radii,
\[
  \roundness{c}=\innerRadius{c}-\outerRadius{c}.
\]
Note that $0 \leq \roundness{c}$,
and we have $\roundness{c}=0$ when the frontier
between inner and outer tiles is a perfect circle.
Two examples of roundness are given on Figure~\ref{fig:roundness}.

\subsection{Base roundnesses}
\label{ss:base}

\begin{figure}
  \centerline{
    \includegraphics[scale=.75]{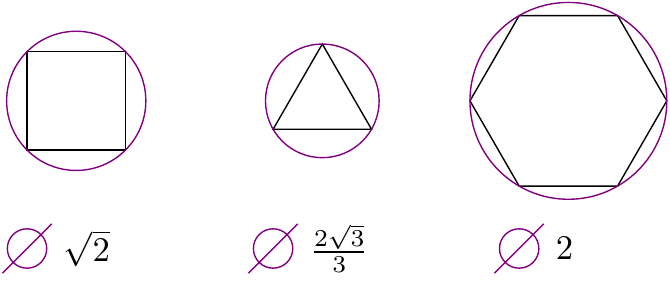} \qquad
    \includegraphics[scale=.75]{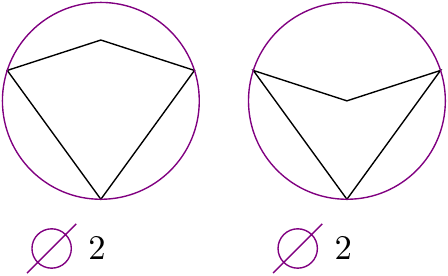} \qquad
    \includegraphics[scale=.75]{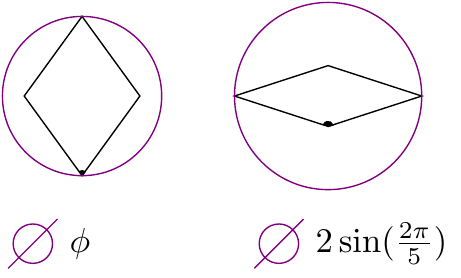}
  }
  \caption{
    Diameters of all tiles, as the diameter of a circumscribed circle (in purple).
    All tiles at the same scale.
  }
  \label{fig:diameters}
\end{figure}

Remark that since all our tiles are polygonal (with three to six sides),
we cannot expect to reach roundness $0$
(except when the stabilization process has converged to $m$).
To get some easy to interpret base values, we consider
the diameter of each tile, 
as the diameter of the circumscribed circle around the tile
(smallest radius of a circle having the tile entirely is its interior).
See Figure~\ref{fig:diameters}.
The greatest diameter of some tiling's tiles can be interpreted as
an upper bound on the best achievable roundness, for any radius.
Indeed, consider some tiling and a circle $C$
(of radius at most equal to the inscribed radius), then
all tiles having the center of their circumscribed circle
on or outside the circle $C$ can be set as outer tiles, and
all tiles having the center of their circumscribed circle
inside the circle $C$ can be set as inner tiles,
which results in a roundness
smaller than twice the radius of the greatest tile's radius,
{\em i.e.} smaller than the greatest tile's diameter.
We obtain the {\em base roundnesses} presented on Table~\ref{table:diameter}.

\begin{table}
  \centerline{
    \begin{tabular}{|c|c|c|c|c|c|}
      \hline
      Tiling &
      \mytwolines{Square}{grids} &
      \mytwolines{Triangular}{grids} &
      \mytwolines{Hexagonal}{grids} &
      P2 tilings &
      P3 tilings
      \\
      \hline
      \rule{0pt}{1.2em} 
      Base roundness &
      $\underset{\approx 1.414}{\sqrt{2}}$ & 
      $\underset{\approx 1.155}{\frac{2\sqrt{3}}{3}}$ & 
      $2$ & 
      $2$ & 
      $\underset{\approx 1.902}{2\sin(\frac{2\pi}{5})}$ 
      \rule[-1.2em]{0pt}{0pt} 
      \\
      \hline
    \end{tabular}
  }
  \caption{Base roundnesses as the greatest diameter of some tiling's tiles.}
  \label{table:diameter}
\end{table}

\subsection{Plots}
\label{ss:plots}

\begin{figure}
  \includegraphics[width=\textwidth]{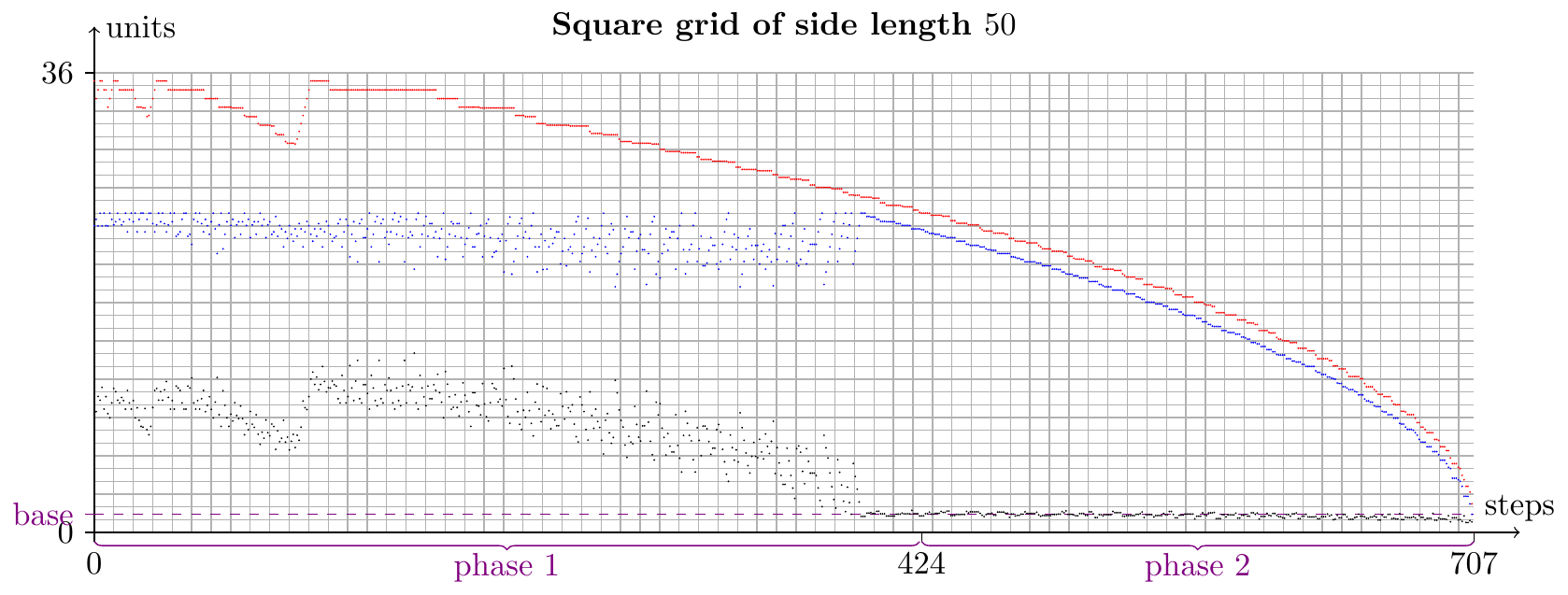}
  \caption{
    Plot of the roundness during the stabilization process $\stable{(m+e)}=m$,
    on the square grid of side length $50$.
    At each step we plot the
    inner radius $\innerRadius{c}$ in red,
    outer radius $\outerRadius{c}$ in blue, and
    roundness $\roundness{c}$ in black.
    Grid has one row per unit and one column per $10$ time steps.
    For example, at step $3$, with $m+e \transitionsync c^1 \transitionsync c^2 \transitionsync c^3$,
    we observe that $\innerRadius{c^3} = 25\sqrt{2} \approx 35.355$
    (the radius of the circumscribed circle around the whole tiling)
    and that $\outerRadius{c^3} = 25$
    (the radius of the inscribed circle inside the whole tiling),
    so that $\roundness{c^3} = 25(\sqrt{2}-1) \approx 10.355$
    (at step $0$ some outer tiles near the $x=0$ and $y=0$ axis give different radii).
    Phase 2 begins at step $424$, and configuration $m$ is reached at step $707$.
    The base roundness of this tiling (dashed) is $\sqrt{2}$.
    See also the companion Figure~\ref{fig:roundness-squaregrid-50-phases}.
  }
  \label{fig:roundness-squaregrid-50-full}
  \vspace*{1em}
  \centerline{$
    \vcenter{\hbox{\includegraphics[scale=.08]{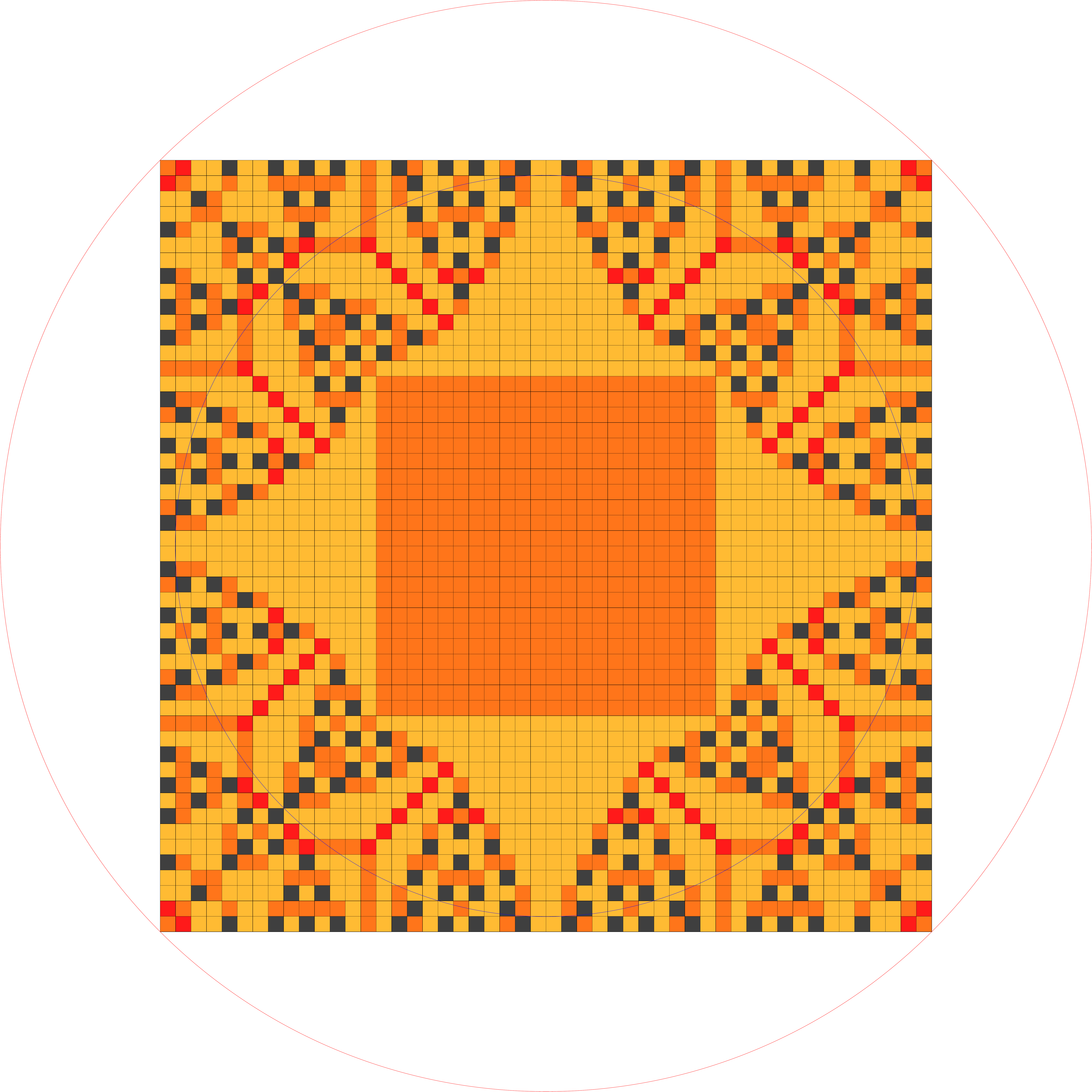}}}\quad
    \vcenter{\hbox{\includegraphics[scale=.08]{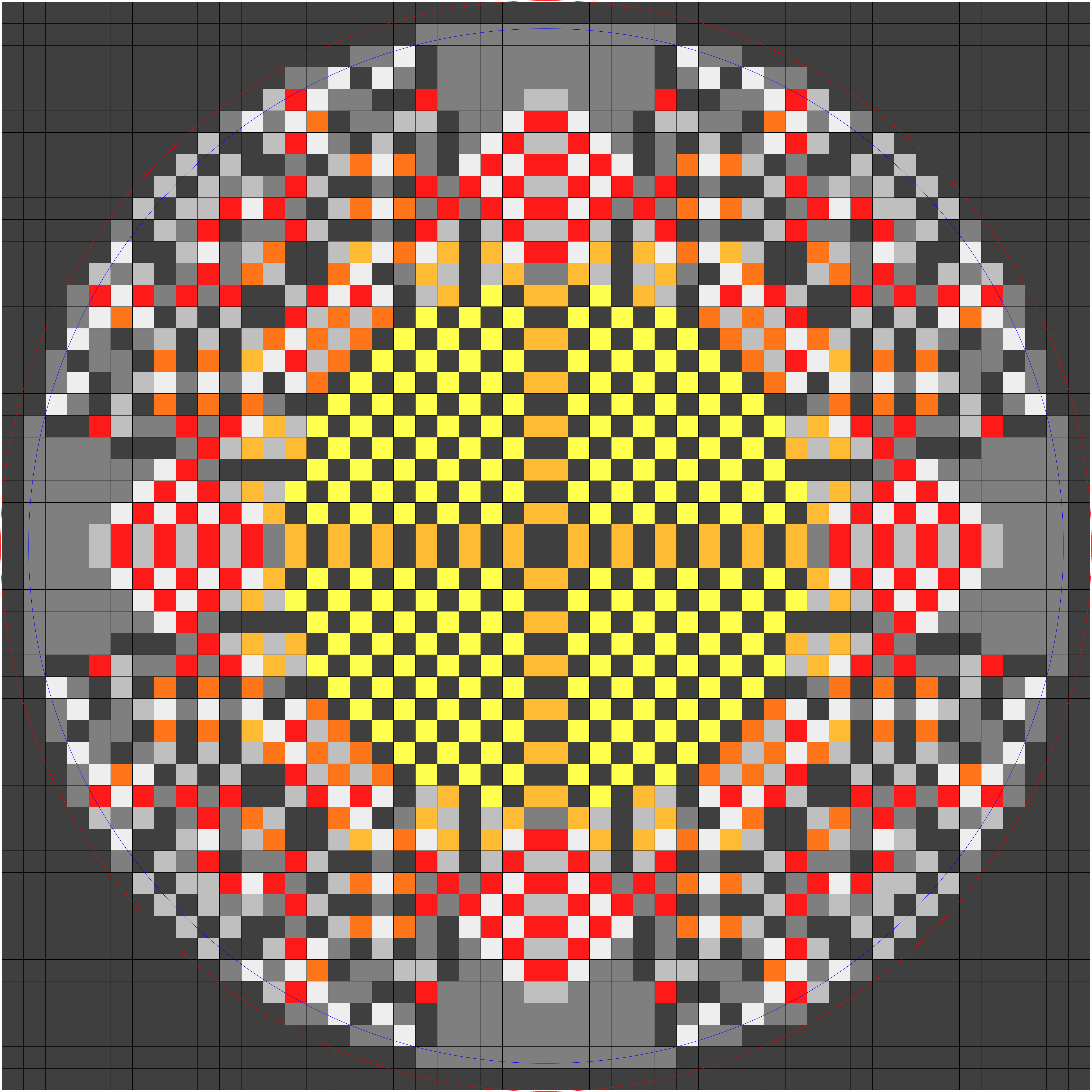}}}\quad
    \vcenter{\hbox{\includegraphics[scale=.08]{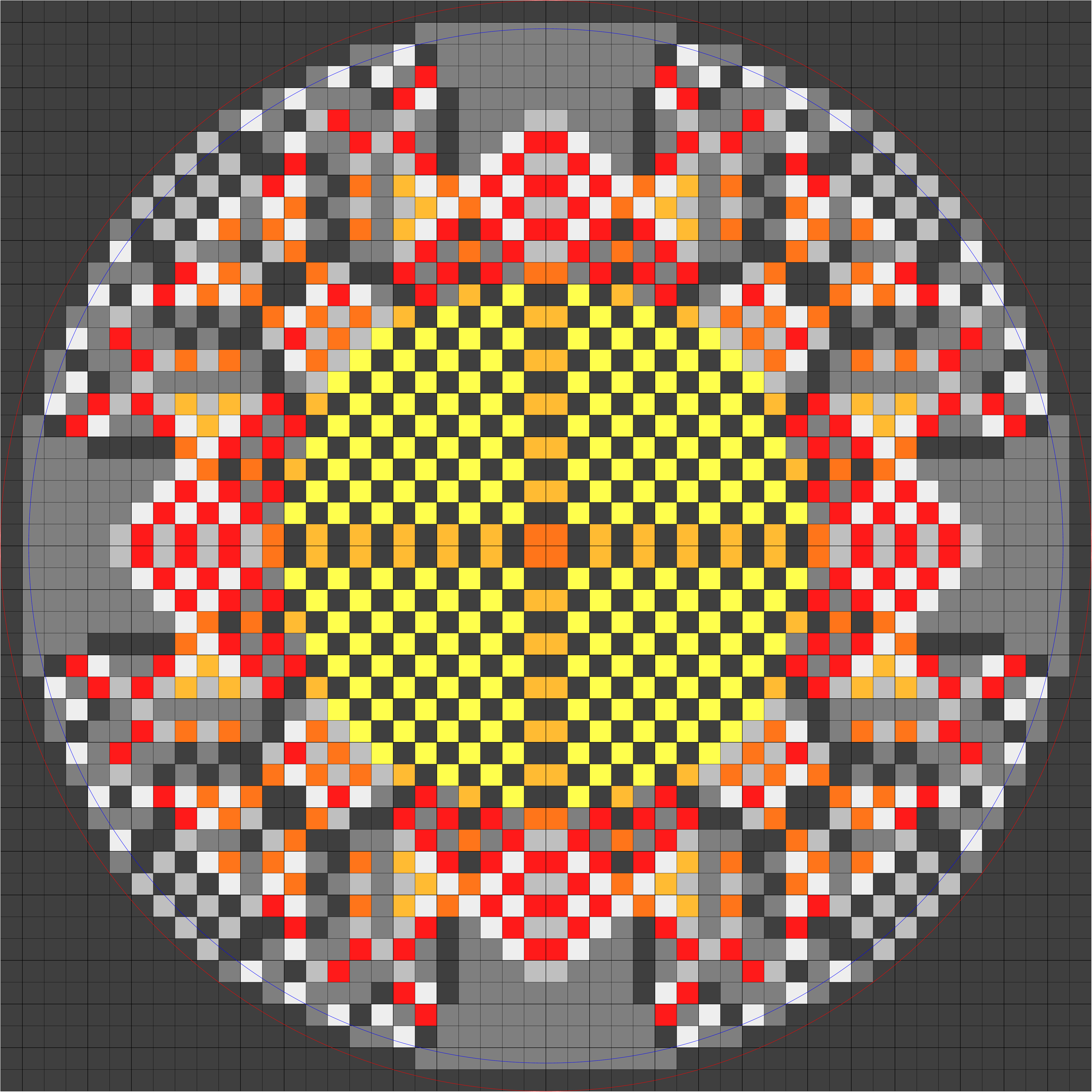}}}
  $}
  \caption{
    Configurations at steps $0$ (this is $m+e$), $423$ and $424$ during
    the stabilization process $\stable{(m+e)}=m$,
    on the square grid of side length $50$.
    Inner radii in red, outer radii in blue.
    One can observe the phase transition occurring at step $424$:
    the inner radius becomes smaller or equal to the inscribed radius of the tiling.
  }
  \label{fig:roundness-squaregrid-50-phases}
\end{figure}

\begin{figure}
  \includegraphics[width=\textwidth]{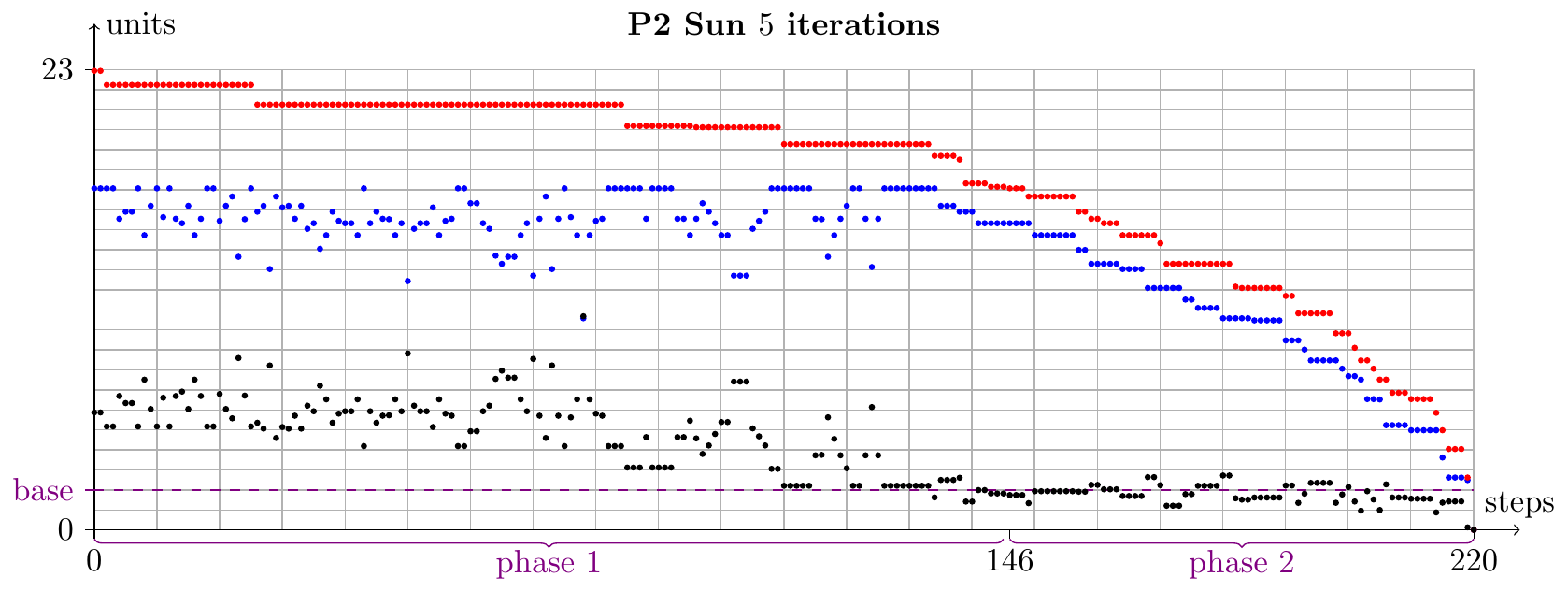}
  \caption{
    Plot of the roundness during the stabilization process $\stable{(m+e)}=m$,
    on the tiling obtained after $5$ iterations of the substitution from a P2 Sun.
    At each step we plot the
    inner radius $\innerRadius{c}$ in red,
    outer radius $\outerRadius{c}$ in blue, and
    roundness $\roundness{c}$ in black.
    Grid has one row per unit and one column per $10$ time steps.
    For example, at step $0$ we observe that $\innerRadius{m+e} \approx 22.940$
    (the radius of the circumscribed circle around the whole tiling)
    and that $\outerRadius{m+e} \approx 17.069$
    (the radius of the inscribed circle inside the whole tiling),
    so that $\roundness{m+e} \approx 5.871$.
    Phase 2 begins at step $146$, and configuration $m$ is reached at step $220$.
    The base roundness of this tiling (dashed) is $2$.
    See also the companion Figure~\ref{fig:roundness-P2-sun-5-phases}.
  }
  \label{fig:roundness-P2-sun-5-full}
  \vspace*{1em}
  \centerline{$
    \vcenter{\hbox{\includegraphics[scale=.1]{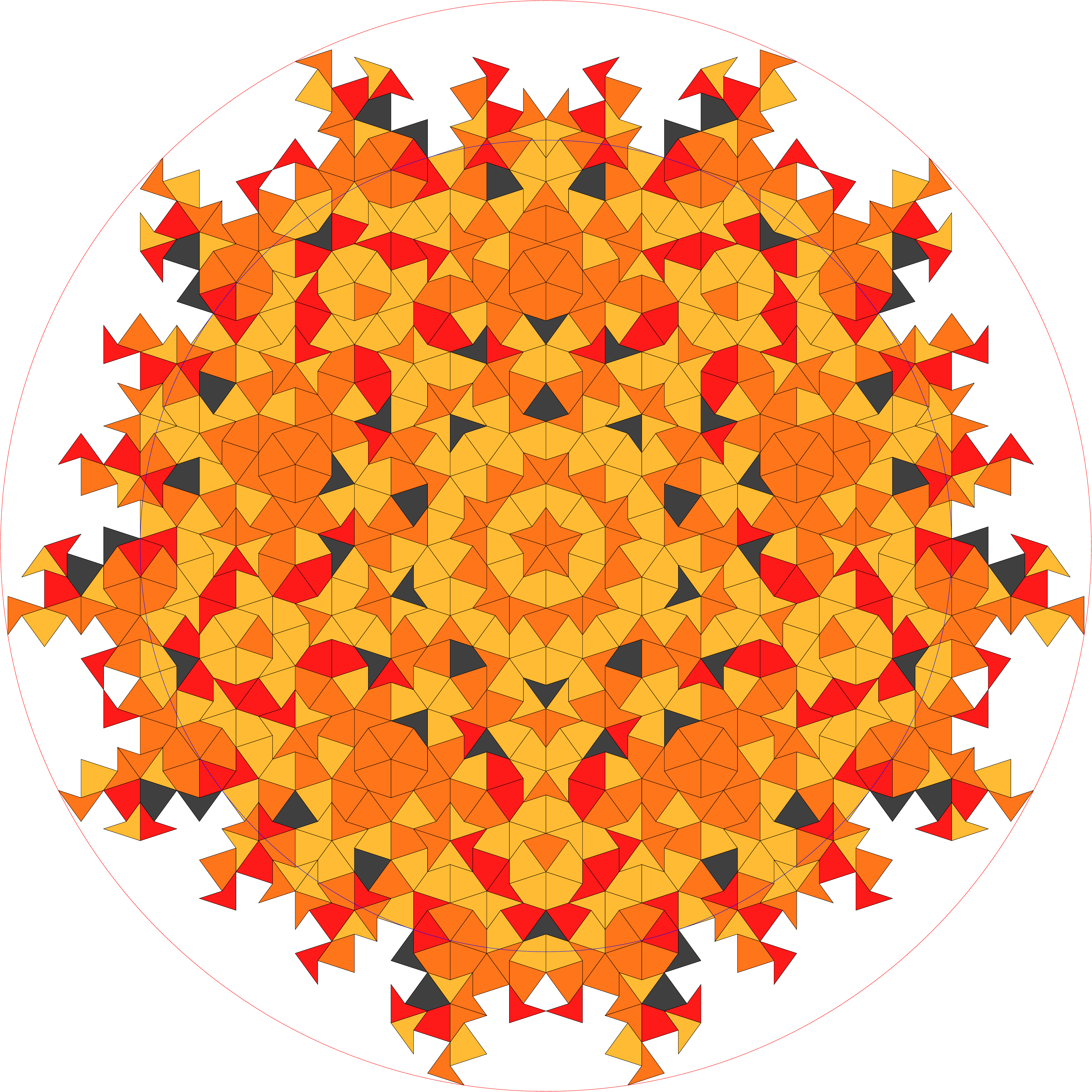}}}\quad
    \vcenter{\hbox{\includegraphics[scale=.1]{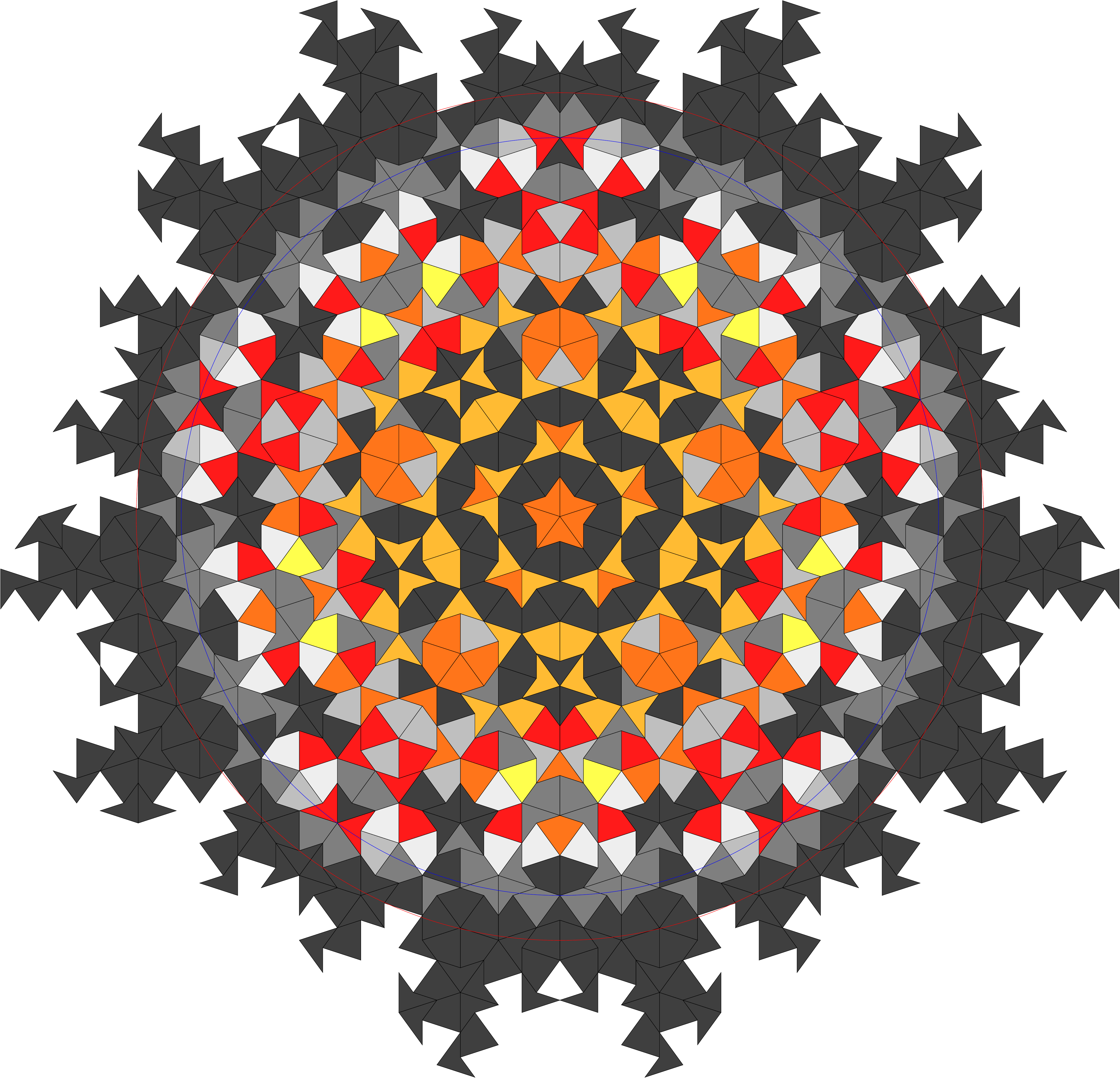}}}\quad
    \vcenter{\hbox{\includegraphics[scale=.1]{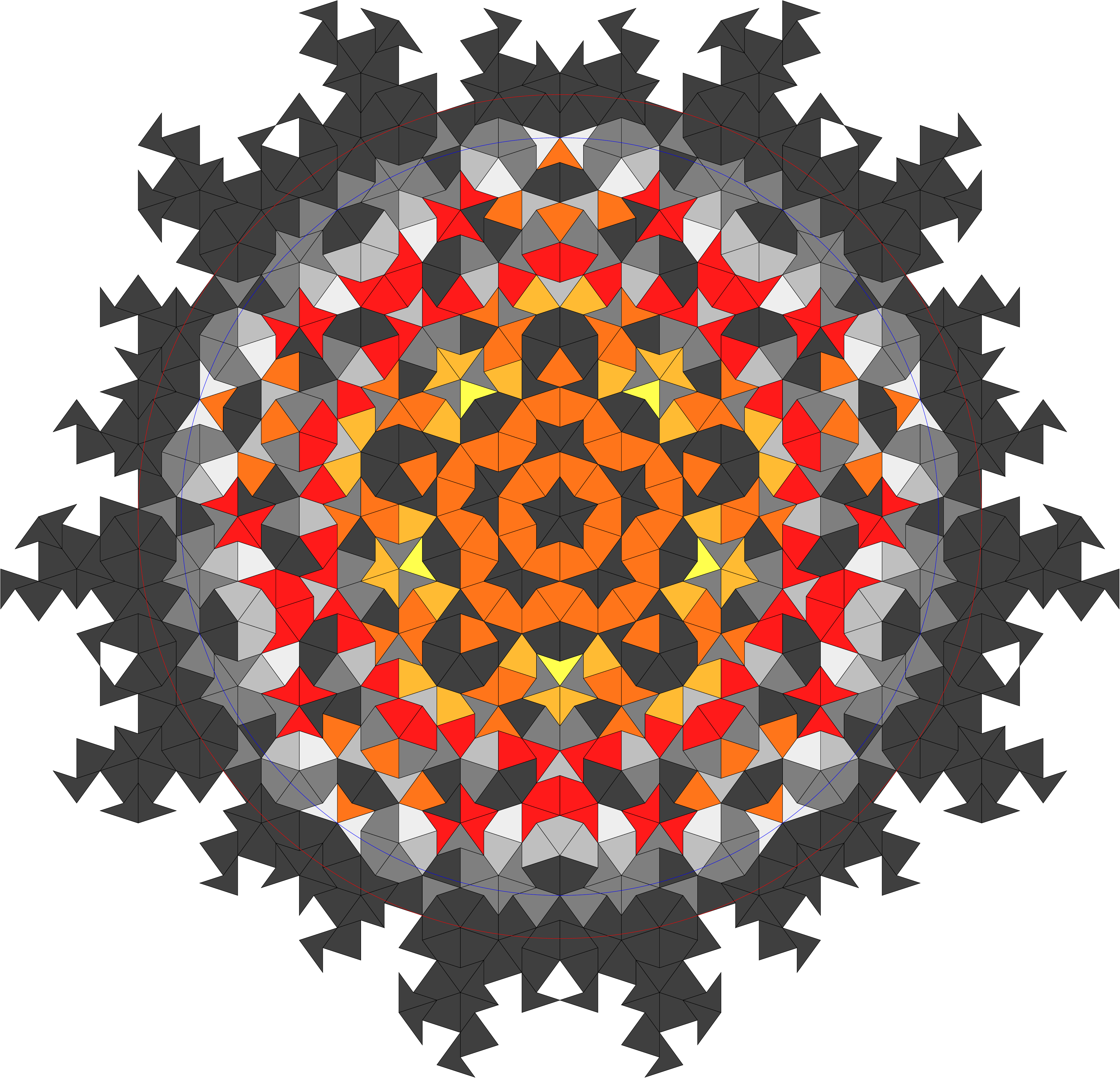}}}
  $}
  \caption{
    Configurations at steps $0$ (this is $m+e$), $145$ and $146$ during
    the stabilization process $\stable{(m+e)}=m$,
    on the tiling obtained after $5$ iterations of the substitution from a P2 Sun.
    Inner radii in red, outer radii in blue.
    One can observe the phase transition occurring at step $146$:
    the inner radius becomes smaller or equal to the inscribed radius of the tiling.
  }
  \label{fig:roundness-P2-sun-5-phases}
\end{figure}

We now present plots of roundness measured during the stabilization process
$\stable{(m+e)}=m$ on the different tilings considered in this article.
We decompose the stabilization process from $m+e$ to $m$ into two phases:
\begin{itemize}[topsep=.5em]
  \item {\bf phase 1:} the dynamics is erratic,
  \item {\bf phase 2:} the set of inner tiles slowly shrinks,
    until reaching $\innerTiles{m}=\emptyset$.
\end{itemize}
The {\em beginning of phase 2} is defined as the first step such that
the inner radius is smaller or equal to the inscribed radius of the tiling
(maximum radius of a circle entirely inside the finite tiling,
and centered at the origin).
Remark that at the beginning of phase 2,
all tiles on the border of the tiling are outer tiles,
because the polygonal shape of any inner tile on the border would otherwise
lead to the inner radius being greater than the inscribed radius of the tiling.
An important observation is that, in all the experiments presented in this article
and performed during its preparation,
once in phase 2 with all border tiles as outer tiles,
then all border tiles {\em remain} outer tiles\footnote{They
all remain stable with $\degree{v}-1$ grains until reaching $m$.
Observe that any outer tile receiving some grain would topple,
and that toppling any outer tile would result in toppling the whole
maximum stable component it belongs to.},
and that the inner radius {\em remains} smaller or equal to the inscribed radius.
Two full examples of roundness plots are given on
Figures~\ref{fig:roundness-squaregrid-50-full}, \ref{fig:roundness-P2-sun-5-full},
and their companion
Figures~\ref{fig:roundness-squaregrid-50-phases}, \ref{fig:roundness-P2-sun-5-phases}.

\begin{figure}
  \includegraphics[width=\textwidth]{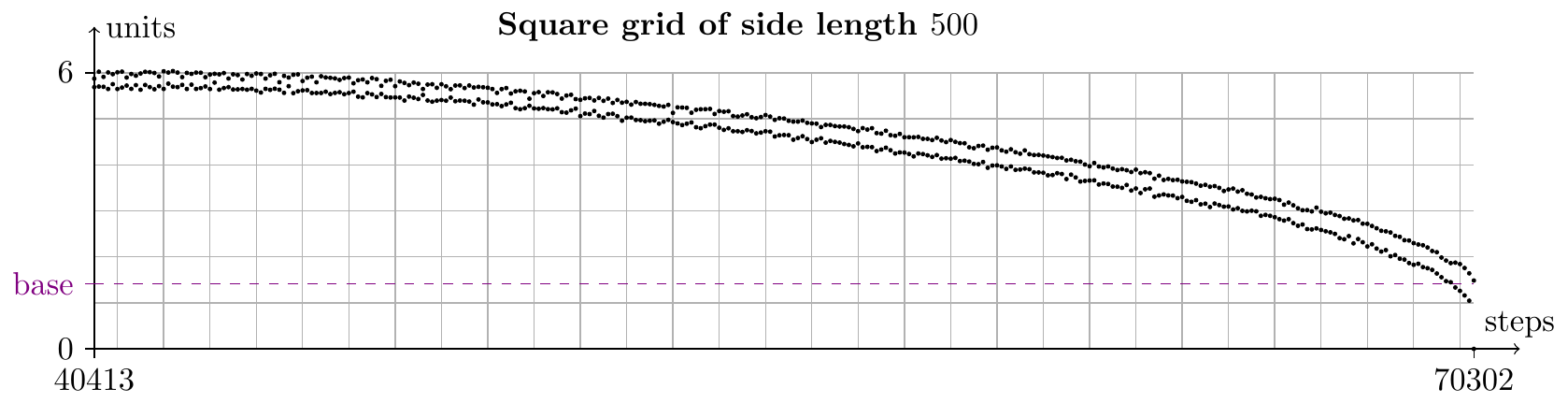}
  \includegraphics[width=\textwidth]{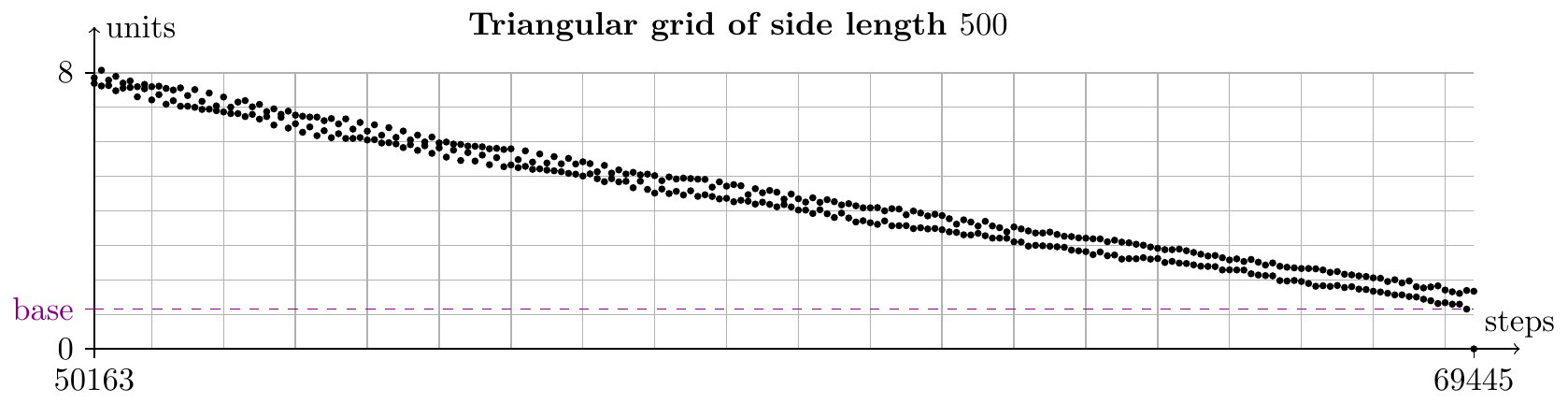}
  \includegraphics[width=\textwidth]{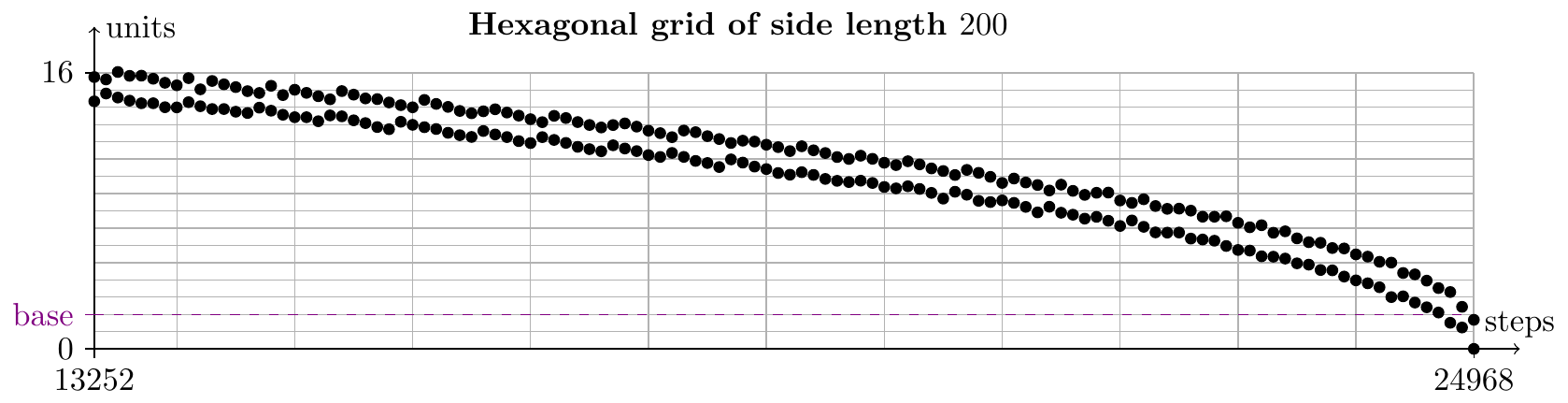}
  \caption{
    Plots of the roundness during phase 2 of the stabilization
    process $\stable{(m+e)}=m$, on the
    square grid of side length $500$,
    triangular grid of side length $500$, and
    hexagonal grid of side length $200$.
    Grids have one row per unit and one column per $1000$ time steps.
    Two points are drawn every $100$ time steps: the maximum and
    minimum roundness values $\roundness{c}$ observed among these
    $100$ steps.
  }
  \label{fig:xp1}
\end{figure}

\begin{figure}
  \includegraphics[width=\textwidth]{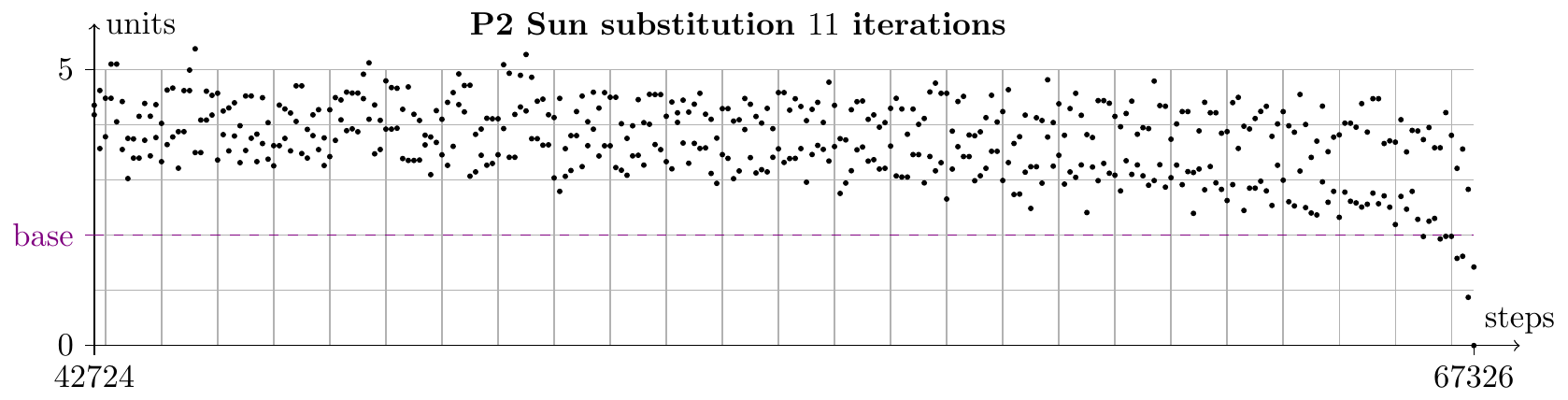}
  \includegraphics[width=\textwidth]{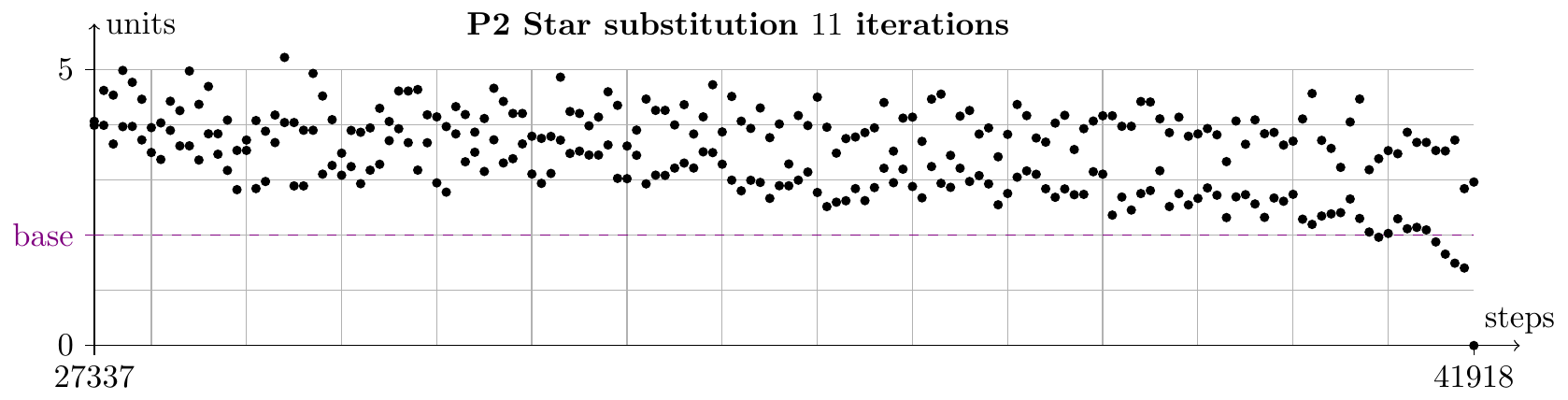}
  \includegraphics[width=\textwidth]{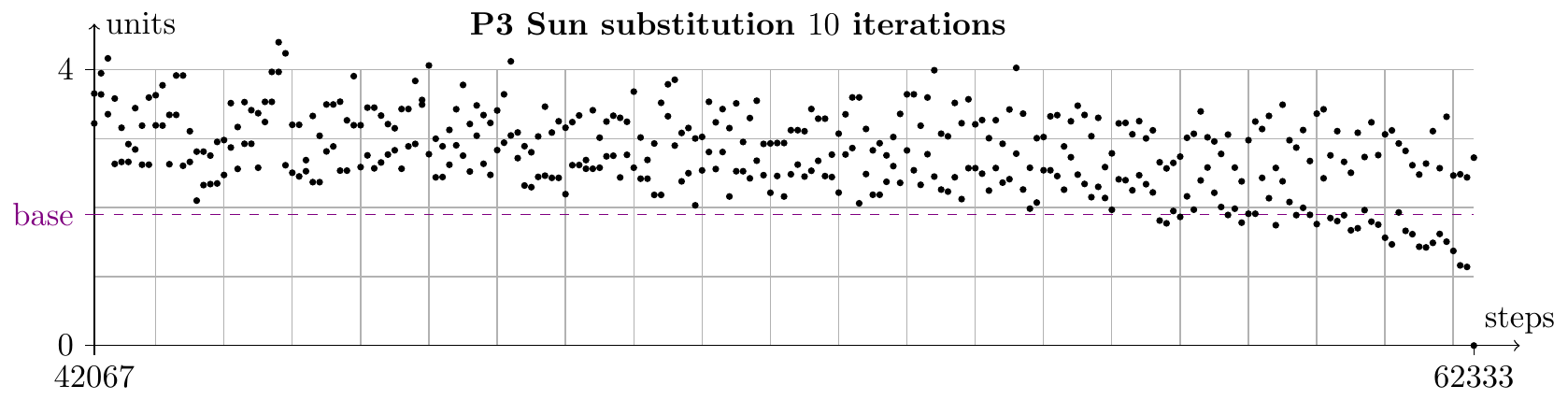}
  \includegraphics[width=\textwidth]{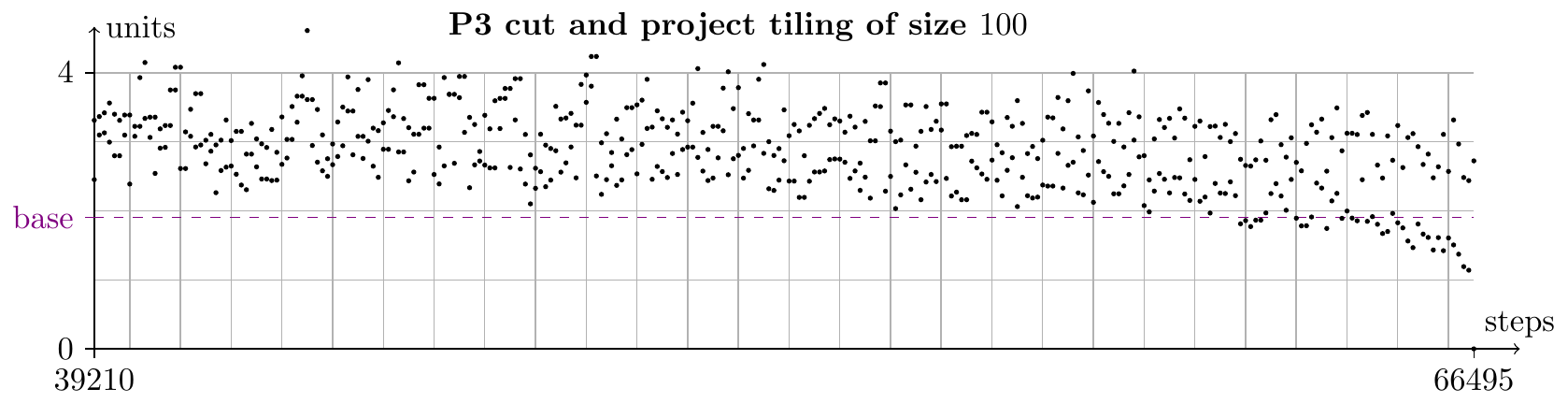}
  \caption{
    Plots of the roundness during phase 2 of the stabilization
    process $\stable{(m+e)}=m$, on the
    P2 Sun after $11$ iterations of the substitution,
    P2 Star after $11$ iterations of the substitution,
    P3 Sun after $10$ iterations of the substitution,
    P3 cut and project tiling of size $100$.
    Grids have one row per unit and one column per $1000$ time steps.
    Two points are drawn every $100$ time steps: the maximum and
    minimum roundness values $\roundness{c}$ observed among these
    $100$ steps.
  }
  \label{fig:xp2}
\end{figure}

Experimental results, picturing only phase 2 of the stabilization process
from $m+e$ to $m$, are presented on Figure~\ref{fig:xp1} and~\ref{fig:xp2}.
Note that during all the experiments we have performed in preparing this article,
we have observed similar behaviors for all other sizes.

\pagebreak

\subsection{Computation times and data}

The graphics were computed on our personal machines
(standard laptops), with simulation times and data
presented on Table~\ref{table:time}.
Details on the implementation (no parallelization) can be found at\\
\centerline{\url{https://github.com/huacayacauh/JS-Sandpile/wiki/Roundness}.}

\begin{table}
  \centerline{
    \begin{tabular}{|c|c|c|c|c|c|c|}
      \hline
      Tiling &
      \mytwolines{Size/}{iter.} &
      Tiles &
      \mytwolines{Identity}{(seconds)} &
      \mytwolines{Roundness}{(seconds)} &
      \mytwolines{Phase 2}{begin step} &
      \mytwolines{Stabilization}{step}
      \\
      \hline
      \hline
      \mytwolines{Square}{grid} &
      $500$ &
      $250000$ &
      $825$ &
      $1294$ &
      $40413$ &
      $70302$
      \\
      \hline
      \mytwolines{Triangular}{grid} &
      $500$ &
      $250000$ &
      $1143$ &
      $689$ &
      $50163$ &
      $69445$
      \\
      \hline
      \mytwolines{Hexagonal}{grid} &
      $200$ &
      $120601$ &
      $271$ &
      $405$ &
      $13252$ &
      $24968$
      \\
      \hline
      \mytwolines{P2 Sun}{subst.} &
      $11$ &
      $327750$ &
      $2931$ &
      $1850$ &
      $42724$ &
      $67326$
      \\
      \hline
      \mytwolines{P2 Star}{subst.} &
      $11$ &
      $234410$ &
      $944$ &
      $640$ &
      $27337$ &
      $41918$
      \\
      \hline
      \mytwolines{P3 Sun}{subst.} &
      $10$ &
      $266860$ &
      $2227$ &
      $1570$ &
      $42067$ &
      $62333$
      \\
      \hline
      \mytwolines{P3 cut}{\& project} &
      $100$ &
      $249610$ &
      $3145$ &
      $2386$ &
      $39210$ &
      $66495$
      \\
      \hline
    \end{tabular}
  }
  \caption{
    Data regarding the computations generating the graphics presented
    on Figures~\ref{fig:xp1} and~\ref{fig:xp2}.
    Running times are given for the computation of identities
    (from Formula~\ref{eq:id}) and roundness measures.
  }
  \label{table:time}
\end{table}

\subsection{Analysis of roundness measures}

Well, it appears clearly that the circular shapes observed during the stabilization
process $\stable{(m+e)}=m$ are not asymptotically
approaching perfect circles.
Indeed, in all experiments conducted on large tilings
(Figures~\ref{fig:xp1} and~\ref{fig:xp2}),
at the beginning of phase 2
the roundness is above the base roundness,
whereas the base roundness is an upper bound on the best achievable
roundness (see Subsection~\ref{ss:base}).

\begin{figure}
  \centerline{\includegraphics{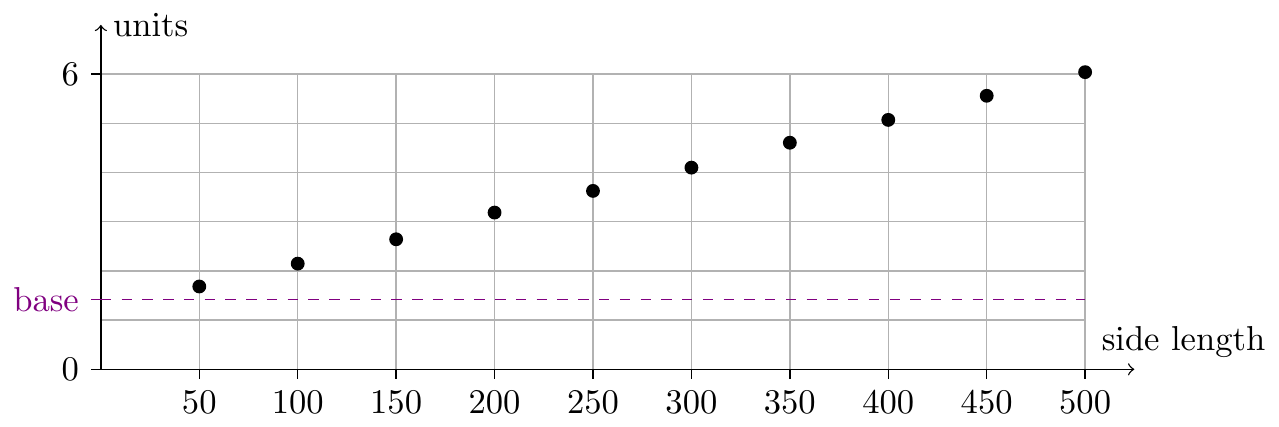}}
  \caption{
    Maximum roundness $\roundness{c}$ measured during phase 2 of
    the stabilization process $\stable{(m+e)}=m$ on square grids
    of various side lengths,
    illustrating the correlation between roundness and size.
  }
  \label{fig:roundness-square-max}
\end{figure}

As the circular shapes shrink, it is normal to see the roundness decrease until
the value $0$ on configuration $m$
(all plots reach the minimum roundness $0$ at stabilization step).
On the other hand, it appears that roundness and tiling size are
correlated, meaning that as the size of the tiling increases, configurations at
the beginning of phase 2 are less round.
This is illustrated on Figure~\ref{fig:roundness-square-max}.

\begin{figure}
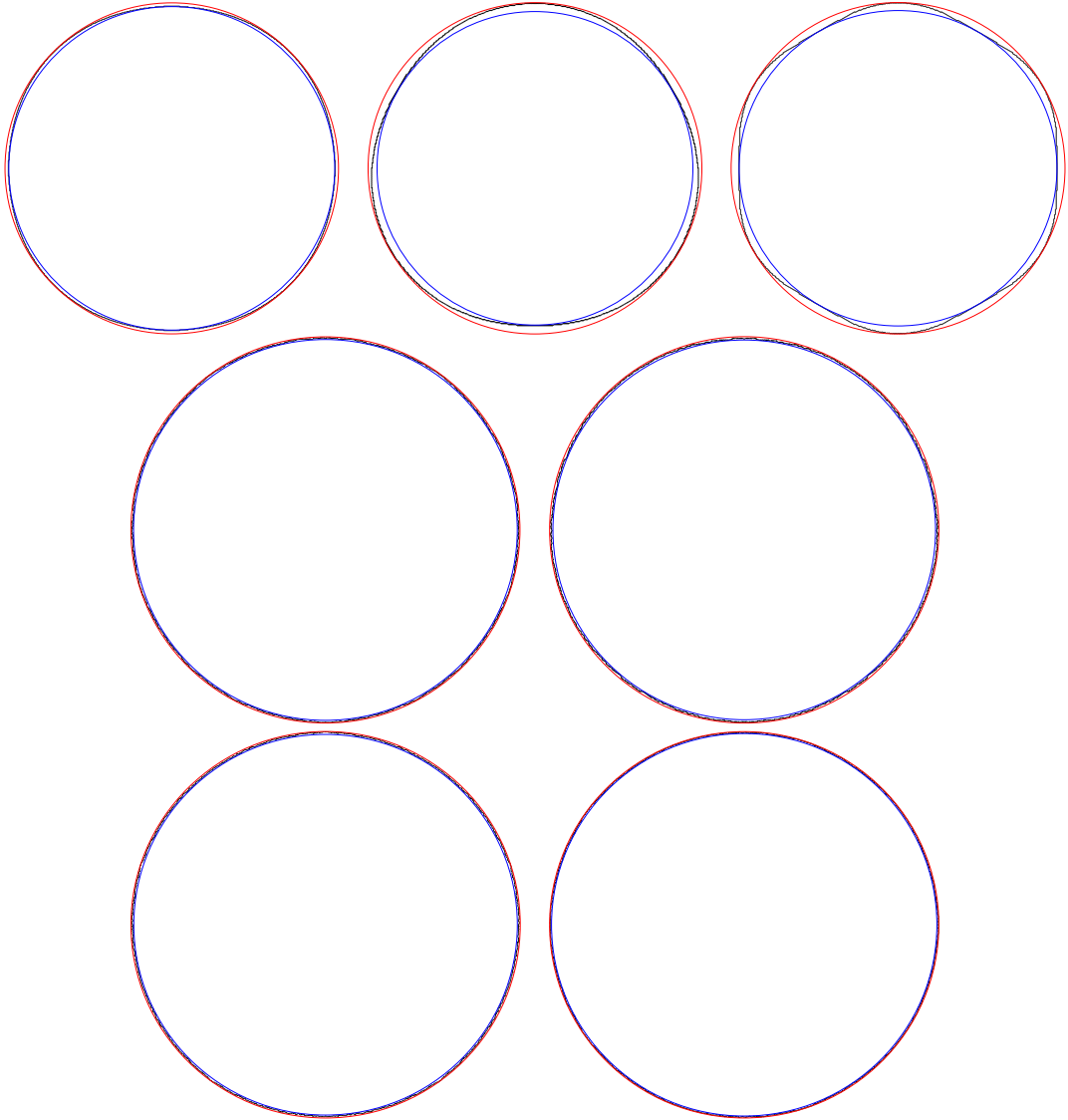

  \centerline{$
    \vcenter{\hbox{\includegraphics[width=.3\textwidth]{pics-squaregrid-500-frontier.pdf}}}\quad
    \vcenter{\hbox{\includegraphics[width=.3\textwidth]{pics-triangulargrid-500-frontier.pdf}}}\quad
    \vcenter{\hbox{\includegraphics[width=.3\textwidth]{pics-hexagonalgrid-200-frontier.pdf}}}
  $}
  \centerline{$
    \vcenter{\hbox{\includegraphics[width=.35\textwidth]{pics-P2-sun-11-frontier.pdf}}}\quad
    \vcenter{\hbox{\includegraphics[width=.35\textwidth]{pics-P2-star-11-frontier.pdf}}}
  $}
  \vspace*{.2em}
  \centerline{$
    \vcenter{\hbox{\includegraphics[width=.35\textwidth]{pics-P3-sun-10-frontier.pdf}}}\quad
    \vcenter{\hbox{\includegraphics[width=.35\textwidth]{pics-P3-cutandproject-100-frontier.pdf}}}
  $}
  \caption{
    Frontier between inner and outer tiles on the configuration
    obtained at the beginning of phase 2,
    as the set of edges shared by one inner tile and one outer tile.
    Outer radius in blue, inner radius in red.
    Top:
    square grid of side length $500$ at step $40413$
    ($\roundness{c}\approx 5.687$), 
    triangular grid of side length $500$ at step $50163$
    ($\roundness{c}\approx 7.755$), 
    hexagonal grid of side length $200$ at step $13252$
    ($\roundness{c}\approx 14.470$). 
    Middle:
    P2 Sun after $11$ iterations of the subsitution at step $42724$
    ($\roundness{c}\approx 4.246$), 
    P2 Star after $11$ iterations of the subsitution at step $27337$
    ($\roundness{c}\approx 4.064$). 
    Bottom:
    P3 Sun after $10$ iterations of the subsitution at step $42067$
    ($\roundness{c}\approx 3.220$), 
    P3 Sun cut and project tiling of size $100$ at step $39210$
    ($\roundness{c}\approx 2.452$). 
  }
  \label{fig:frontier}
\end{figure}

Although they do not tend to perfect circles, we may admit that these shapes
are close to be, in regard of tiling sizes (Table~\ref{table:time}).
Increases of roundness are visible to the naked eye on hexagonal and triangular grids
which deviate largely from a perfect circle, but are hardly noticeable without
measurements on other tilings.
Figure~\ref{fig:frontier} illustrates this with the frontier between outer and
inner tiles, on the configuration obtained at the beginning of phase 2 during
the experiments from Subsection~\ref{ss:plots} (since these configurations are
quite large, we picture only the frontier in order to reduce the numerical weight
of the present document).
If they are not perfect circles, then we may naturally ask: what characterize these frontier shapes for each tiling?

\begin{figure}
  \centerline{
    \includegraphics{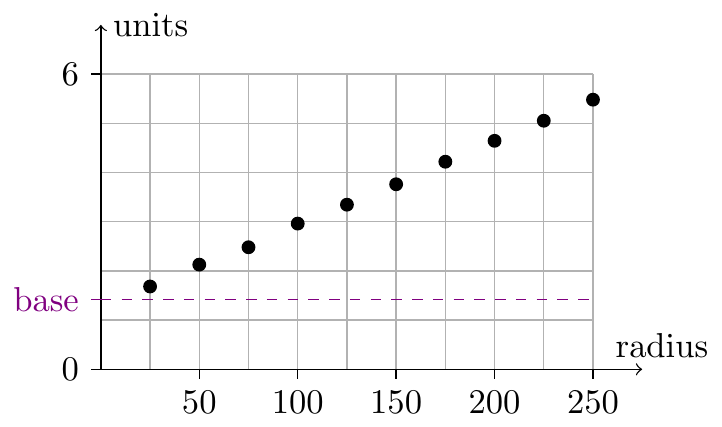}
    \includegraphics[width=6cm]{pics-squaregridcrop-250-frontier.pdf}
  }
  \caption{
    Left: maximum roundness $\roundness{c}$ measured during phase 2 of
    the stabilization process $\stable{(m+e)}=m$ on square grids
    cropped to circles of various radii.
    Right: frontier between inner and outer tiles on the configuration
    obtained at the beginning of phase 2 (step $32131$,
    $\roundness{c} \approx 4.073$)
    on a square grid cropped to a circle of radius $250$
    (the border of the tiling is pictured in green).
  }
  \label{fig:squarecrop}
\end{figure}

Finally, we see on Figure~\ref{fig:frontier} (first line)
that the frontier on grids 
(respectively square, triangular and hexagonal)
reflect the shape on the border of the tilings, somehow ``rounded''.
Indeed, it appears that the number of corners of the grids are
equal to the number of parts where the frontier deviates from a circle
(the number of times it goes from the inner radius to the outer radius).
Theses symmetries are expected, they come from the symmetry of the grids
and therefore the symmetries of the dynamics.
A natural attempt would be to experiment the roundness of a square grid
cropped to a circle, in order to remove the effect of the border's shape.
Despite the fact that the difference between the circumscribed
and inscribed radius is smaller than the base roundness,
this does not lead to significantly smaller roundness measurements
(frontiers are not closer to perfect circles)
during phase 2 of the stabilization process $\stable{(m+e)}=m$,
as shown on Figure~\ref{fig:squarecrop}.
It feels that the frontier reflects the anisotropy of the square
grid itself rather than that of its border's shape.

\section{Conclusions and perspectives}

The experiments presented in this article are reproducible with
{\em JS-Sandpile} (links in Preamble). The software implements
no parallelization mechanism, which would allow to perform
larger simulations ({\em e.g.} using GPU). Nevertheless we believe that
this would not lead to qualitatively different observations.

We have presented some identity elements on Penrose tilings,
revealing no obvious structure related to these famous aperiodic
tilings.
Identities are highly sensitive to the shape of the tiling,
and it may be the case that other
finite croppings of (infinite) Penrose tilings
lead to different observations.
We tried to build the most ``natural'' finite Penrose tilings:
Suns and Stars obtained by substitution,
along with cut and project from a $5$-dimensional hypercube.

The apparent isotropy, observed during the stabilization of
the maximum stable configuration plus the identity element
of the sandpile group, has been measured through the notion
of roundness. Experiments revealed that these frontiers are actually
not approaching perfect circles on any tiling under consideration.
Two further directions may be investigated.
First, if these shapes are not perfect circles,
then how to characterize them on each tiling
(especially on grids)?
Second, which tilings would lead to frontiers approaching
perfect circles (if any)?
The modest attempt to crop a square grid to a circle fails,
suggesting that this may not be easy to achieve on tilings which
are intrinsically anisotropic at the tile level.

Penrose tilings exhibit frontiers closer to perfect circles,
though they also deviate significantly from
their best achievable roundness.
It is not very surprising to observe on Penrose tilings
a behavior similar to regular lattices, as they are quasi-periodic,
and in some sense the ``most regular'' aperiodic tilings.
Let us open the large perspective of considering other tilings,
for example tilings with higher quasi-periodicity functions
(above affine~\cite{bj10,cd04,d99}\footnote{The
difficulty may be to find constructions from non Wang tiles,
because Wang tiles would lead to square grids for the sandpile model
to play on (as we remove tile decorations).}),
Cayley graphs~\cite{r94},
or hyperbolic planes ({\em e.g.} with Poincar\'e disk models).

Finally, could the stabilization process $\stable{(m+e)}=m$
shed some light on the enigmatic identity elements on grids?
This is really out of reach for our present knowledge,
but we hope that the recent progresses of Levine
{\em et al.}~\cite{lps16,lps17,lp17,ps13,ps20}
are breaking some scientific locks in the domain of sandpiles.


\section*{Acknowledgments}

The authors are thankful
to Valentin Darrigo for his contributions to {\em JS-Sandpile},
to Victor Poupet for stressing
that the apparent isotropy of the $\stable{(m+e)}$ process
is surprising at the occasion of a talk given by KP
during AUTOMATA'2014 in Himeji,
to Thomas Fernique for sharing his expertise (and code!)
regarding the cut and project method, and
to Christophe Papazian for useful comments on
quasi-periodicity functions.

The work of JF was conducted while a Master student at Aix-Marseille Université, doing an internship at the LIS laboratory (UMR 7020), both in Marseille, France.
The work of KP was funded mainly by his salary as a French
State agent and therefore by French taxpayers' taxes,
affiliated to
Aix-Marseille Univ., Univ. de Toulon, CNRS, LIS, UMR 7020, Marseille, France
and
Univ. C\^{o}te d'Azur, CNRS, I3S, UMR 7271, Sophia Antipolis, France.
Secondary financial support came from
ANR-18-CE40-0002 FANs project,
ECOS-Sud C16E01 project, and
STIC AmSud CoDANet 19-STIC-03 (Campus France 43478PD) project.

\bibliographystyle{plain}
\bibliography{biblio}

\end{document}